\documentclass[12pt]{article}
\usepackage{amssymb,amsmath}
\usepackage{graphicx}
\usepackage{epsfig}
\usepackage{epstopdf}
\usepackage{latexsym}
\usepackage{psfrag}
\usepackage{booktabs}
\usepackage{bbm}
\usepackage[toc]{appendix}
\usepackage{color}
\usepackage{datetime}
\usepackage[
      colorlinks=true,
      linkcolor=darkblue,  
      urlcolor=blue,    
      filecolor=blue,     
      citecolor=red,
      linktocpage=true,
      pdfstartview=FitV,
      bookmarksopen=true    
      ]{hyperref}

\definecolor{darkblue}{rgb}{0.2, 0, 0.8}

\numberwithin{equation}{section}

\newcommand{\ds}{\displaystyle}

\newcommand{\s}{\sigma}

\newcommand{\im}{{\text{i}}}

\newcommand{\HH}{\mathbb{H}}

\newcommand{\abs}[1]{\left\lvert #1 \right\rvert}

\newcommand{\Z}{\mathbb{Z}}

\newcommand{\R}{\mathbb{R}}

\newcommand{\es}[2] {\begin{equation} \label{#1} \begin{split} #2 \end{split} \end{equation}}

\usepackage{cite,dsfont}
\usepackage{amsfonts}
\usepackage{amssymb}
\usepackage{amsmath}
\usepackage{graphicx}
\usepackage{slashed}
\usepackage{setspace}

\usepackage{color}

\newcommand{\reef}[1]{(\ref{#1})}

\newcommand{\be}{\begin{equation}}
\newcommand{\ee}{\end{equation}}

\def\be{\begin{equation}}
\def\ee{\end{equation}}
\def\bea{\begin{eqnarray}}
\def\eea{\end{eqnarray}}
\def\ba{\begin{array}}
\def\ea{\end{array}}
\def\bd{\begin{displaymath}}
\def\ed{\end{displaymath}}

\def\Tr{{\rm Tr}}
\def\tr{{\rm tr}}

\def\b{\beta}

\def\g{\gamma}

\def\s{\sigma}                                   

\def\pa{\partial}                              
\def\>{\rangle} 
\def\<{\langle} 
\def\Dsl{D \hskip-.6em \raise1pt\hbox{$ / $ } }
\def\to{\rightarrow}
\def\pa{\partial}

\newcommand{\eps}{\epsilon}
\newcommand{\lra}{\leftrightarrow}

\def\tz{\tilde{z}}

\def\mL{\mathcal{L}}

\def\pa{\partial}

\def\g5{\gamma_5}

\def\ep{\epsilon}

\def\lam{\lambda}
\def\lamt{\tilde{\lambda}}
\def\chit{\tilde{\chi}}
\def\Zt{\tilde{Z}}
\def\ept{\tilde{\ep}}
\def\sigmab{\bar{\sigma}}
\def\Ft{\tilde{F}}

\def\mt{\tilde{m}}

\def\s2{\sigma_2}

\def\mN{\mathcal{N}}

\newcommand{\zt}{\tilde{z}}
\newcommand{\als}[1]{\begin{align} \begin{split} #1 \end{split} \end{align}}

\begin{document}  


\begin{titlepage}

\begin{flushright}
\texttt{PUPT-2502}\\
\texttt{MCTP-16-12}
\end{flushright}

\vspace*{2cm}

\begin{center}
{\LARGE \bf Holography for ${\cal N}=1^*$ on $S^4$} \\

\vspace*{1.2cm}

{\bf Nikolay Bobev$^{1}$, Henriette Elvang$^{2}$, Uri Kol$^{2}$, \\[2mm] Timothy Olson$^{2}$, and Silviu S. Pufu$^{3}$}
\medskip

$^{1}$ Instituut voor Theoretische Fysica, K.U. Leuven \\
Celestijnenlaan 200D, B-3001 Leuven, Belgium
\bigskip

$^{2}$ Michigan Center for Theoretical Physics, \\
Randall Laboratory of Physics, Department of Physics,\\
University of Michigan, Ann Arbor, MI 48109, USA
\bigskip

$^{3}$ Joseph Henry Laboratories, Princeton University\\
Princeton, NJ 08544, USA
\bigskip

\bigskip
\texttt{nikolay@itf.fys.kuleuven.be, elvang@umich.edu, urikol@umich.edu, timolson@umich.edu, spufu@princeton.edu} \\
\end{center}

\vspace*{0.1cm}

\begin{abstract}  
\noindent  We construct the five-dimensional supergravity dual of the $\mathcal{N}=1^*$ mass deformation of the $\mathcal{N} =4$ supersymmetric Yang-Mills theory on $S^4$ and use it to calculate the universal contribution to the corresponding $S^4$ free energy at large 't Hooft coupling in the planar limit. The holographic RG flow solutions  are smooth and preserve four supercharges.  
As a novel feature compared to the holographic duals of $\mathcal{N} = 1^*$ on $\mathbb{R}^4$, in our backgrounds the five-dimensional dilaton has a non-trivial profile, and the gaugino condensate is fixed in terms of the mass-deformation parameters.  Important aspects of the analysis involve characterizing the ambiguities in the partition function of non-conformal $\mathcal{N}=1$ supersymmetric theories on $S^4$ as well as the action of S-duality on the $\mathcal{N}=1^*$ theory. 
\end{abstract}

\end{titlepage}

\setcounter{tocdepth}{2}
{\small
\setlength\parskip{-0.5mm}
\tableofcontents
}

\newpage

\section{Introduction}
\label{sec:Introduction}

The gauge/gravity duality \cite{Maldacena:1997re, Witten:1998qj, Gubser:1998bc} has provided many insights into the physics of strongly-coupled field theories.  While originally proposed as a correspondence between conformal field theories and anti-de Sitter vacua of string/M theory, the duality was soon extended to non-conformal situations, one of the main motivations being to gain a better understanding of confinement in four-dimensional gauge theories -- see, for instance, \cite{Witten:1998zw, Klebanov:2000hb, Polchinski:2000uf}.  A rather simple setup which exhibits confinement is a generic ${\cal N} = 1$-preserving mass deformation of ${\cal N} = 4$ supersymmetric Yang-Mills theory.  In relation to holography, this theory, known as ${\cal N} = 1^*$, was studied in~\cite{Girardello:1999bd,Polchinski:2000uf,Pilch:2000fu,Aharony:2000nt,Dorey:2000fc,Dorey:2001qj}.

Our goal here is to gain insights into the physics of the ${\cal N} = 1^*$ theory at strong coupling by placing it on a four-dimensional sphere with an Einstein metric and constructing its holographic dual.  The interest in studying supersymmetric field theories on curved manifolds is rather recent and stems from the fact that, in certain situations, one can calculate explicitly the partition function, as well as other BPS observables, using the technique of supersymmetric localization (see, for instance, \cite{Pestun:2007rz,Kapustin:2009kz,Jafferis:2012iv,Benini:2012ui,Doroud:2012xw} for a few of the recent localization results in various dimensions).  These results then can be further related to properties of the theory on flat $\R^{1,3}$.  For the ${\cal N} = 1^*$ theory on $S^4$, however, there are currently no known supersymmetric localization results.  At strong coupling, the gauge/gravity duality remains our only tool.  This provides the main motivation for the present work.

The existence of the $\mathcal{N}=1^{*}$ theory can be most easily inferred by viewing the field content of $\mathcal{N}=4$ SYM as an $\mathcal{N}=1$ vector multiplet\footnote{In this paper we will mostly take the gauge group to be $SU(N)$ and will often omit writing explicitly the gauge indices of various operators.} and three $\mathcal{N}=1$ chiral multiplets, $\Phi_i$, transforming in the adjoint representation of the gauge group and interacting through a certain cubic superpotential.  On $\R^4$, the $\mathcal{N}=1^{*}$ theory is then obtained by giving masses $m_1$, $m_2$, and $m_3$ to the chiral multiplets via superpotential mass terms. A well-known feature of Euclidean field theories is that the chiral fields $\Phi_i$ and their Lorentzian conjugates $\tilde{\Phi}_i$ should be treated as independent in Euclidean signature and not as complex conjugates of each other. Therefore, one can introduce different independent mass parameters $m_i$ and $\tilde{m}_i$ for the chiral multiplets and their conjugates, respectively. The superpotential mass terms induce the usual scalar and fermion mass terms for the components of the superfields $\Phi_i$ and $\tilde \Phi_i$. To study the $\mathcal{N}=1^*$ theory on an $S^4$ of radius $a$, one has to be careful when coupling the theory to curvature while preserving supersymmetry.  A systematic way to do so was presented in \cite{Pestun:2007rz,Festuccia:2011ws}. An important consequence of the supersymmetric coupling to curvature is that the action on $S^4$ contains new bilinear terms in the lowest components, $Z_i$ and $\tilde Z_i$, of the chiral superfields that are of the form $\im m_i Z_i^2/a$ and $\im \tilde{m}_i\tilde Z_i^2/a$.  Our main interest  here is 
 to compute the partition function (or free energy) of the $\mathcal{N}=1^*$ theory on $S^4$ as a function of the dimensionless parameters $\{m_ia,\tilde{m}_ia\}$.

One may question whether a supersymmetric observable, like the partition function, computed on a curved manifold is free of scheme dependence. Indeed, for instance in the case of superconformal $\mathcal{N}=1$ theories in four dimensions it was shown in \cite{Gerchkovitz:2014gta} that the partition function on $S^4$, seen as a function of the exactly marginal couplings, is completely scheme dependent. The situation is different for ${\cal N} = 2$ superconformal theories where the $S^4$ partition function can be expressed in terms of the K\"ahler potential of the Zamolodchikov metric and thus contains physically interesting information \cite{Gerchkovitz:2014gta} (see also \cite{Papadodimas:2009eu,Baggio:2014sna,Baggio:2014ioa}). As a precursor to understanding the physics of the ${\cal N} = 1^*$ theory on $S^4$, we thus have to understand the possible finite local 
counterterms that are compatible with supersymmetry. We present this analysis in Section \ref{sec:FE} where we show that the possible supersymmetric finite local 
counterterms can be at most cubic in the dimensionless masses $m_i a$. One can in addition utilize the $SL(2,\mathbb{Z})$ S-duality of the $\mathcal{N}=4$ SYM theory to argue that the coefficient of the cubic counterterm should be zero. The upshot of this analysis is that taking three derivatives of the $S^4$ free energy in the $\mathcal{N}=1^*$ theory with respect to the dimensionless mass yields a scheme independent result.

To study the partition function of the $\mathcal{N}=1^*$ theory we will employ holography. The justification for this approach is twofold. On one hand the study here is a natural continuation of our work  \cite{Bobev:2013cja} on holography for the $\mathcal{N}=2^{*}$ theory on $S^4$. This in turn was motivated by the explicit results available in the large $N$ limit of this theory using supersymmetric localization \cite{Russo:2012ay,Buchel:2013id,Russo:2013qaa,Russo:2013kea,Russo:2013sba,Buchel:2013fpa,Chen:2014vka}.\footnote{See also \cite{Russo:2012kj,Russo:2014nka,Chen-Lin:2015dfa,Chen-Lin:2015xlh,Zarembo:2014ooa,Karch:2015vra,Karch:2015kfa} for recent related work.} These authors  found that the free energy of $\mathcal{N}=2^{*}$ on $S^4$ at large $N$ and large t'Hooft coupling $\lambda = g_\text{YM}^2 N$ is 
\be
  \label{N2freeE}
  F_{\mathcal{N}=2^{*}} = -\frac{N^2}{2} (1+m^2 a^2) \log \big[\lambda (1+m^2 a^2) C\big] \,,
\ee
where $m$ is the mass-deformation parameter, $a$ is the radius of the $S^4$, and
$C= e^{2\gamma+1/2}/(16\pi^2)$ is a non-universal constant that depends on the choice of the UV regularization scheme. In  \cite{Bobev:2013cja} it was shown that the universal part of the free energy \reef{N2freeE} of ${\cal N} = 2^*$ can be reproduced precisely in holography.  In addition to this success in the case of the ${\cal N} = 2^*$ theory on $S^4$, further motivation for using holography to study the ${\cal N} = 1^*$ on $S^4$ comes from the well-known results on holography for the $\mathcal{N}=1^*$ theory on flat space \cite{Girardello:1999bd,Pilch:2000fu,Polchinski:2000uf}.  In particular, in \cite{Polchinski:2000uf}  the rich vacuum structure of the theory was studied holographically and many of the known field theory features discussed in \cite{Vafa:1994tf,Donagi:1995cf,Dorey:1999sj,Dorey:2000fc,Aharony:2000nt} were qualitatively and quantitatively reproduced.  

To construct the gravitational dual of the $\mathcal{N}=1^*$ theory on $S^4$ we utilize the five-dimensional maximal $SO(6)$ gauged supergravity \cite{Gunaydin:1984qu,Gunaydin:1985cu,Pernici:1985ju}. This approach offers a considerable technical simplification compared to a direct search for supersymmetric solutions of type IIB supergravity. The construction is nevertheless complicated.  It turns out that in order to capture the most general $\mathcal{N}=1^*$ theory on $S^4$ with arbitrary masses, $m_i$ and $\tilde{m}_i$, one has to work with a consistent truncation of the maximal supergravity theory to an $\mathcal{N}=2$ gauged supergravity with two vector and four hyper multiplets. This truncated model contains 18 scalar fields, and so constructing explicit solutions in it is still technically challenging. To render the problem more manageable, we observe that massive deformations with $\tilde{m}_i=m_i$ are captured by a simpler supergravity model with only 10 real scalars, which we discuss in great detail.  We focus on three subsectors of this 10-scalar model that capture the holographic dual of three different limits of the $\mathcal{N}=1^*$ theory:
\begin{itemize}
 \item $m_1 = m_2 \ne 0 ~~\text{and}~~ m_3 = 0$: in this limit, the supersymmetry is enhanced to $\mathcal{N}=2$ and the theory is  $\mathcal{N}=2^{*}$. The holographic dual of this theory on flat space was previously constructed by Pilch and Warner \cite{Pilch:2000ue}. The dual of the theory on $S^4$ was studied in \cite{Bobev:2013cja}.
\item $m_1 = m_2 = 0 ~~\text{and}~~ m_3 \ne 0$: we refer to this limit as the one-mass model. In flat space this theory flows to a Leigh-Strassler fixed point \cite{Leigh:1995ep} in the infrared. The flat sliced holographic dual was studied by Freedman-Gubser-Pilch-Warner (FGPW) \cite{Freedman:1999gp}.  See also \cite{Girardello:1998pd,Pilch:2000fu}.
\item $m_1 = m_2 = m_3 \ne 0$: We refer to this limit of the $\mathcal{N}=1^{*}$ theory as the equal-mass model. In flat space, the five-dimensional holographic dual was constructed by 
Girardello-Petrini-Porrati-Zaffaroni (GPPZ) \cite{Girardello:1999bd}.  See also \cite{Pilch:2000fu}. 
\end{itemize}
The holographic  RG flows we present here for the duals of the one-mass and equal-mass $\mathcal{N}=1^*$ theories on $S^4$ have not previously appeared in the literature. 

For all these three models we can construct explicit supersymmetric solutions of the equations of motion of the maximal supergravity theory.  These solutions are smooth geometries with $S^4$ radial slices and radial profiles for the scalar fields. They are completely regular and cap off smoothly in the IR\@. Upon careful holographic renormalization we can extract the finite part of the on-shell action of the maximal supergravity. Via the standard holographic dictionary the on-shell action should be identified with the free energy of the $\mathcal{N}=1^*$ theory. There are two notable aspects of this construction that are not 
present for the theory on flat space. The first is that the radius of $S^4$ provides an infrared cutoff for the dynamics of the field theory. On the supergravity side this IR cutoff is manifested in the smooth cap-off of the solutions in the IR\@. This is a notable difference from the singular solutions constructed for the equal mass model in flat space in GPPZ 
\cite{Girardello:1999bd}. Such singular solutions are typical for holographic RG flows and present a challenge in understanding which supergravity singularities describe acceptable IR physics \cite{Gubser:2000nd}. This problem is alleviated on $S^4$, so in some sense our holographic RG flows are better behaved and easier to analyze than similar solutions with flat slicing.  The second notable aspect of our holographic construction is that there is an important subtlety in the holographic renormalization procedure, namely the possible appearance of finite local 
counterterms.  Indeed, it turns out that for the solutions of interest here, there are such finite counterterms needed and their coefficients are fixed by supersymmetry. To determine them, we use the ``Bogomolnyi trick" recently employed in other similar holographic solutions \cite{Freedman:2013ryh,Bobev:2013cja}.  It would be interesting, however, to have a better understanding of these counterterms from more basic principles, and we leave such a discussion for future work.

It is curious to highlight an interesting feature of the supergravity solution dual to the equal mass model on $S^4$. We find that this solution necessarily involves a non-trivial radial profile of the five-dimensional dilaton. To the best of our knowledge, this feature has not been observed before in holographic RG flows in the five-dimensional maximal supergravity. In addition in this model, we find that the vacuum expectation value (vev) for the gaugino bilinear operator is non-zero and it is fixed by the IR smoothness conditions. This situation is to be contrasted with the singular flat-sliced solutions in \cite{Girardello:1999bd}, where the gaugino bilinear vev was a free parameter.

It should be emphasized from the outset that our supergravity backgrounds are obtained numerically.  Unfortunately, for the general $\mathcal{N}=1^*$ theory we will not be able to provide an explicit analytic formula for the free energy as a function of the masses similar to the one in \eqref{N2freeE}. Nevertheless, from our numerical solutions in the three limits outlined above we are able to extract certain 
results about the general features and some of the analytic structure of the function $F_{S^4}(m_i,\tilde{m}_i)$ for general values of $m_i$ and $\tilde{m}_i$.

We begin in the next section with a general discussion of the  $\mathcal{N}=1^*$ gauge theory on $S^4$. 
In Section \ref{sec:FE} we analyze in detail how the supersymmetric partition function of $\mathcal{N}=1$ theories can depend on the relevant couplings and identify the possible finite counterterms which may lead to scheme ambiguities. We then turn to the construction of the supergravity dual of $\mathcal{N}=1^*$ on $S^4$ in Section \ref{sec:SG}. Following a general broad discussion of the supergravity solutions, in Section \ref{sec:num} we present detailed analytic and numerical results for the one-mass and  equal-mass limits of  $\mathcal{N}=1^*$. In addition we also summarize the implications of our results for the general $\mathcal{N}=1^*$ theory with arbitrary mass parameters. Readers in a hurry can immediately skip ahead to Section \ref{sec:discuss} where we provide a discussion of our results and some interesting open questions. The Appendices contain various technical results about 
the Lagrangian of $\mathcal{N}=1^*$ on $S^4$, details of the supergravity construction, a summary of the holographic renormalization needed for our model, and aspects of the numerical analysis.

\section{Field theory}
\label{sec:FT}

In flat space, the ${\cal N} = 1^*$ deformation of ${\cal N} = 4$ SYM is well known \cite{Polchinski:2000uf}.  To construct it, one starts with the ${\cal N} = 4$ SYM theory written in ${\cal N} = 1$ notation in terms of a vector multiplet ${\cal V}^a$, $a$ being an adjoint gauge index, and three chiral multiplets $\Phi_i^a$, $i = 1, 2, 3$, also transforming in the adjoint representation of the gauge group.  The Lagrangian of the ${\cal N} = 1^*$ theory consists of the standard kinetic term for the chiral multiplets, the Yang-Mills term for the vector multiplet, and a superpotential interaction
 \es{WN1star}{
  W =  \sqrt{2} g_\text{YM} f^{abc} \Phi_1^a \Phi_2^b \Phi_3^c +  \frac{1}{2} \sum_{i=1}^3 m_i \Phi_i^a\Phi_i^a\,,
 }
where $g_\text{YM}$ is the Yang-Mills coupling, $f^{abc}$ are the structure constants of the gauge group normalized in a way independent of $g_\text{YM}$.  Repeated gauge indices are being summed over.  To be concrete about the normalization of the fields that enter in the ${\cal N} = 1^*$ action, let us exhibit the kinetic and mass terms for the theory in flat space:
 \es{Lagrangian}{
\mathcal{L}_{\mathbb{R}^4}  &= 
 \frac{1}{4 g_\text{YM}^2} \left(  F_{\mu\nu}^{a}  \right)^2
  + \frac{\theta}{ 16 \pi^2} \epsilon^{\mu\nu\rho\sigma} F_{\mu\nu}^a F_{\rho \sigma}^a
   - \lamt^{a\; T}_{i} \sigma_2 \bar{\sigma} ^{\mu} D_{\mu} \lam_i^a 
   + D^{\mu}\Zt ^a _i D_{\mu} Z ^a _i 
   - \chit^{a\; T}_{i} \sigma_2 \bar{\sigma} ^{\mu} D_{\mu} \chi_i^a \\
 &{}+
\mt_i m_i \Zt_i^a Z_i^a
 -\frac{1}{2} m_i \left(  \chi^{a\; T}_i \s2 \chi_i^a  \right)
 - \frac{1}{2} \mt_i \left(  \chit^{a\; T}_i \s2 \chit_i^a  \right) 
 + {\cal L}_\text{interactions}\,.
}
Here, $F_{\mu\nu}^a$ is the gauge field strength, $\lambda_\alpha^a$ and $\tilde \lambda^{a \dot \alpha}$ are the left-handed and right-handed components of the gauginos, $Z_i^a$ are the bottom components of the chiral multiplets and $\tilde Z_i^a$ are their conjugates, and $\chi_{i \alpha}^a$ and $\tilde \chi_i^{a \dot \alpha}$ are the left-handed and right-handed components of the fermions in the chiral multiplets.  It should be emphasized that while in Lorentzian signature the fields with a tilde are related by complex conjugation to the ones without, in Euclidean signature these fields should be considered as independent, and this is why we wrote $\tilde Z_i^a$ or $\tilde m_i$ instead of $\bar Z_i^a$ or $\bar m_i$, and similarly for the fermion fields.

The ${\cal N} = 1^*$ theory can be placed on $S^4$ while preserving supersymmetry as explained in the work of Festuccia and Seiberg \cite{Festuccia:2011ws} where the general approach to coupling ${\cal N} = 1$ supersymmetric theory to background curvature is outlined.   Their construction involves coupling a given supersymmetric theory to an off-shell background supergravity in the old minimal formalism and taking the gravitational constant to be small.   We discuss these ideas in the context of the analysis of counterterms in the next section; here, it suffices to say that apart from covariantizing the derivatives in \eqref{Lagrangian}, 
when going from $\R^4$ to $S^4$ one should also add to the Lagrangian the term
 \es{Wprimeterms}{
  \delta {\cal L} =  \pm \frac{\im}{a} \bigg(3W(Z) - \frac{\partial W(Z)}{\partial Z_i^a} Z_i^a \bigg) \pm \frac{\im}{a} \bigg(3 \widetilde{W}(\tilde Z) - \frac{\partial \widetilde{W}(\tilde Z)}{\partial \tilde Z_i^a} \tilde Z_i^a\bigg)
  + \frac{2}{a^2} \tilde Z_i^a Z_i^a \,,
 }
where $W$ is the superpotential in \eqref{WN1star}. This result was explicitly derived in Appendix A of \cite{Bobev:2013cja} (see also \cite{Festuccia:2011ws}).  Here, $a$ is the radius\footnote{We hope that using the same symbol for the $S^4$ radius and the adjoint index of the various fields will not cause confusion.} of $S^4$, 
and the $\pm$ corresponds to a discrete choice of sign in the Killing spinor equation on $S^4$ that the supersymmetry parameters have to obey.\footnote{The theory on $S^4$ preserves the ${\cal N} = 1$ supersymmetry algebra $\mathfrak{osp}(1|4)$.  This algebra is a subalgebra of the ${\cal N}=1$ superconformal algebra $\mathfrak{su}(4|1)$, which in turn is a subalgebra of the ${\cal N} = 4$ superconformal algebra $\mathfrak{su}(4|4)$.  The choice of sign in the Killing spinor equation on $S^4$ corresponds to two different embeddings of $\mathfrak{osp}(1|4)$ into $\mathfrak{su}(4|1)$.}  The last term in \eqref{Wprimeterms} is the usual conformal coupling of the scalars on a curved manifold.  For a conformal theory in which there are only cubic terms in the superpotential $W$
 the first two terms vanish, in harmony with the fact that the round $S^4$ is a conformally flat manifold.    Since $\mathcal{N}=1^*$ is not a conformal theory, there are non-trivial contributions from the first two terms of \eqref{Wprimeterms}, namely new boson mass terms 
 \es{mZExtra}{
   \pm \frac{\im}{2a}  \Big( m_i Z_i^2 + \tilde{m}_i \tilde{Z}_i^2  \Big) \,.
 }
It should be noted that in the Euclidean theory we can regard $m_i$ and $\tilde{m}_i$ as completely independent mass parameters.  
The result for the full $\mN=1^*$ Lagrangian on $S^4$ is presented in component form in Appendix \ref{app:FT}. 

We will be interested in three particular limits of the  $\mN=1^*$  theory on $S^4$ as described in the bullet points in the Introduction. Briefly summarizing the limits here, we have  $m_i = \tilde m_i$ and

\vspace{1mm}
\begin{tabular}{ll}
  \text{$\mN=2^*$  theory:} & $m_1 = m_2 \ne 0$ ~~\&~~  $m_3 = 0$\,,\\
  \text{$\mN=1^*$  one mass:} & $m_1 = m_2 = 0$ ~~\&~~  $m_3 \ne 0$\,,\\
  \text{$\mN=1^*$  equal mass:} & $m_1 = m_2 = m_3$\,.\\
\end{tabular}

\vspace{1mm}
\noindent As we shall see, the study of these limits will also allow us to draw conclusions for general aspects of the  $\mN=1^*$ theory with general values of the masses.

\section{Universal contributions to the $S^4$ free energy }
\label{sec:FE}

Like many other quantities in quantum field theory, $F_{S^4}$ is UV divergent and requires regularization.  In this section we will discuss the regularization-independent information encoded in $F_{S^4}$, and in the following sections we will extract this information from holography.

In general, it is convenient to think of placing a QFT on $S^4$ in terms of coupling the flat space theory to a background metric and turning on additional sources, such as mass terms, as desired.   The regularization of the free energy $F_{\cal M}$ on a 4-manifold ${\cal M}$ is achieved by adding local terms in these background sources with arbitrary coefficients that could depend on the UV length cutoff $\epsilon$.  Such terms must be consistent with all the symmetries of the theory that are preserved by the regularization scheme.  In particular, the terms involving the background metric must be invariant under local diffeomorphisms.  If the theory is supersymmetric or has other symmetries one wishes to preserve, then these local terms must obey these symmetries as well.

For instance, one can remove a quartic divergence in $F_{\cal M}$ by adding a cosmological constant counterterm $\frac{1}{\eps^4} \int d^4x\, \sqrt{g}$.   As another example, in a mass-deformed theory, where the (complex) mass parameter $m$ can be promoted to a background field $\mathfrak{m}$ (and its conjugate $\tilde{\mathfrak{m}}$), one can remove a quadratic divergence by adding to $F_{\cal M}$ a linear combination of $\frac{1}{\eps^2} \int d^4x\, \sqrt{g} R$ and  $\frac{1}{\eps^2} \int d^4x\, \sqrt{g}\, \mathfrak{m} \tilde{\mathfrak{m}}$.  Of course, one should evaluate all these counterterms on the background of interest.  In a non-supersymmetric theory, a quartic divergence is the largest divergence one can remove by this procedure.  In a supersymmetric theory, the largest divergence one can remove is only quadratic, because the cosmological constant counterterm used to remove the quartic divergence is not supersymmetric.  In both supersymmetric and non-supersymmetric theories $F_{\cal M}$ contains a term proportional to $\log \epsilon$ whose coefficient is a linear combination of the conformal anomaly coefficients $\mathfrak{a}$ and $\mathfrak{c}$;  the coefficient of the logarithmic term is universal, as it cannot be removed by a local counterterm. Since the sphere is conformally flat, on $S^4$ the coefficient of $\log \epsilon$ is proportional to the coefficient $\mathfrak{a}$ only.\footnote{The anomaly coefficient $\mathfrak{c}$ multiplies a scalar constructed out of the Weyl tensor which vanishes on conformally flat manifolds.}

This discussion shows that the finite part of $F_{\cal M}$ is ambiguous if there exist finite counterterms that are invariant under all the symmetries one wishes to preserve and that are non-vanishing on the background of interest.  In supersymmetric theories, such finite counterterms can be quite restricted.  For instance, in a generic ${\cal N} = 2$ superconformal field theories (SCFTs) placed on $S^4$, Gerchkovitz, Gomis, and Komargodski \cite{Gerchkovitz:2014gta} showed\footnote{See Ref.~\cite{Knodel:2014xea} for a related analysis.} that, in a regularization scheme preserving supersymmetry, the finite part of $F_{S^4}$ only has a shift ambiguity given by a sum of a holomorphic and an anti-holomorphic function of the exactly marginal couplings $\tau_i$, 
 \es{N2SCFTAmbiguity}{
  \text{$\mathcal{N}=2$ SCFT:}~~~~~F_{S^4} \to F_{S^4} + f(\tau_i) + \bar f (\tilde \tau_i)\,.
 }
Here, $\tilde \tau_i$ represents the Euclidean continuation of the complex conjugate of the marginal coupling $\tau_i$;  while in Lorentzian signature we would have had $\tilde \tau_i = \tau_i^*$ in a unitary theory, in Euclidean signature we relax this requirement. It is easy to see from \eqref{N2SCFTAmbiguity} that for an ${\cal N} = 2$ SCFT on $S^4$, the second derivative $\partial_{\tau_i} \partial_{\tilde \tau_j}F_{S^4}$ is unambiguous and therefore universal in a supersymmetric regularization scheme---the quantity $\partial_{\tau_i} \partial_{\tilde \tau_j}F_{S^4}$  is, in fact, the Zamolodchikov metric on the conformal manifold \cite{Gerchkovitz:2014gta}. 

If instead we were to consider an ${\cal N} = 1$ SCFT with no other symmetries, then the finite counterterms would be less constrained, and the finite part of $F_{S^4}$ would now be ambiguous even in a regularization scheme that preserves supersymmetry, in that different regularization schemes would differ by an arbitrary function of the marginal couplings \cite{Gerchkovitz:2014gta}, 
\be
  \text{$\mathcal{N}=1$ SCFT:}~~~~~F_{S^4} \to F_{S^4} + f_1(\tau_i,\tilde \tau_i)\,.  \label{N1SCFT}
\ee
This ambiguity means that the only universal information in $F_{S^4}$ is the coefficient of $\log \epsilon$, which is proportional to the anomaly coefficient $\mathfrak{a}$.  Everything else can be changed by adding local counterterms in the background marginal sources. 

In the situation of interest to us, namely the ${\cal N} = 1^*$ deformation of ${\cal N} = 4$ SYM by complex mass parameters $m_i$, we will show that if $F_{S^4}$ is computed in a supersymmetric regularization scheme that is consistent with the $SL(2, \Z)$ duality of the ${\cal N} = 4$ super-Yang Mills theory,\footnote{We will assume such a scheme exists, at the very least in the large $N$ limit and strong 't Hooft coupling.  As we will see, from our supergravity analysis in the next section, the ambiguities we find on the supergravity side are indeed of the form \eqref{N1starAmbiguity}.} then $F_{S^4}$ is ambiguous up to shifts of the form
 \be
  \text{$\mathcal{N}=1^*$:}~~~~~F_{S^4} \to F_{S^4} + f_1(\tau,\tilde \tau) + a^2 \sum_{i=1}^3 m_i \tilde m_i f_2(\tau, \tilde \tau) \,,  \label{N1starAmbiguity}
 \ee
where $f_1$ and $f_2$ are arbitrary functions of the complexified UV gauge coupling\footnote{The functions $f_1$ and $f_2$ can in general depend on the other marginal couplings of the ${\cal N} = 4$ SYM theory that preserve ${\cal N}=1$ supersymmetry \cite{Leigh:1995ep}.  In our supergravity construction, those marginal couplings are fixed to their ${\cal N} = 4$ values.} 
 \es{tauDef}{
  \tau \equiv \frac{\theta}{2 \pi} + \im \frac{4 \pi}{g_\text{YM}^2} \,, \qquad   \tilde \tau \equiv \frac{\theta}{2 \pi} - \im \frac{4 \pi}{g_\text{YM}^2} \,.
 }
If one were to not insist on $SL(2, \Z)$, then $F_{S^4}$ could also be shifted by 
 \be
  \im a^3 \left[ {m_1 m_2 m_3} h(\tau) + {\tilde m_1 \tilde m_2 \tilde m_3} \bar h(\tilde \tau)  \right]\,,
 \ee
where $h$ and $\bar h$ are holomorphic and anti-holomorphic functions, respectively.  If one would further not insist on preserving supersymmetry, then $F_{S^4}$ could also be shifted by a more general cubic term in the masses and also by an arbitrary quartic term in the masses with coefficients that are arbitrary functions of $\tau$ and $\tilde \tau$.

From \eqref{N1starAmbiguity} it follows that one can extract universal information from $F_{S^4}$ by taking at least two derivatives w.r.t.~the mass parameters $m_i$ or at least two derivatives w.r.t.~$\tilde m_i$, or one derivative w.r.t.~$m_i$ and one derivative w.r.t.~$\tilde m_j$, with $j \neq i$, such that the ambiguous terms in \eqref{N1starAmbiguity} drop out.  In the limit where $\tilde m_i = m_i$, we just take three mass-derivatives to eliminate any ambiguities. 
We will now proceed to a derivation of \eqref{N1starAmbiguity}.

\subsection{Supersymmetric counterterms in the ${\cal N} = 1^*$ theory}

One can derive the possible supersymmetric counterterms that can affect $F_{S^4}$ by generalizing the analysis performed in \cite{Gerchkovitz:2014gta} for general ${\cal N} = 1$ SCFTs to the case of the non-conformal ${\cal N} = 1^*$ theory.\footnote{Our analysis easily generalizes to relevant deformations of other ${\cal N} = 1$ SCFTs, but for concreteness we focus on the ${\cal N} = 1^*$ case.}  The analysis we are about to perform relies on the fact that the superconformal symmetry $\mathfrak{su}(4|4)$ is broken explicitly to the massive $S^4$ supersymmetry algebra $\mathfrak{osp}(1|4)$ through turning on certain background sources.  These sources can be studied in the context of the old minimal formulation of off-shell 4d supergravity, where the off-shell gravity multiplet consists of the metric $g_{\mu\nu}$, the gravitino, a complex auxiliary scalar field $M$ (with conjugate $\overline{M}$), and an auxiliary vector field $b_\mu$.  This gravity multiplet is treated as a background used to place the theory on curved space, i.e.~it is not dynamical.   From it, one can construct the following two chiral superfields
 \es{ChiralMetric}{
  {\cal E} &= \frac 12 \sqrt{g} - \frac 12 \sqrt{g} \overline{M} \Theta^2 \,, \\
  {\cal R} &= -\frac 16 M - \frac 16 \Theta^2 \left( -\frac 12 R + \frac 23 M \overline{M} + \frac 13 b_\mu b^\mu - \im \nabla_\mu b^\mu \right) \,,
  }
whose expressions are written after having set the gravitino field to zero.  The superfield ${\cal E}$ is the chiral superspace measure that represents a supersymmetric completion of $\sqrt{g}$, and the superfield ${\cal R}$ is a chiral superfield containing the Ricci scalar $R$.  One can also construct the anti-chiral fields $\overline{{\cal E}}$ and $\overline{{\cal R}}$ by taking the conjugates of \eqref{ChiralMetric}.  We are eventually interested in evaluating these superfields on an $S^4$ of radius $a$, where preserving supersymmetry requires that \cite{Festuccia:2011ws}
 \es{bMValues}{
   M = \overline{M} = -\frac{3\im}{a}\,, \qquad b_\mu = 0 \,, \qquad R = \frac{4}{3} M \overline{M} = -\frac{12}{a^2} \,.
  } 

 In addition to the background gravity multiplet, we introduce a background chiral multiplet, $\mathcal{T}$, whose lowest component is the complexified gauge coupling $\tau$ promoted to a space-dependent field, as well as three background chiral multiplets, $\mathfrak{m}_i$, whose lowest components are proportional to the mass parameters $m_i$, again promoted to space-dependent fields.  When promoting $m_i$ to the bottom component of a chiral multiplet, one has to be careful because the non-holomorphic quantity $g_\text{YM}$ appears explicitly in the superpotential \eqref{WN1star} in the first term.  If we rescale the superfields $\Phi_i^a$ such that $g_\text{YM}$ does not appear in the first term by defining $\Phi_i^{a\prime} \equiv \Phi_i^a g_\text{YM}^{1/3}$, then the superpotential becomes 
  \es{SuperpotAgain}{
   W = \sqrt{2} f^{abc} \Phi_1^{a\prime} \Phi_2^{b\prime} \Phi_3^{c\prime} +  \frac{1}{2} \sum_{i=1}^3 \frac{m_i}{g_\text{YM}^{2/3}} \Phi_i^{a\prime} \Phi_i^{a\prime} \,.
  }
 This expression shows that the holomorphic quantity that can be promoted to the bottom component of a chiral superfield is $m_i / g_\text{YM}^{2/3}$.   After setting the fermion background fields in the $\mathcal{T}$ and $\mathfrak{m}_i$ multiplets to zero, we have
  \es{BackgroundPhi}{
   \mathcal{T}= \tau + O(\Theta^2) \,, \qquad
    \mathfrak{m}_i = \frac{m_i}{g_\text{YM}^{2/3}} + O(\Theta^2) \,.
  }
We also have the corresponding anti-chiral superfields $\overline{\mathcal{T}} = \tilde \tau + O(\overline{\Theta}^2)$ and $\overline{\mathfrak{m}}_i = \frac{\tilde m_i}{g_\text{YM}^{2/3}} + O(\overline{\Theta}^2)$.
 
Using these ingredients, we can write all possible counterterms.  The counterterms must be 
integrals of local functions of the background fields that are locally supersymmetric, and whose form is dictated by the symmetries of the ${\cal N} = 4$ theory they are breaking.  Our local supersymmetry superfield construction above only relies on an ${\cal N} = 1$ supersymmetry algebra, and in ${\cal N} = 1$ notation the ${\cal N} = 4$ theory has an $SU(3)$ flavor symmetry that rotates the chiral multiplets $\Phi_i$ as a fundamental vector.  The mass deformation we consider breaks this $SU(3)$ symmetry.  Since we only turn on $3$ distinct complex mass parameters, it is impossible to assign to $m_i$ and $\tilde m_i$ any $SU(3)$ (spurionic) transformation properties in such a way that the deformed Lagrangian would preserve this $SU(3)$.   It is possible, however, to preserve a 
 \es{Subgroup}{
  U(1)^2 \times \Z_3 \subset SU(3)\;,
 }
subgroup of $SU(3)$.  Under $U(1)^2$, we have 
 \es{TransfSubgroup}{
  U(1)^2: \qquad \Phi_i \to e^{\im \alpha_i} \Phi_i\;, \qquad \overline{\Phi}_i \to e^{-\im \alpha_i} \overline{\Phi}_i\;, \qquad 
   \mathfrak{m}_i \to e^{-2 \im \alpha_i} \mathfrak{m}_i\;, \qquad \overline{\mathfrak{m}}_i \to e^{ 2\im \alpha_i}\overline{\mathfrak{m}}_i\;,
 } 
with $\alpha_1 + \alpha_2 + \alpha_3 = 0$, and under the $\Z_3$ the three sets of $\Phi_i$, $\overline \Phi_i$, $\mathfrak{m}_i$, and $\overline{\mathfrak{m}}_i$ are cyclically permuted simultaneously.  The counterterms must obey this $U(1)^2 \times \Z_3$ symmetry.

We thus have the following possible finite counterterms that are locally supersymmetric and $U(1)^2 \times \Z_3$ invariant:
 \es{CounterSphere}{
   I_0 &= - \frac{1}{4 \pi^2} \int d^4 x \int d^2\Theta\, \mathcal{E} 
    (\overline{D}^2 - 8 {\cal R}) \, {\cal R} \overline{{\cal R}} \, f_1(\mathcal{T},\overline{\mathcal{T}}) \,, \\
   I_2 &= \frac{1}{\pi^2} \int d^4 x \int d^2\Theta\, \mathcal{E} 
    (\overline{D}^2 - 8 {\cal R}) \,  f_2(\mathcal{T},\overline{\mathcal{T}}) \sum_{i=1}^3 \mathfrak{m}_i\overline{\mathfrak{m}}_i \,, \\
   I_3 &=\frac{1}{4 \pi^2} \int d^4 x \int d^2\Theta\, \mathcal{E} \, \mathfrak{m}_1 \mathfrak{m}_2 \mathfrak{m}_3 h(\mathcal{T}) + \frac{1}{4 \pi^2} \int d^4 x \int d^2 \overline{\Theta}\, \overline{\mathcal{E}} \,  \overline{\mathfrak{m}}_1 \overline{\mathfrak{m}}_2 \overline{\mathfrak{m}}_3 \bar h(\overline{\mathcal{T}})
   \,. 
 }
Here, $(\overline{D}^2 - 8 {\cal R})$ is the covariant generalization of the chiral projection from rigid supersymmetry constructed such that $(\overline{D}^2 - 8 {\cal R})$ acting on any Lorentz-index-free superfield will produce a chiral field.\footnote{It is possible to write down another term contributing to $I_0$ and one contributing to $I_2$ that involve an integral over $\overline{\Theta}$, the anti-chiral measure $\overline{{\cal E}}$, and the anti-chiral projector.  On the supersymmetric sphere, these anti-chiral terms are redundant in the sense that they evaluate to expressions of the same form as those given by the chiral terms in \eqref{CounterSphere}.}  The counterterms in \eqref{CounterSphere} are supersymmetric because they are half-superspace integrals of chiral quantities, and they are marginal by dimensional analysis using the counting that each derivative has mass dimension $1$ and the superfields $\cal E$, ${\cal R}$, $\mathcal{T}$, and $\mathfrak{m}_i$ have mass dimensions $0$, $2$, $0$, and $1$, respectively.   The functions $f_1$ and $f_2$ are arbitrary complex functions, while $h$ and $\bar h$ are restricted to be holomorphic and anti-holomorphic, respectively, because otherwise the integrands in $I_3$ would not be, respectively, chiral and anti-chiral.  It is impossible to write non-vanishing finite supersymmetric counterterms as in \eqref{CounterSphere} that are linear or higher order than cubic in the~$\mathfrak{m}_i$.

Upon using \eqref{bMValues} and \eqref{BackgroundPhi} in order to evaluate the finite counterterms \eqref{CounterSphere} on the supersymmetric sphere, we obtain
 \es{CounterSphereEvaluated}{
  I_0 &=  f_1(\tau, \tilde \tau)  \,, \\
   I_2 &=  \,
   f_2(\tau,\tilde \tau) \sum_{i=1}^3 m_i \tilde m_i a^2 \,, \\
   I_3 &=  \im a^3  \left[ m_1 m_2 m_3 h(\tau)  + \tilde m_1 \tilde m_2 \tilde m_3 \bar h(\tilde \tau) \right] \,.
 }
The ambiguity of the $S^4$ partition function thus amounts to a shift of the form
 \es{Ambiguity}{
  F_{S^4} \to F_{S^4} + I_0 + I_2 + I_3 \;,
 }
and is parameterized by the functions $f_{1, 2}$, $h$, and $\bar h$.  Note that these ambiguities amount to shifts in $F_{S^4}$ that are either mass-independent or that are proportional to  $\sum_{i=1}^3 m_i \tilde m_i $ or $m_1 m_2 m_3$ or $\tilde m_1 \tilde m_2 \tilde m_3$.   One can thus construct derivatives of $F_{S^4}$ with respect to the masses that are free of these ambiguities.

\subsection{Constraints on counterterms from $SL(2, \Z)$}

One can impose additional constraints on the finite counterterms representing the ambiguity \eqref{Ambiguity} in $F_{S^4}$ by recalling that the ${\cal N} = 4$ SYM theory with gauge group $SU(N)$ is also invariant under an $SL(2, \Z)$ duality group.\footnote{We will assume it is possible to work in a regularization scheme preserving $SL(2, \Z)$ in the sense that $\log Z$, computed as a function of the various background fields $(g_{\mu\nu}, b_\mu, M, \overline{M}, m, \tilde m, \tau, \tilde \tau, \text{etc.})$ is invariant under $SL(2, \Z)$ when these backgrounds are assigned the $SL(2, \Z)$ transformation properties in this section.  In fact, we only require $\log Z$ to have this property at large $N$ and large 't Hooft coupling, where the supergravity description provides a good approximation.}  At large $N$ and large 't Hooft coupling, where the theory has a dual supergravity description, this $SL(2, \Z)$ is enhanced to an $SL(2, \R)$ duality group that is present in supergravity as a global symmetry. An $SO(2)$ subgroup of $SL(2, \R)$ acts as a ``bonus symmetry'' of the ${\cal N} = 4$ SYM theory in this limit \cite{Intriligator:1998ig}.

Even though the $SL(2, \Z)$ duality group acts in a highly non-trivial way on the fields of the ${\cal N} = 4$ SYM Lagrangian, it acts relatively simply on the operators in the stress tensor multiplet, which contain $\tr F_{\mu\nu} F^{\mu\nu}$ and $\tr F_{\mu\nu} \tilde F^{\mu\nu}$.\footnote{Recall that electromagnetic duality in a 4d Abelian theory with gauge field $A_\mu$ and field strength $F_{\mu\nu}$ also acts simply on $F_{\mu\nu}$ and $\tilde F_{\mu\nu}$, even though its action on $A_\mu$ is quite complicated.  In the context of the ${\cal N} = 4$ SYM theory, the analog of electromagnetic duality is the discrete $S$-transformation of $SL(2, \Z)$.}  In particular, in the normalization where $1/g_\text{YM}^2$ appears in front of $\tr F_{\mu\nu}^2$ but not in front of the scalar kinetic terms, as in Appendix~\ref{app:FT}, the operators $\tr \left[ X_{(i} X_{j)} - \frac 16 \delta_{ij} X_k X_k \right] $ in the ${\bf 20'}$ of $SO(6)$ are $SL(2, \Z)$ invariant.  The transformation properties of all the other operators in the ${\cal N} = 4$ stress tensor multiplet can be obtained from the fact that the supercharges $Q_\alpha^A$ and $\bar Q_{A \dot \alpha}$ can be thought of as $SL(2, \Z)$ modular forms of weights $(1/4, -1/4)$ and $(-1/4, 1/4)$, respectively. As in \cite{Intriligator:1998ig}, by an $SL(2, \Z)$ modular form of weights $(w, \bar w)$ we mean a complex function $f(\tau, \tilde \tau)$ that under an $SL(2, \Z)$ transformation obeys
 \es{ModularForm}{
  f(\tau', \tilde \tau') = (c \tau + d)^w (c \tilde \tau + d)^{\bar w} f(\tau, \tilde \tau) \,, \qquad
   \tau' \equiv \frac{a \tau + b}{c \tau + d} \,, \qquad
    \tilde \tau' \equiv \frac{a \tilde \tau + b}{c \tilde \tau + d} \,,
 }
where $\begin{pmatrix} a & b \\ c & d \end{pmatrix} \in SL(2, \Z)$.  When $\bar w = - w$, 
as is the case for the supercharges, $f$ simply gets multiplied by a phase:\footnote{This is true when $\tilde \tau$ is the complex conjugate of $\tau$, e.g.~in the Lorentzian theory.}  
$f(\tau', \tilde \tau') = e^{\im w \theta} f(\tau, \tilde \tau)$, where $\theta = 2 \arg (c \tau + d)$.  More generally, the possibility of multiplying the supercharges by a phase when moving on the conformal manifold (the $SL(2, \Z)$ transformations are a particular case) is available because this multiplication by a phase is an outer automorphism of the ${\cal N}=4$ superconformal algebra.  In fact, the supercharges are sections of a holomorphic line bundle of non-vanishing curvature on the conformal manifold (see for example \cite{Papadodimas:2009eu,Baggio:2014ioa}).     In the ${\cal N} = 1$ superspace notation used above, if one assigns modular weights $(-1/4, 1/4)$ and $(1/4, -1/4)$ to the superspace coordinates $\Theta$ and $\bar \Theta$, respectively, then each superfield will transform under $SL(2, \Z)$ as a modular form with the same weights as its lowest component.

The mass deformation of the ${\cal N} = 4$ SYM theory as well as the additional background fields needed to place the theory on $S^4$ break $SL(2, \Z)$.  The $SL(2, \Z)$ can be restored if in addition to transforming the fields of the deformed ${\cal N} = 4$ SYM theory, we also transform the deformation parameters.  

The $SL(2, \Z)$ transformation properties of the various background fields can be worked out straightforwardly as follows.  The fermion bilinears $\chi_i \chi_j$ are part of the ${\bf 10}$ of $SO(6)$ and are obtained by acting with two $Q$'s on operators in the ${\bf 20'}$, so they should transform under $SL(2, \Z)$ as modular forms with weights $(1/2, -1/2)$.  The bilinears $\tilde \chi_i \tilde \chi_j$ are part of the $\overline{{\bf 10}}$, so they  transform under $SL(2, \Z)$ as modular forms of weights $(-1/2, 1/2)$.  Consequently, in order for the deformed action to be invariant, we should require that $m_i$ transform as modular forms with weights $(-1/2, 1/2)$, and  that the $\tilde m_i$ transform with weights $(1/2, -1/2)$.  Using that $1/g_\text{YM}^2  = 4 \pi \text{Im}\, \tau$ transforms with modular weights $(-1, -1)$, from  \eqref{BackgroundPhi} it follows that $\mathfrak{m}_i$ transforms with modular weights $(-5/6, 1/6)$, and $\overline{\mathfrak{m}}_i$ transforms with modular weights $(1/6, -5/6)$.    From \eqref{ChiralMetric}, using the fact that the background metric is a modular invariant, we see that $M$ and $\overline{M}$ should have modular weights $(-1/2, 1/2)$ and  $(1/2, -1/2)$ respectively.  The superfields $\mathcal{R}$ and $\overline{\mathcal{R}}$ should also have modular weights $(-1/2, 1/2)$ and $(1/2, -1/2)$, respectively, while ${\cal E}$ is modular invariant.

We can now see that in order for the counterterms \eqref{CounterSphere} to be $SL(2, \Z)$ invariant, we must require that $f_1$ should be invariant under $SL(2, \Z)$, $f_2$, $h$, and $\bar h$ should transform with modular weights $(2/3, 2/3)$, $(2, 0)$, and $(0, 2)$, respectively.  However, in \eqref{CounterSphere}, $h$ and $\bar h$ are restricted to be holomorphic and anti-holomorphic functions, respectively, and there exist no such non-vanishing modular forms of these weights (see, for example, \cite{zzproc-houches89}).  Consequently, the counterterm $I_3$ is not consistent with the $SL(2, \Z)$ duality.  It follows that, in a scheme consistent with $SL(2, \Z)$, the ambiguity in the $S^4$ partition function of the ${\cal N}=1^*$ theory is given by
 \es{AmbiguitySL}{
  F_{S^4} \to F_{S^4} + I_0 + I_2 \,.
 }
We have thus proven \eqref{N1starAmbiguity}.  As we will see shortly, we will be able to reproduce the ambiguity \eqref{AmbiguitySL} from the supergravity side.

\subsection{Constraints on the dependence of $F_{S^4}$ on the masses}
\label{sec:Fgen}

The counterterm analysis above also gives us an insight into the dependence of the universal (ambiguity-free) part of $F_{S^4}$ on the masses.  While the ambiguity in the partition function is parameterized by local counterterms in the background sources that, in an appropriate regularization scheme, are invariant under the various symmetries we considered (local supersymmetry, $U(1)^2 \times \Z_3$, and $SL(2, \Z)$), the full partition function is expected to be a highly non-local expression in the background sources, but it must be invariant under the same symmetries.
  
In particular, when we specify the background fields to take the appropriate values for the mass deformation of the theory on $S^4$, one can see that 
the $U(1)^2 \times \Z_3$ action in \eqref{TransfSubgroup} is a symmetry, and consequently $F_{S^4}$ must be invariant under it.  For instance, only the following combinations of $m_i$ and $\tilde m_i$ are invariant under $U(1)^2$:
\be 
  \label{msyms}
  m_i \tilde{m}_i ~~\text{for each $i = 1, 2, 3$} \,, 
  ~~~\text{and}~~~
  m_1 m_2 m_3\,,
  ~~~\text{and}~~~
  \tilde{m}_1 \tilde{m}_2 \tilde{m}_3\,.
\ee
In addition, $m_i \tilde m_i$ must enter the expression for $F_{S^4}$ in a $\Z_3$ invariant way.  This  explains for instance, why, as derived in \cite{Russo:2012kj,Russo:2013qaa,Russo:2012ay,Buchel:2013id}, in the $\mN=2^*$  theory the free energy depended only on even powers in the mass $m \equiv m_1 = m_2$. 

To illustrate a further consequence of this analysis, consider the $O(m^4)$ terms in the free energy of the $\mN=1^*$ theory: it can depend only on two different combinations
\be
    F_{S^4}  \xrightarrow{O(m^4)} 
    A \Big(\sum_{i=1}^3 (m_i \tilde{m}_i)^2 \Big)
    +
    B \Big(\sum_{i=1}^3 m_i \tilde{m}_i \Big)^2\,.
\ee
The known result for $\mN=2^*$ \eqref{N2freeE} fixes a relation between the constants $A$ and $B$, namely $A + 2B = -\frac{1}{8}$. This leaves only one undetermined constant, and calculating the $O(m^4)$ terms in $F_{S^4}$ for both the LS limit $m_1 = m_2 = 0$, $m_3 \neq 0$, and the equal mass limit $m_1 = m_2 = m_3$ not only fixes the  $O(m^4)$ result for the full $\mN=1^*$ theory, but also provides a non-trivial consistency check of our results in the two limits. As will be shown in Section \ref{sec:compare}, 
this consistency check is fulfilled by our analyses. Of course we are interested in more information about the free energy, not just the  $O(m^4)$-terms, and we indeed are able to say a lot more.

\section{Supergravity}
\label{sec:SG}

We construct the holographic dual of the ${\cal N} = 1^*$ theory within the context of the $SO(6)$, ${\cal N} = 8$ gauged supergravity theory in five dimensions \cite{Gunaydin:1984qu,Gunaydin:1985cu,Pernici:1985ju}.  This gauged supergravity theory has been believed for many years to be a consistent truncation of maximal type IIB supergravity compactified on $S^5$.  Formulas for the uplift of the metric and the dilaton were provided in \cite{Pilch:2000ue,Pilch:2000fu}, and recent work \cite{Lee:2014mla, Baguet:2015sma} presented uplift formulas for the type IIB form fields.  These formulas should thus provide a way of uplifting our holographic duals of ${\cal N} = 1^*$ to ten dimensions, but we leave the details of such a computation for future work.

In order to construct our ${\cal N} = 1^*$ holographic duals, we will not need the full details of the ${\cal N} = 8$ gauged supergravity theory, because, as we will explain, our holographic duals lie entirely within certain consistent truncations of the ${\cal N} = 8$ theory of a more manageable complexity.  These consistent truncations can be obtained by the well-known method of keeping only the fields that are invariant under certain symmetries of the ${\cal N} =8$ theory \cite{Warner:1983vz}.  We present the full details of these truncations in Appendix \ref{app:SG}, and here we just summarize the results.

The most general ${\cal N} = 1^*$ deformation corresponds to the case where the $m_i$ and $\tilde m_i$ are all distinct. Compared to ${\cal N} = 4$ SYM, this theory has sources for the fermion bilinears $\Tr \chi_i \chi_i$ and $\Tr \tilde \chi_i \tilde \chi_i$, sources for the scalar bilinears $\Tr Z_i\tilde Z_i$ of which the diagonal $\Tr \left( Z_1\tilde Z_1 + Z_2\tilde Z_2+Z_3\tilde Z_3 \right)$ is the Konishi operator which does not have a dual supergravity field, as well as sources for $\Tr Z_i^2$ and $\Tr \tilde Z_i^2$ when the theory is placed on $S^4$---see \eqref{Lagrangian} and \eqref{mZExtra}.  At the very least, the corresponding truncation of the ${\cal N} = 8$ theory should contain the $14$ bulk fields dual to these operators.  It turns out that the largest symmetry of the ${\cal N} = 8$ theory under which these operators are invariant is a $(\Z_2)^3$ subgroup of the gauged $SO(6)_R$.  The bulk scalar fields invariant under this $(\Z_2)^3$ are not just the $14$ fields mentioned above, but also the duals of the gaugino bilinears $\Tr \lambda \lambda$ and $\Tr \tilde \lambda \tilde \lambda$, of $\Tr F_{\mu\nu}^2$ and of $\epsilon^{\mu\nu\rho\sigma} \Tr F_{\mu\nu} F_{\rho \sigma}$, for a total of $18$ bulk scalar fields.  The scalar manifold parameterized by these $18$ scalars is
 \es{M18}{
\mathcal{M}_{18}= \left[SO(1,1)\times SO(1,1)\right] \times \ds\frac{SO(4,4)}{SO(4)\times SO(4)} \,,
 }
which is obtained as the subgroup of the full scalar manifold of the ${\cal N} = 8$ theory, $E_{6(6)}/USp(8)$, that commutes with $(\Z_2)^3$.  In addition to the $18$ scalar fields, the $(\Z_2)^3$ invariant sector also contains fields of spin larger than $0$ which are linked together by two supersymmetry parameters, yielding an ${\cal N} = 2$ gauged supergravity theory with two vector multiplets and four hypermultiplets.  The $SO(1,1)\times SO(1,1) \backsimeq \R^2$ factor in \eqref{M18} is a very special manifold describing the scalars in the two vector multiplets, and the $\ds\frac{SO(4,4)}{SO(4)\times SO(4)}$ coset is a quaternionic K\"ahler manifold parametrized by the scalars in the four hypers.

The truncation mentioned above is still fairly unwieldy, and we will truncate further by keeping only the fields that are invariant under an additional $\mathbb{Z}_2$ symmetry of the supergravity theory.  In the field theory, this further truncation amounts to setting $m_i = \tilde m_i$ for all $i$.  The resulting truncation of the 18-scalar model above contains only 10 scalars parameterizing the manifold
 \es{M10}{
\mathcal{M}_{10} =  \left[SO(1,1)\times SO(1,1)\right] \times \left[\ds\frac{SU(1,1)}{U(1)} \right]^4 \,.
 }
This truncation is \textit{not} a fully fledged supergravity theory, but it is of course a consistent bosonic theory that can be utilized to construct explicit solutions. The 10 scalars include the dilaton $\varphi$, dual to $\Tr F_{\mu\nu} F^{\mu\nu}$, four scalars $\phi_i$, $i=1, \ldots, 4$, dual to fermion bilinear operators that are part of the $\bf{10}\oplus\overline{\bf{10}}$ irreducible representation of $SU(4)_R$, and 5 scalars, $\{\beta_1,\beta_2,\alpha_{1,2,3}\}$, dual to scalar bilinears that are part of the $\bf{20}'$ of $SU(4)_R$: 
 \es{Scalars}{
  \varphi &\quad \leftrightarrow \quad \text{Tr} F_{\mu\nu} F^{\mu\nu} \,, \\
   \phi_j &\quad \leftrightarrow \quad \text{Tr}(\chi_j\chi_j+ \tilde \chi_j \tilde \chi_j)\,, \qquad j=1, 2, 3 \,, \\
   \phi_4&\quad \leftrightarrow \quad \text{Tr}(\lambda \lambda + \tilde \lambda \tilde \lambda)\,, \\
   \beta_1 &\quad \leftrightarrow \quad \text{Tr}(Z_1\tilde Z_1+Z_2 \tilde Z_2 -2 Z_3 \tilde Z_3 )\,,\\
   \beta_2 &\quad \leftrightarrow \quad \text{Tr}(Z_1\tilde Z_1-Z_2 \tilde Z_2)\,,\\
   \alpha_j &\quad \leftrightarrow \quad \text{Tr}(Z_j^2+\tilde{Z}_j^2) \,, \qquad j=1, 2, 3\,.
 }
The scalars $\{\beta_1,\beta_2\}$ parametrize the $\left[SO(1,1)\times SO(1,1)\right]$ factor in $\mathcal{M}_{10}$, while the rest parameterize $\left[\ds\frac{SU(1,1)}{U(1)} \right]^4 \backsimeq (\mathbb{H}^2)^4$ in a rather mixed up way, as we will see.  We will make the identification between these scalar fields and the supergravity fields clearer in the next section.  

\subsection{The $10$-scalar model in Lorentzian signature}
\label{LORENTZIAN}

We first present the $10$-scalar model mentioned above in Lorentzian $(+----)$ signature, following the conventions of \cite{Gunaydin:1985cu}, and afterwards we will Wick rotate it to Euclidean signature.  It is convenient to write the explicit Lagrangian for our 10-scalar model using $\beta_{1, 2}$ and four complex fields $z^a$, $a = 1, \ldots, 4$, that parameterize $(\HH^2)^4$ as a K\"ahler manifold with K\"ahler potential 
 \es{Kahler}{
   \mathcal{K} = -\ds\sum_{a=1}^4 \log(1-z^{a}\bar{z}^{a}) \;,
 }
and K\"ahler metric $\mathcal{K}_{a\bar{b}} \equiv \frac{\partial^2 {\cal K}}{\partial {z^a} \partial {\bar{z}^{\bar b}}}$.  The relation between the $z^a$ and the scalar fields in \eqref{Scalars} that have a direct interpretation in terms of the field theory operators is:
\be
  \label{ztophysUV}
  \begin{split}
     z^1 &= \tanh \Big[ \frac{1}{2} \big( \alpha_1 + \alpha_2 + \alpha_3 + \varphi 
       -\im \phi_1 -\im \phi_2 -\im \phi_3 +\im \phi_4 \big) \Big] \,,\\
     z^2 &= \tanh \Big[ \frac{1}{2} \big( \alpha_1 - \alpha_2 + \alpha_3 - \varphi 
       -\im \phi_1 +\im \phi_2 -\im \phi_3 -\im\phi_4 \big) \Big] \,,\\
     z^3 &= \tanh \Big[ \frac{1}{2} \big( \alpha_1 + \alpha_2 - \alpha_3 - \varphi 
       -\im\phi_1 -\im\phi_2 +\im \phi_3 -\im \phi_4 \big) \Big] \,,\\
     z^4 &= \tanh \Big[ \frac{1}{2} \big( \alpha_1 - \alpha_2 - \alpha_3 + \varphi 
       -\im \phi_1 +\im \phi_2 +\im\phi_3 +\im\phi_4 \big) \Big] \,, \\
      \bar z^1 &= \tanh \Big[ \frac{1}{2} \big( \alpha_1 + \alpha_2 + \alpha_3 + \varphi 
       +\im \phi_1 +\im \phi_2 +\im \phi_3 -\im \phi_4 \big) \Big] \,,\\
     \bar z^2 &= \tanh \Big[ \frac{1}{2} \big( \alpha_1 - \alpha_2 + \alpha_3 - \varphi 
       +\im\phi_1 -\im \phi_2 +\im \phi_3 +\im \phi_4 \big) \Big] \,,\\
     \bar z^3 &= \tanh \Big[ \frac{1}{2} \big( \alpha_1 + \alpha_2 - \alpha_3 - \varphi 
       +\im \phi_1 +\im \phi_2 -\im \phi_3 +\im \phi_4 \big) \Big] \,,\\
     \bar z^4 &= \tanh \Big[ \frac{1}{2} \big( \alpha_1 - \alpha_2 - \alpha_3 + \varphi 
       +\im \phi_1 -\im \phi_2 -\im \phi_3 -\im \phi_4 \big) \Big] \,.  
  \end{split}
\ee
In terms of the $z^a$ and $\bar z^{\bar a}$, the gravity-scalar part of the Lagrangian is thus written as
 \es{bulkLag}{
\mathcal{L} = -\frac{1}{4}R +3(\partial\beta_1)^2+(\partial\beta_2)^2+ \frac{1}{2}\mathcal{K}_{a\bar{b}}\partial_{\mu}z^{a}\partial^{\mu}\bar{z}^{\bar{b}} - \mathcal{P}  \,,
 }
where  ${\cal P}$ is the scalar potential.   The scalar potential can be conveniently derived from a ``holomorphic superpotential"\footnote{Clearly this superpotential is only holomorphic 
in the $z_a$ and is a real function of the real scalars $\{\beta_{1},\beta_2\}$.} 
\begin{equation}\label{suppotdef}
\begin{split}
\mathcal{W} \equiv ~& \frac{1}{L}e^{2\beta_1+2\beta_2}\left(1+z_1z_2+z_1z_3+z_1z_4+z_2z_3+z_2z_4+z_3z_4+z_1z_2z_3z_4\right)\\
&+ \frac{1}{L}e^{2\beta_1-2\beta_2}\left(1-z_1z_2+z_1z_3-z_1z_4-z_2z_3+z_2z_4-z_3z_4+z_1z_2z_3z_4\right)\\
& +\frac{1}{L}e^{-4\beta_1}\left(1+z_1z_2-z_1z_3-z_1z_4-z_2z_3-z_2z_4+z_3z_4+z_1z_2z_3z_4\right)
\end{split}
\end{equation}
through the formula
 \es{Ppot}{
   \mathcal{P} = \frac{1}{8}e^{\mathcal{K}}\left[\frac{1}{6}\partial_{\beta_1}\mathcal{W}\partial_{\beta_1}\overline{\mathcal{W}}
     +\frac{1}{2}\partial_{\beta_2}\mathcal{W}\partial_{\beta_2}\overline{\mathcal{W}}+\mathcal{K}^{\bar{b} a}    
      \nabla_{a}\mathcal{W}\nabla_{\bar{b}}\overline{\mathcal{W}} -\frac{8}{3}\mathcal{W}\overline{\mathcal{W}}\right]\,,
 }
where $\mathcal{K}^{\bar{b}a}$ is the inverse of ${\cal K}_{a \bar b}$.  Here, the K\"ahler covariant derivative of a function $\mathcal{F}$ is defined as $\nabla_{a} \mathcal{F}\equiv \partial_{a}\mathcal{F} + \mathcal{F}\partial_{a}\mathcal{K}$.   In terms of the fields appearing on the RHS of \eqref{ztophysUV}, the scalar kinetic term and potential have the following expansion at small field values
\begin{equation}
 \label{Sbulkexpanded}
  \begin{split}
  \mathcal{L}_\text{kin} 
  &=
  + \frac{1}{2}\Big[ 
  6(\partial_\mu \beta_1)^2
  +2(\partial_\mu \beta_2)^2
  + \sum_{i=1}^3 (\partial_\mu \alpha_i)^2 + \sum_{i=1}^4  (\partial_\mu \phi_i)^2
  + (\partial_\mu \varphi)^2\Big] + \dots \,, \\
 \mathcal{P} =& \frac{1}{L^2} \Biggl[ - 3 - \frac{1}{2} \Bigg( 4 \cdot 6 \beta_1^2 +4 \cdot 2 \beta_2^2
    + 4 \sum_{i=1}^3 \alpha_i^2 
    + 3 \sum_{i=1}^4 \phi_i^2  \Bigg)
    \\
    &
    + \frac{1}{2} \sum_{i=1}^4 \phi_i^4 
    - 5 \sum_{i <j} \phi_i^2 \phi_j^2 
       + 12 \phi_1 \phi_2 \phi_3 \phi_4 + \dots \Biggr] \,, 
  \end{split}
\end{equation}
where the  ``$+ \dots$" stand for terms that are higher order when $\beta_{1, 2}$, $\alpha_{1, 2, 3}$ and $\phi_{1, 2, 3, 4}$ are small.  
This expansion shows that the fields $\beta_1$, $\beta_2$, and $\alpha_{1,2,3}$ all have mass-squared $m^2 L^2 = -4$, as appropriate for scalar fields dual to dimension-2 operators.  Similarly, the fields $\phi_{1,2,3,4}$ have  mass-squared $m^2 L^2 = -3$, as required for bulk duals of the dimension-3 fermion bilinear operators.  Finally, the dilaton $\varphi$ is massless and does not appear at all in the potential, as expected for the bulk dual of a dimension $4$ operator.  See~\eqref{Scalars}.

Although not a supergravity theory, the $10$-scalar model admits supersymmetric field configurations when embedded into the larger ${\cal N} = 8$ gauged supergravity theory.  As we show in Appendix~ \ref{app:susyvarE}, a field configuration is supersymmetric provided that one can find a pair of symplectic Majorana conjugate spinors $(\varepsilon_1, \varepsilon_2)$, with $\varepsilon_2 = \gamma_5 \varepsilon_1^*$, that obey\footnote{We use conventions where the five-dimensional gamma matrices obey $\{\gamma_m, \gamma_n\} = 2 \eta_{mn} = 2 \text{diag}\, \{1, -1, -1, -1, -1\}$, and take $\gamma_m$ to be pure imaginary for $0 \leq m \leq 3$ and $\gamma_4$ to be real.  We also have $\gamma_5 = -\im \gamma_4$. }
 \es{Vars}{
    \nabla_\mu \varepsilon_1 
    - \frac{1}{4} \big[ \pa_a \mathcal{K} \pa_\mu z^a 
          -   \pa_{\bar b} \mathcal{K} \pa_\mu \bar{z}^{\bar b}  \big] \varepsilon_1
          - \frac{1}{6} e^{\mathcal{K}/2} \overline{\mathcal{W}} \gamma_\mu \varepsilon_2
    &= 0\,,\\
\gamma^{\mu}\partial_{\mu}z^{a} \varepsilon_1 +\frac{1}{2} e^{\mathcal{K}/2}\mathcal{K}^{\bar{b} a}\left(\nabla_{\bar{b}}\overline{\mathcal{W}}\right) \varepsilon_2 &= 0\;, \\
 3 \gamma^{\mu}\partial_{\mu}\beta_1 \varepsilon_1+\frac{1}{4}e^{\mathcal{K}/2}\left(\partial_{\beta_1}\overline{\mathcal{W}}\right) \varepsilon_2 &= 0\;, \\
  \gamma^{\mu}\partial_{\mu}\beta_2 \varepsilon_1+\frac{1}{4}e^{\mathcal{K}/2}\left(\partial_{\beta_2}\overline{\mathcal{W}}\right) \varepsilon_2 &= 0\;.
 }
In addition to being differential equations for $\varepsilon_{1, 2}$, these equations also impose restrictions (i.e.~BPS equations) on the supergravity scalars and metric coming from their integrability conditions.

\subsection{The $10$-scalar model in Euclidean signature and its BPS equations} 
\label{EUCLIDEAN}

In Section~\ref{LORENTZIAN}, we presented the scalar$+$gravity theory we will be working with in Lorentzian signature.   Since we are interested in constructing the holographic dual of the ${\cal N} = 1^*$ theory on $S^4$, which is a Euclidean space, we should first continue the $10$-scalar model presented in the previous section to Euclidean signature, and then investigate whether it admits supersymmetric solutions with $S^4$ slicing.  The continuation to Euclidean signature is realized by taking
  \es{Cont}{
  \bar{z}^{\bar a} \to  \tilde{z}^{\bar a}\,,~~~~
  \gamma_\mu \to \im \gamma_\mu \,,~~~~
  \gamma^\mu \to -\im \gamma^\mu \,,~~~~
  \gamma_5 \to \gamma_5 \,.
 }
The tilde now indicates that quantities such as $z^a$ and $\tilde{z}^{\bar a}$ are no longer each other's complex conjugates, but instead should be treated as  independent.  When looking for supersymmetric solutions in Euclidean signature, one should require that both \eqref{Vars} and their Lorentzian conjugates, transformed to Euclidean signature using \eqref{Cont}, should hold.  These are two sets of independent equations in Euclidean signature.

For solutions with $S^4$ radial slices we use the metric Ansatz 
 \es{MetricAnsatz1}{
  ds^2 = dr^2 + e^{2A} ds_{S^4}^2\,,
 }
where $ds_{S^4}^2$ is the line element on a unit radius $S^4$, and assume that all scalars and the warp factor $A$ are functions only of the radial coordinate $r$,
 \es{ScalarsR}{
  A = A(r) \,, \qquad z^a = z^a(r)\,,\qquad \tilde{z}^{\bar a} = \tilde{z}^{\bar a}(r) \,, \qquad
   \beta_i = \beta_i(r) \,.
 }
In \eqref{MetricAnsatz1} and henceforth we set $L=1$.  A standard analysis of the integrability condition of \eqref{Vars} and their complex conjugates implies that supersymmetric solutions of this kind exist only provided that the following BPS equations are obeyed:
\be
  \label{BPSAll}
  \begin{array}{rclcrcl}
    \pa_r z^a 
    &=& \displaystyle
    6 \pa_r \beta_1 \,\frac{\mathcal{K}^{a \bar b} \nabla_{\bar b} \widetilde{\mathcal{W}}}{\pa_{\beta_1}\widetilde{\mathcal{W}}}\,,
    &~~~&
    \pa_r \tilde{z}^{\bar b} 
    &=& \displaystyle
    6 \pa_r \beta_1 \,\frac{\mathcal{K}^{a \bar b} \nabla_{a} {\mathcal{W}}}{\pa_{\beta_1}{\mathcal{W}}}\,,
    \\[3mm]
    (\pa_r \beta_1)^2 
    &=&\displaystyle
    \frac{1}{144} e^{\mathcal{K}} \pa_{\beta_1} \mathcal{W} \pa_{\beta_1} \widetilde{\mathcal{W} }\,,
        &~~~&
    (\pa_r \beta_2)^2 
    &=&\displaystyle
    \frac{1}{16} e^{\mathcal{K}} \pa_{\beta_2} \mathcal{W} \pa_{\beta_2} \widetilde{\mathcal{W} }\,,
    \\[3mm]
    \pa_r \beta_1 \pa_r \beta_2
    &=& \displaystyle
    \frac{e^{{\cal K}}}{48}  \pa_{\beta_1} \mathcal{W} \pa_{\beta_2} \widetilde{\mathcal{W} } \,, \\
  e^{-2A} &=&  \displaystyle \frac{e^{\mathcal{K}}}{36} \frac{\big(\widetilde{\mathcal{W}} \pa_{\beta_1}  \mathcal{W}-{\mathcal{W}} \pa_{\beta_1}  \widetilde{\mathcal{W}}\big)^2} {\pa_{\beta_1} \mathcal{W} \pa_{\beta_1}\widetilde{\mathcal{W}} }\;, \qquad 
       &~~~&
  \displaystyle
    \frac{1}{48}  \pa_{\beta_1} \mathcal{W} \pa_{\beta_2} \widetilde{\mathcal{W} } &=& \frac{1}{48}  \pa_{\beta_2} \mathcal{W} \pa_{\beta_1} \widetilde{\mathcal{W} }\,,
  \end{array}
\ee
The first three lines contain differential equations for the scalars, while the last line contains algebraic constraints.  The algebraic constraints can be solved for $A$ and $\beta_1$, for instance.  One can verify by direct computation that the algebraic constraints are consistent with the other BPS equations, and that in turn all these BPS equations are consistent  with 
(and imply) the second order equations of motion that can be derived from varying the action with respect to the dynamical fields.

In summary, for the Ansatz in \eqref{MetricAnsatz1}-\eqref{ScalarsR} the 10 scalar model yields a system of $9$ first order equations and two algebraic constraints.  Solving them leads to 9 integration constants, of which one is a trivial shift in the radial coordinate.  As we show below, the other eight integration constants are easily identified in the UV:  we have three mass deformation parameters $\mu_{1,2,3}$, three vev-rates $v_{1,2,3}$, the gaugino condensate vev $w$, and the source $s$ of the dilaton.  Smoothness in the bulk of the solution will impose further constraints by relating $4$ of these parameters to the other $4$.

Before we proceed, let us make a comment that will be relevant later.  If one uses \eqref{ztophysUV} to pass to the fields $(\alpha_i, \phi_i, \varphi)$ that have a more direct holographic interpretation, one finds that the differential equations in the first line of \eqref{BPSAll} take the schematic form 
\be
  \partial_r \alpha_i = F_i (\alpha_i,\phi_i)\,,
  ~~~~
   \partial_r \phi_i = G_i (\alpha_i,\phi_i)\,,
  ~~~~
   \partial_r \varphi = H (\alpha_i,\phi_i)\,.
\ee
In addition, the BPS equations for the $\beta_i$'s and the warp factor $A$ are independent of $\varphi$, so the dilaton field ${\varphi}$ completely drops out from most scalar equations, and it only appears in one equation through $\partial_r \varphi$.  This means that the dilaton has a shift symmetry 
\be
  \label{shiftsym}
  \varphi \to \varphi + \text{constant}\,,
\ee
in the sense that if we construct a supersymmetric solution with a given $\varphi(r)$ profile, we can find another supersymmetric solution for which $\varphi(r)$ is shifted by a constant.

We also note that the 10-scalar model \eqref{ztophysUV} and its BPS equations \eqref{BPSAll} are completely invariant under the transformation
 \es{InvariancezInv}{
  z^a \to \frac{1}{z^a} \,, \qquad \tilde z^{\bar b} \to \frac{1}{\tilde z^{\bar b}} \,.
 }
Therefore supergravity backgrounds 
related by \eqref{InvariancezInv} describe the same physics.  This transformation maps the interiors of the Poincar\'e disks parameterized by $(z^a, \tilde z^{\bar b})$ to their exteriors.  Another invariance of the 10-scalar model \eqref{ztophysUV} and of its BPS equations \eqref{BPSAll} is the sign flip symmetry 
 \es{zTomz}{
  z^a \to -z^a\,,\qquad \tilde z^{\bar b} \to - \tilde z^{\bar b} \,.
 }
Again, we will see that supergravity backgrounds related by \eqref{zTomz} describe the same physics.

\subsection{UV expansion}
\label{sec:UV}

The BPS equations \eqref{BPSAll} admit smooth solutions that approach the hyperbolic space $\HH^5$ asymptotically as $r \to \infty$.  Before we construct such solutions numerically, we  
solve the BPS equations for the fields with a systematic iterative expansion in the large-$r$ limit. 
Such an expansion will enable us to make more precise contact between the bulk fields and the field theory.

The metric of $\HH^5$ is $ds^2 = dr^2 + \sinh^2 r\, ds_{S^4}^2$, corresponding to taking $e^{2A} = \sinh^2 r$ in \eqref{MetricAnsatz1}.  This is a solution to the BPS equations with arbitrary constant $\varphi$ and with $\alpha_{1, 2, 3} = \phi_{1, 2, 3, 4} = \beta_{1, 2} = 0$.  The more general solutions of the BPS equations that have non-vanishing scalars and are asymptotically $\HH^5$ have the form
\be
\label{UVexp2}
\begin{split}
   e^{2A} &\,=\, \frac{1}{4} e^{2r} - \frac{1}{2} + \frac{1}{12}\big({\mu}_1^2+{\mu}_2^2+{\mu}_3^2\big)  + O\big(r^2 e^{-4r}\big)\,, \\
\alpha_i &\,=\, (2 \mu_i r +v_i ) e^{-2r} + O\big(r^2 e^{-4r}\big) \,, \\[1mm]
-\im\phi_i &\,=\, -\mu_i \, e^{-r}+ O\big(r e^{-3r}\big) \,, \\[1mm]
-\im\phi_4 &\,=\, w \, e^{-3r} + O\big(r e^{-5r}\big) \,, \\[1mm]
\varphi &\,=\, 2\, \text{arctanh}(s) + O\big(r e^{-4r}\big)\,,\\[1mm]
\beta_1 &\,=\, 
  - \frac{1}{6}
  \Big(
  2({\mu}_1^2+{\mu}_2^2 -2 {\mu}_3^2)\,r +{\mu}_1({v}_1+{\mu}_1)+{\mu}_2({v}_2+{\mu}_2)-2{\mu}_3({v}_3+{\mu}_3)
  \Big)
  \,e^{-2r}+ O\big(r^2 e^{-4r}\big)\,,\\[1mm]
  \beta_2 &\,=\, 
  -  \frac{1}{2}
  \Big(
   2({\mu}_1^2-{\mu}_2^2)\,r +{\mu}_1({v}_1+{\mu}_1)-{\mu}_2({v}_2+{\mu}_2)
   \Big)
  \,e^{-2r}+ O\big(r^2 e^{-4r}\big)\,.
\end{split}
\ee
This UV expansion is determined by 8 parameters: ${\mu}_1$, ${\mu}_2$, ${\mu}_3$ (which have the interpretation of sources in the field theory), ${v}_1$, ${v}_2$, and ${v}_3$ (which can be interpreted as expectation values in the field theory),\footnote{As usual one has to perform careful holographic renormalization to determine precisely what is the vev in the dual field theory. We therefore choose to refer to subleading coefficients in the UV expansion as vev-rates.} ${w}$ (interpreted as the gaugino expectation value), and a parameter $s$ related to the Yang-Mills coupling $g_\text{YM}$.

Note that a shift in the $r$ coordinate changes ${v}_i$ by $2{\mu}_i$ times a constant. This means that ${v}_i$ cannot directly be an observable, one has to account for the  
ambiguity $v \sim v + \mu$. 

The uplift of this five-dimensional solution to type IIB supergravity in ten dimensions is in general not easy to find. However in Appendix \ref{app:uplift} we provide an expression for the ten-dimensional dilaton $\Phi_{10}$ and axion $C_{10}$ as rather complicated functions of the ten five-dimensional scalars in our truncation; see, for instance, \reef{cMgeneral}.  In the UV limit, only the ten-dimensional dilaton is sourced, the axion is not, and it is precisely $s$ that provides the source:
 \es{PhiCBdry}{
   e^{-\Phi_\text{10}} ~\xrightarrow{r \to \infty}~\frac{4 \pi}{g_\text{YM}^2}= \frac{(1+s)^2}{(1-s)^2} \,, \qquad
    C_{10}  ~\xrightarrow{r \to \infty}~ \frac{\theta}{2\pi} = 0 \,.
 }
That means we should think of the UV parameter $s$ as encoding the dependence of the Yang-Mills coupling. 
The shift symmetry \reef{shiftsym}  in the 5d dilaton means that certain physical quantities are independent of $g_\text{YM}$.  

Note that the expression \eqref{PhiCBdry} for $g_\text{YM}^2$ is invariant under $s \to 1/s$, so physical observables should be invariant under this transformation.  The transformation $s \to 1/s$ is just the boundary limit of the symmetry in \eqref{InvariancezInv} of the full 10-scalar model.  Indeed, one can check that in terms of the UV parameters appearing in \eqref{UVexp2}, the $z^a \to 1/z^a$ and $\tilde z^{\bar b} \to 1/ \tilde z^{\bar b}$ symmetry in \eqref{InvariancezInv} imposes $s \to 1/s$ and leaves all the other parameters invariant.

From \eqref{PhiCBdry}, one can see that under the sign flip symmetry in \eqref{zTomz}, we have $s \to -s$ and consequently $\frac{4\pi}{g_\text{YM}^2} \to \frac{g_\text{YM}^2}{4 \pi}$.  As we show in Appendix~\ref{app:uplift}, this symmetry corresponds precisely to the $S$-transformation that is part of $SL(2, \Z)$ under which, more generally, $\tau \to -1/\tau$.  

To take the analysis further, we also need to fix boundary conditions in the IR, i.e.~one should require that the solutions to the BPS equations are smooth away from the boundary.  These IR boundary conditions will fix the $v_i$ and $w$ as functions of the mass-deformation parameters $\mu_i$.  We defer the discussion of the IR boundary conditions to Section~\ref{sec:num} (with more details in Appendix~\ref{app:numdetails}), and we now discuss the field theory interpretation of the UV parameters while performing a careful holographic renormalization analysis.

\subsection{Holographic renormalization}
\label{sec:holoren}

A careful holographic study involves holographic renormalization of various divergences realized by adding to the bulk action given above boundary terms that ensure that the on-shell action is finite and supersymmetric.\footnote{See \cite{Skenderis:2002wp} for a review on holographic renormalization.} We write the Euclidean regularized 
 action as the sum of terms
\be
   \label{Sreg}
   {\cal S}_\text{reg} = 
      \frac{1}{4 \pi G_5} \left[ {\cal S}_\text{bulk} + {\cal S}_\text{GH} + {\cal S}_\text{ct} + {\cal S}_\text{finite} \right]\,,
\ee
where $G_5 = \pi L^3 / (2 N^2)$ is the 5d Newton constant, ${\cal S}_\text{bulk} = \int d^5 x \sqrt{G} \mathcal{L}_E$ with $\mathcal{L}_E$ given by the Euclidean continuation of the Lagrangian in  \reef{bulkLag}.  The last three terms in \eqref{Sreg} are boundary terms  evaluated at a cutoff surface near the UV boundary;  they  will be defined shortly. The renormalized action ${\cal S}_\text{ren}$ is then the limit of ${\cal S}_\text{reg}$ as this cutoff is removed. 

While we are interested in supergravity solutions with $SO(5)$ invariance as exhibited in the  metric Ansatz \reef{MetricAnsatz1}, 
for the purpose of performing holographic renormalization it is necessary to be more general and write the 5d metric as
 \es{MetricAnsatz2}{
  ds^2 = \frac{d\rho^2}{4 \rho^2} + \frac 1\rho g_{ij}(x, \rho) dx^i dx^j
  \,, 
 }
where $\rho$ is defined as $\rho = e^{-2r}$ in terms of the radial coordinate $r$ appearing in \eqref{UVexp2}.  The conformal boundary is at $\rho = 0$.  In an expansion at small $\rho$, the solutions to the second order equations of motion following from varying the Euclidean Lagrangian ${\cal L}_E$ with respect to the fields take the form
 \es{Expansion}{
  g_{ij}(x, \rho) &= g_{0}(x) + \rho g_{2ij}(x) + \cdots \,, \\
  \alpha_i (x, \rho) &= \rho \log \rho\, \alpha_{0i}(x) + \tilde \alpha_{0i}(x) \rho + \cdots \,, \\
  \beta_i (x, \rho) &= \rho \log \rho\, \beta_{0i}(x) + \tilde \beta_{0i}(x) \rho + \cdots \,, \\
  \phi_i (x, \rho) &= \rho^{1/2} \phi_{0i}(x) + \rho^{3/2} \log \rho\, \phi_{2i}(x) + \rho^{3/2} \tilde \phi_{2i}(x) + \cdots \,, \\
  \varphi (x, \rho) &= \varphi_{0}(x) + \rho^{2} \log \rho\, \varphi_{2}(x) + \rho^{2} \tilde \varphi_{2}(x) + \cdots \,.
 }
The second order equations of motion impose various constraints between the coefficients of the small $\rho$ expansion.  

In order to define the boundary terms in \eqref{Sreg} precisely, one has to impose a cutoff at a small value of $\rho$, say $\rho = \eps$,  and write the boundary terms as integrals of local expressions on the cutoff surface.   A constant-$\rho$ 
surface is parameterized by $x^i$ and has induced metric 
$\gamma_{ij} =  \frac{1}{\rho} g_{ij}$.  
In terms of this data, the Gibbons-Hawking term is 
 \es{SGH}{
  {\cal S}_\text{GH} = \int d^4 x \sqrt{\gamma}\, \mathcal{K}
  \,,
 }
where ${\cal K} =  -2\rho \partial_\rho \log \sqrt{\gamma}$ is the extrinsic curvature of the cutoff surface.  This term ensures a well-defined variational principle. 

The next boundary term in \eqref{Sreg}, namely ${\cal S}_\text{ct}$, ensures that the action is finite when evaluated on any solution to the second order equations of motion.  This boundary term can be computed using the Hamilton-Jacobi approach to holographic renormalization proposed and discussed in 
 \cite{deBoer:1999tgo,Kalkkinen:2001vg,Martelli:2002sp,Papadimitriou:2004ap}  and  recently developed to easily-applicable form in \cite{Elvang:2016tzz}.   While details are given in  Appendix \ref{app:holoren}, here we simply present the result for the 10-scalar model.  The infinite counterterms are 
\be
  \label{Sctinf}
  \begin{split}
   {\cal S}_\text{ct} = &\int_{\rho = \eps} d^4x \sqrt{\gamma} 
   \bigg\{
      \frac{3}{2} + \frac{1}{8}R[\gamma] 
      +  \frac{1}{2} \sum_{i=1}^4 \phi_i^2 
 + \bigg( 1+ \frac{1}{\log\rho} \bigg) \Big(  6\beta_1^2 +2 \beta_2 + \sum_{k=1}^3 \alpha_k^2 \Big) 
 \\[1mm]
 & \hspace{2cm}
   -\log\rho 
   \bigg[
     \frac{1}{32} \bigg(  R[\gamma]_{ij}R[\gamma]^{ij} - \frac{1}{3} R[\gamma]^2\bigg) 
     - \frac{1}{24} R[\gamma]  \sum_{i=1}^4 \phi_i^2 
  \\[1mm]
 & \hspace{3.6cm}
     - \frac{2}{3}  \sum_{i=1}^4 \phi_i^4 
     + \frac{2}{3}  \sum_{1\le  i<j\le 4} \phi_i^2 \phi_j^2 
   \bigg]
   \bigg\}\;,
  \end{split}  
\ee
for the case when 
 the scalar fields do not depend on the $S^4$-coordinates. 

The last boundary term in \eqref{Sreg}, ${\cal S}_\text{finite}$, is a finite counterterm that is needed in order to ensure supersymmetry.   Note that this counterterm may not be unique.  Indeed, once a finite counterterm that preserves supersymmetry has been found, one has the freedom to add any other finite counterterms that by themselves preserve supersymmetry, if any such terms exist.  In the case of the ${\cal N} = 1^*$ theory on $S^4$, our field theory analysis shows that the $S^4$ partition function has the ambiguity in \eqref{AmbiguitySL} corresponding to two supersymmetric finite counterterms.  We expect that this field theory ambiguity is also present in the holographic dual.  We thus expect that one should be able to construct two independent supersymmetric finite counterterms, but we leave such a construction for future work.  

We now determine a possible ${\cal S}_\text{finite}$ that ensures supersymmetry by employing the Bogomolnyi method\footnote{This method has been proven fruitful in previous work on holographic duals to supersymmetric
theories on the round sphere \cite{Freedman:2013ryh,Bobev:2013cja}.}  for the 10-scalar theory reduced to flat space via the consistent truncation \reef{genGPPZ} that gives the generalization of the GPPZ model (see \reef{genGPPZ}). 
The details are described in Appendix \ref{app:holoren}. 
Since the finite counterterms are universal, they take the same form 
for the theory on $S^4$: here we simply present the result
\be
\label{Sfinite}
   \begin{split}
    {\cal S}_\text{finite} = &\int_{\rho = \eps} d^4x \sqrt{\gamma} \,
    \bigg\{
     \frac{1}{3} \big( \phi_1^4 +  \phi_2^4  + \phi_3^4 \big) + \phi_4^4
    + 2\big( \phi_1^2 +  \phi_2^2  + \phi_3^2 \big) \phi_4^2 
    \\& \hspace{2.1cm}
    + \big( \phi_1^2  \phi_2^2+  \phi_2^2 \phi_3^2 + \phi_3^2 \phi_1^2\big) 
    - 6\phi_1\phi_2\phi_3\phi_4
    \bigg\} \,.
    \end{split}
\ee
In terms of these boundary terms, one can then define the renormalized action as 
$\mathcal{S}_\text{ren} \equiv\lim_{\eps \to 0} \mathcal{S}_\text{reg}$, as indicated below \reef{Sreg}.

Before applying these results to compute physical observables, let us comment on the structure of the finite counterterms. It is clear from \reef{Sfinite} that the four scalars $\phi_i$ dual to the fermion bilinears in the field theory do not enter ${\cal S}_\text{finite}$ on an equal footing, as one could have expected for universal counterterms of the full $\mathcal{N}=8$ supergravity model. The reason is that the Bogomolnyi trick leads to flat-space finite counterterms that preserve the particular supersymmetry of the chosen truncation. In the $\mathcal{N}=1^*$ field theory we made a choice to add sources for three of the four available fermion bilinears.  Given this choice, it is then incompatible with supersymmetry to additionally turn on a source for the fourth fermion bilinear, namely the gaugino condensate dual to $\phi_4$.  For this reason it is not unreasonable that the Bogomolnyi procedure singles out  $\phi_4$ from the three other dimension-3 scalars. However, this means we cannot claim this result as {\em the} universal finite counterterms for the $\mathcal{N}=8$ supergravity model. Such a counterterm would have to respect the full $SU(4)$ R-symmetry of the $\mathcal{N}=4$ theory. 
In practice this has no consequence for our holographic renormalization: the terms 
$\phi_4^4$ and $\phi_i^2\phi_4^2$ in \eqref{Sfinite} never contribute to the 1-point functions since $\phi_4$ is not sourced but has only the vev-rate falloff.  The last term with $\phi_1\phi_2\phi_3\phi_4
$ in \reef{Sfinite} is  $SU(4)$-invariant by itself and it does contribute, but only to the gaugino condensate $ \<\Tr\, (\lambda  \lambda + \tilde \lambda \tilde \lambda)\>  \propto \< \mathcal{O}_{\phi_4}\>$. 

One may also worry that the Bogomolnyi result for the finite counterterms for solutions with $S^4$ radial slicing comes with the potential ambiguity of finite counterterms that depend on the spatial curvature of the cutoff surface. However, such finite counterterms are either independent of $\mu$ or $\mathcal{O}(\mu^2)$ and therefore they can never contribute to the universal part of $F$, therefore we can safely ignore them when discussing scheme-independent quantitites. These counterterms would be holographic incarnations of $I_0$ and $I_2$ in \reef{CounterSphereEvaluated}.

Let us finally comment on the third possible  finite local counterterm $I_3$ identified in the field theory analysis in Section \ref{sec:FE}. We have argued that $I_3$, which is cubic in the masses, cannot be modular invariant, so if there is a renormalization scheme that respects the $SL(2,\mathbb{Z})$ inherited from $\mathcal{N}=4$ SYM, then such a term would be disallowed. It is a natural question if we can identify this finite local counterterm on the supergravity side. The answer is yes: $I_3$ is captured by the local term $\frac{1}{\log\eps} \alpha_1 \phi_2 \phi_3 + \text{perms}$, which is finite at the boundary of the holographic space. The Bogomolnyi-approach can never produce this term: in the flat space limit where the Bogomolnyi trick is valid, the dimension-2 scalars $\alpha_i$ (dual to the mass-deformations $\tfrac{\im m_i}{a} \Tr\, (Z_i^2 + \tilde{Z}_i^2)$ required on $S^4$) are absent.  We will also not worry about this shortcoming of the Bogomolnyi approach, because we suspect that supergravity picks a regularization scheme that is consistent with the $SL(2, \Z)$ duality of ${\cal N} = 4$ SYM (which on the supergravity side is enhanced to $SL(2, \R)$ and is a global symmetry) and in which the coefficient of $I_3$ is required to vanish.

To conclude: for the practical purposes of our calculations, the Bogomolnyi results adopted from flat space to  $S^4$ are sufficient for fixing the finite counterterms by supersymmetry in our applications.  In the following section, we confirm that ${\cal S}_\text{reg}$ preserves supersymmetry by verifying that the 1-point functions computed with ${\cal S}_\text{reg}$ satisfy the expected supersymmetric Ward identities. 

\subsection{Free energy, one-point functions, and supersymmetric Ward identities}
\label{sec:1pt}

From the regularized action \reef{Sreg} 
we can easily calculate the 1-point functions for each of the 10 field theory bilinear operators that are dual to our scalars. For a dimension-$\Delta$ operator $\mathcal{O}_\Delta$ dual to a bulk scalar $\Phi_\Delta$, we have for $\Delta =2,3,4$:
 \es{VEVs}{
   \< \mathcal{O}_2 \> = \lim_{\eps\to 0}
   \frac{\log\eps}{\eps} \frac{1}{\sqrt{\gamma}} 
   \frac{\delta \mathcal{S}_\text{reg}}{\delta \Phi_2}\,,
   \qquad
   \< \mathcal{O}_3 \> = \lim_{\eps\to 0}
   \frac{1}{\eps^{3/2}} \frac{1}{\sqrt{\gamma}} 
   \frac{\delta \mathcal{S}_\text{reg}}{\delta \Phi_3}\,,
   \qquad
   \< \mathcal{O}_4 \> = \lim_{\eps\to 0}
   \frac{1}{\eps^{2}} \frac{1}{\sqrt{\gamma}} 
   \frac{\delta \mathcal{S}_\text{reg}}{\delta \Phi_4}\,.
 }
Up to an overall sign, we can be very precise about the normalizations of our operators dual to the 10 bulk scalars in our model.  Indeed, from \cite{Bobev:2013cja}, we know that in the ${\cal N} = 2^*$ theory the parameter $\mu = \pm \im m a$, where the $\pm$ refers to an overall sign we could not determine.  Our normalization of the $\mu_i$ parameters in the UV asymptotics \eqref{UVexp2} is consistent with that used in \cite{Bobev:2013cja} in the case where any two of the masses are equal and the third is set to zero, so we must have $\mu_i = \pm \im m_i a$.  From \eqref{UVexp2}, one can then deduce that the sources for the operators dual to $\alpha_i$ and $\phi_i$ are $\pm \frac{2 \im m_i}{a}$ and $\pm m$, respectively, because these are the coefficients of the leading terms in the large $r$ expansions of $\alpha_i$ and $\phi_i$. Comparing with the action \eqref{Lagrangian} and \eqref{mZExtra}, we see that these coefficients multiply, up to an overall sign,
 \es{OpList}{
  {\cal O}_{\alpha_i} &= \pm \frac 14 \Tr \left( Z_i Z_i + \tilde Z_i \tilde Z_i \right) \,, \qquad i = 1, 2, 3 \,,\\
  {\cal O}_{\phi_i} &= \pm \frac 12 \Tr \left( \chi_i^T \sigma_2 \chi_i + \tilde \chi_i^T \sigma_2 \tilde \chi_i 
   + \text{cubic in $Z_j$ and $\tilde Z_j$} \right)  \,, \qquad i=1, 2,3 \,.
 } 
A similar exercise involving $\beta_1$ and $\beta_2$ gives that the sources for the dual operators are $\frac 13 \left( m_1^2 + m_2^2 - 2 m_3^2\right) $ and $\frac 12 \left( m_1^2 - m_2^2 \right)$, respectively.  Also taking into account the Konishi operator and comparing to the $\sum_{i=1}^3 m_i Z_i \tilde Z_i$ term in the action \eqref{Lagrangian}, we obtain
 \es{OpBeta}{
  {\cal O}_{\beta_1} &= \frac 12 \Tr \left( Z_1 \tilde Z_1 + Z_2 \tilde Z_2 - 2 Z_3 \tilde Z_3 \right) \,, \\
  {\cal O}_{\beta_2} &= \frac 12 \Tr \left( Z_1 \tilde Z_1 - Z_2 \tilde Z_2 \right) \,.
 }
In order to determine the normalization of the operator dual to $\varphi$, we notice that $e^{2 \varphi} \to \frac{(1 + s)^2}{(1 - s)^2}$, so from \eqref{PhiCBdry} and the Lagrangian \eqref{Lagrangian}, we see that the operator appearing in the Lagrangian is $\frac{1}{16 \pi} e^{2 \varphi \vert_\text{bdy}} \Tr F_{\mu\nu} F^{\mu\nu}$.  The expectation value computed using \eqref{VEVs} is obtained by differentiating this expression with respect to the boundary value of $\varphi$, so 
 \es{OpDilat}{
  {\cal O}_\varphi = \frac 1{2 g_\text{YM}^2} \Tr F_{\mu\nu} F^{\mu\nu} \,.
 }
Lastly, we can determine the normalization of the operator dual to ${\cal O}_{\phi_4}$ using group theory:  as a state in the ${\bf 10} + \overline{\bf 10}$ of $SU(4)_R$, all the $\phi_i$ have the same norm, as can be seen from \eqref{SigmaComp}.  Since the kinetic term for the gaugino in the Lagrangian \eqref{Lagrangian} has the same normalization as that of the fermions $\chi_i$, we conclude that, up to a sign,
 \es{Opphi4}{
   {\cal O}_{\phi_4} &= \pm \frac 12 \Tr \left( \lambda^T \sigma_2 \lambda + \tilde \lambda^T \sigma_2 \tilde \lambda + \text{cubic in $Z_j$ and $\tilde Z_j$} \right) \,.
 }

To evaluate \eqref{VEVs}, we have to calculate the variation of the regularized action ${\cal S}_\text{reg}$, which is a sum of the bulk and counterterm actions.  The variations of the counterterm actions are straightforward to compute. The variation of the bulk action gives the equation of motion and therefore vanishes up to a boundary term. Writing generically the kinetic terms as $\frac{1}{2} \mathcal{G}_{IJ} \partial_r \Phi^I \partial_r \Phi^J$, we have  ($\partial_r = -2\rho \partial_\rho$)
\be
  \frac{1}{\sqrt{\gamma}} \frac{\delta S_\text{bulk}}{\delta \Phi^I}= 
  - 2\rho  \,\mathcal{G}_{IJ} \partial_\rho \Phi^J \bigg|_{\rho = \eps} \,.
\ee
Using these results and the fact that our leading order boundary metric $g_0{}_{ij}$ is the metric on a round $S^4$ with radius 1/2 (as can be seen from comparing \eqref{Expansion} with \eqref{UVexp2}), and therefore has $R_0 = 4 \times R_\text{unit $S^4$} = 48$, we have \be
  \begin{split}
    \label{1ptfct-aphi}
      \< \mathcal{O}_{\alpha_1}\> 
      &= \frac{N^2}{ \pi^2}  v_1\,,\\
      \< \mathcal{O}_{\phi_1}\> 
      &= \im \frac{N^2}{2 \pi^2} \bigg[ -4 v_1 + 2 \mu_1 + \frac{4}{3}\mu_1 \Big( 2 \mu_1 (v_1 + \mu_1)
      - \mu_2 (v_2 + \mu_2) - \mu_3 (v_3 + \mu_3)\Big)\bigg]
      \,,\\ 
  \end{split}
\ee
and similarly with obvious permutations for $\< Z_i Z_i \> $ and $\< \chi_i  \chi_i \> $ with $i=2,3$.

For the gaugino condensate, we find 
\be
    \label{1ptfct-phi4}
      \< \mathcal{O}_{\phi_4}\> 
      = \im \frac{N^2}{\pi^2}  w\,,
 \ee
and for the dilaton 
\be
    \label{1ptfct-varphi}
  \< \mathcal{O}_{\varphi}\> 
      = \frac{3 N^2}{\pi^2} (w - 2 \mu_1 \mu_2 \mu_3) \,.
\ee
The two remaining dimension-2 scalar mass operators give
\be
    \label{1ptfct-beta}
  \begin{split}
      \< \mathcal{O}_{\beta_1}\> 
      &= -\frac{N^2}{\pi^2} \Big(  \mu_1 (v_1 + \mu_1) +  \mu_2 (v_2 + \mu_2)  - 2 \mu_3 (v_3 + \mu_3)  \Big) \,,\\
      \< \mathcal{O}_{\beta_2}\> 
      &= - \frac{N^2}{\pi^2} \Big(  \mu_1 (v_1 + \mu_1) -  \mu_2 (v_2 + \mu_2) \Big)\,.
  \end{split}
\ee

Let us now consider the supersymmetric Ward identities that these 1-point functions have to obey. Consider the $\mathcal{N}=1$ supersymmetry transformations for the chiral multiplet $(Z_i, \chi_i)$ on $S^4$. From eq (A.10) in \cite{Bobev:2013cja}, we have
\be
 \delta Z_i = - \eps^T \sigma_2 \chi_i \,, 
 ~~~~~
 \delta \chi_i = \sigma^\mu \partial_\mu Z_i \tilde{\eps} + F_i \eps + \frac{\im}{a} Z_i \eps\,.
\ee
Thus  we have the supersymmetric Ward identity
\be
  \label{susyWI1}
  \delta \< \Tr\, Z_i \chi_i \>  = -\< \Tr\, \chi_i^T \sigma_2 \chi_i\> +  \frac{\im}{a} \< \Tr\, Z_i Z_i \> + \<\Tr\,  Z_i F_i \>\;.
\ee
Solving the field equation for the auxiliary field, we have $F_i = d\tilde{W}/d\tilde{Z}_i$. 
When there is a mass term $\frac{1}{2}m_i Z_i^2$ in the superpotential, the F-term part of \reef{susyWI1} therefore  contributes and we get
\be
  \label{susyWI2}
  \< \Tr\, \chi_i^T \sigma_2 \chi_i \> = \frac{\im}{a} \<\Tr\, Z_i Z_i \> + m_i \< \Tr\, Z_i \tilde{Z}_i \>\,.
\ee
The cubic term in the superpotential goes into the full definition of the 1-point function of the completion of  $\chi_i^T \sigma_2 \chi_i$ to a chiral primary. Therefore we do not write it explicitly, but it will be included in the  1-point functions $\< {\cal O}_{\phi_i}\>$.

Let us check the supersymmetric Ward identity \reef{susyWI2}  in supergravity. First, we have to identify properly what $\< \Tr\,  Z_i \tilde{Z}_i \>$ is in terms of $\< \mathcal{O}_{\beta_1}\>$ and $\< \mathcal{O}_{\beta_2}\>$. We note that from \eqref{OpBeta}, up to a term proportional to the expectation value of the  Konishi operator ${\cal O}_K \equiv \Tr (Z_1 \tilde{Z}_1  +  Z_2 \tilde{Z}_2 +  Z_3 \tilde{Z}_3) $, whose dual mode is absent in supergravity, we have
 \es{ZZToO}{
  \< \Tr\, Z_1 \tilde{Z}_1 \> 
   &=
  \frac{1}{3} \< \mathcal{O}_{\beta_1}\> + \< \mathcal{O}_{\beta_2}\> + \< {\cal O}_K \> \,,
  \\
   \< \Tr\, Z_2 \tilde{Z}_2 \> 
   &=
  \frac{1}{3} \< \mathcal{O}_{\beta_1}\> - \< \mathcal{O}_{\beta_2}\> + \< {\cal O}_K \> \,,
  \\
   \<\Tr\, Z_3 \tilde{Z}_3 \> 
   &=
  -\frac{2}{3} \< \mathcal{O}_{\beta_1}\> + \< {\cal O}_K \> \,.
 }
Using the holographic results \reef{1ptfct-aphi}-\reef{1ptfct-beta}, 
we then find that the following relation holds among our supergravity 1-point functions
 \es{Ward}{
  \< \mathcal{O}_{\phi_1}\> 
  &~=~ - 2 \im \< \mathcal{O}_{\alpha_1}\> 
  -  \im \mu_1
    \Big[ \frac{1}{3} \< \mathcal{O}_{\beta_1}\> + \< \mathcal{O}_{\beta_2}\> \Big]\,,
  \\
  \< \mathcal{O}_{\phi_2}\> 
  &~=~ - 2 \im \< \mathcal{O}_{\alpha_2}\> 
  -  \im \mu_2
    \Big[ \frac{1}{3} \< \mathcal{O}_{\beta_1}\> - \< \mathcal{O}_{\beta_2}\> \Big]\,,
 \\
  \< \mathcal{O}_{\phi_3}\> 
  &~=~ - 2 \im \< \mathcal{O}_{\alpha_3}\> 
  -  \im \mu_3
    \Big[ -\frac{2}{3} \< \mathcal{O}_{\beta_1}\> \Big]\,,
 }
up to a term linear in $\mu_i$. We see from \eqref{OpList} and \eqref{ZZToO} that \eqref{Ward} explicitly realizes the supersymmetric Ward identity \reef{susyWI2} plus its conjugate, with $m_i a \propto \pm \im \mu_i$.  It also indicates that there is a relative factor of $-1$ between the normalization of ${\cal O}_{\phi_i}$ and ${\cal O}_{\alpha_i}$ in \eqref{OpList}.
 
The $\mu_i$-derivative of the free energy can  be expressed in terms of the 1-point functions. We have
\be
  \frac{\pa F}{\pa \mu_i} = \int d^4 x \sqrt{g_0} 
  \bigg(
     \sum_{j=1}^4 \< \mathcal{O}_{\phi_j} \> \frac{\pa \phi_{0j}}{\pa \mu_i}
     +\sum_{k=1}^3 \< \mathcal{O}_{\alpha_k} \> \frac{\pa \alpha_{0k}}{\pa \mu_i}
     +\sum_{l=1}^2\< \mathcal{O}_{\beta_l} \> \frac{\pa \beta_{0l}}{\pa \mu_i}
  \bigg) \,,
\ee
using the UV expansion \reef{Expansion}.
(Details of the derivation can be found in \cite{Bobev:2013cja}.)
 
Using our results for the UV expansion of the fields \reef{UVexp2} and the 
1-point functions \reef{1ptfct-aphi} and  \reef{1ptfct-beta}, we find 
\be
  \label{dFdmu}
  \frac{\pa F}{\pa \mu_i} = \frac{N^2}{2\pi^2} \text{vol($S^4$)} \Big[ - 6 v_i + 2 \mu_i \Big]
   = - \frac{N^2}{2}  \Big[ v_i - \frac{1}{3} \mu_i \Big]
  \,.
\ee 
Here we have used that the volume of the 4-sphere with radius 1/2 is $\text{vol($S^4$)} = \frac{1}{2^4} \frac{8\pi^2}{3} = \frac{\pi^2}{6}$. For the case of the truncation to the 3-scalar $\mathcal{N}=2^*$ model on $S^4$, we differentiate $F$ with respect to $\mu_{12} \equiv \mu_1 = \mu_2$, and this produces $\frac{\pa F}{\pa \mu_{12}} =  - N^2  \Big[ v_{12} - \frac{1}{3} \mu_{12} \Big]$, which agrees with the result in \cite{Bobev:2013cja}.

Inspecting \reef{dFdmu}, we note that the shift ambiguity $v_i \sim v_i + \mu_i$ discussed below \reef{UVexp2} is equivalent to the statement that the free energy $F$ is ambiguous at quadratic order in $\mu$. This is precisely what we expect from the field theory finite counterterm $I_2$ in \reef{CounterSphereEvaluated}. Note that we are not sensitive to the mass-independent finite counterterm $I_0$ in \reef{CounterSphereEvaluated} since we are directly evaluating $\frac{\pa F}{\pa \mu_i}$.

Once we fix smooth IR boundary conditions for our holographic flows, we find that $v_i$ and $w$ become functions of the $\mu_i$. From the analysis in Section \ref{sec:FE} we know that we have to take further derivatives of $F$ with respect to the mass parameters $\mu_i$ in order to obtain a universal result. Therefore the linear term in $\mu_i$ in $\frac{\pa F}{\pa \mu_i} $ is non-universal and will drop out. An example of a universal quantity is given by
\be
  \label{freeE}
  \frac{\pa^3 F}{\pa \mu_i^3} = - \frac{N^2}{2}   \frac{\pa^2 v_i}{\pa \mu_i^2}\,.
\ee
To proceed, we need to know $v_i$ in terms of $\mu_i$. To find this we have to resort to a numerical analysis for the truncations of the theory to the $\mathcal{N}=1^*$ one mass and equal mass models, which are described in  Sections \ref{sec:holoLS} and \ref{sec:holoEqM}.

\subsection{Further truncations of the $10$-scalar model}
\label{sec:truncs}

As a further simplification, we focus only on certain consistent truncations of the 10-scalar model:
\begin{itemize}
\item {\bf $\mathcal{N}=2^*$ model}: this 3-scalar model  is obtained by keeping ${\alpha}_1 = {\alpha}_2$,  ${\phi}_1={\phi}_2$, and $\beta_1$ non-zero, while imposing ${\alpha}_3 = {\phi}_3={\phi}_4={\varphi}=\beta_2=0$.
This is achieved by setting 
\be
   \text{$\mathcal{N}=2^*$ :}~~~
   z_4=z_2=0\,,~~~ \tilde{z}_4=\tilde{z}_2=0\,,~~~z_3 = z_1\,,~~~\tilde{z}_3 = \tilde{z}_1\,,~~~
   \beta_2 = 0\,.
\ee
One recovers the action and the BPS equation of our previous work \cite{Bobev:2013cja} after identifying $z_{\text{there}}= z_1$, $\tilde{z}_{\text{there}} = \tilde{z}_1$, and $\eta_{\text{there}}=e^{-\beta_1}$.

\item {\bf $\mathcal{N}=1^*$ one mass model}: This model is obtained by taking the limit where two of the masses vanish. It can be obtained by consistent truncation from the 10 scalar model in three equivalent ways, depending on which mass is kept non-zero:
\als{
\label{LStruncSec5}
\text{$m_1\neq 0$:} \quad & \quad z=z_1 = z_2 = z_3 = z_4\,~~\text{and}~~ \b= 2\b_1 = \frac{2}{3}\b_2\;,
\\
\text{$m_2\neq 0$:} \quad & \quad z=z_1 = -z_2 = z_3 = -z_4\,~~\text{and}~~ \b= 2\b_1 = -\frac{2}{3}\b_2\;,
\\
\text{$m_3\neq 0$:} \quad & \quad z=z_1 = z_2 = -z_3 = -z_4\,~~\text{and}~~ \b= -\b_1,~~ \b_2 = 0\,,
}
and in each case the $\zt_i$'s are truncated the same way as the $z_i$'s. The result is a 3-scalar model with fields $z$, $\tz$, and $\beta$, which we analyze in further detail in Section \ref{sec:holoLS}.
\item {\bf $\mathcal{N}=1^*$ equal-mass model}: in this model, we must have ${\alpha}_1 = {\alpha}_2= {\alpha}_3$ and  ${\phi}_1={\phi}_2={\phi}_3$ non-zero to account for the sources of the associated field operators. It turns out to be inconsistent to turn off the gaugino condensate ${\phi}_4$ and the five-dimensional dilaton/axion $\varphi$. The resulting 4-scalar model with $(z_1,\tilde{z}_1,z_2,\tilde{z}_2)$ is obtained from the 10-scalar model by setting 
\be
\label{eqmasstrunc}
   \text{$\mathcal{N}=1^*$ equal-mass:}~~~
   z_4=-z_3=-z_2\,,~~~ \tilde{z}_4=-\tilde{z}_3=-\tilde{z}_2\,,~~~
   \beta_1=\beta_2 = 0\,.
\ee
We study this model in Section \ref{sec:holoEqM}. It should be noted that for this case, some of the BPS equations \eqref{BPSAll} are not quite adequate to use since in deriving them we assumed that $\beta_1 \ne 0$. Instead one extracts the BPS equations from the matrix equations \reef{matrixBPS1} and \reef{matrixBPS2} without using the $\beta_1$ or $\beta_2$ equations.  

\item {\bf 7-scalar model}: There is a truncation of the 10-scalar model that encompasses all of the models discussed  above, namely the 7-scalar model obtained by setting 
\be
   \text{$\mathcal{N}=1^*$ 7-scalar model:}~~~
   z_4=-z_2\,,~~~ \tilde{z}_4=-\tilde{z}_2\,,~~~
   \beta_2 = 0\,.
\ee
\item {\bf 6-scalar model for $\mathcal{N}=2^*$ on $S^4$ with condensate}: 
Setting $\tilde{z}_2 = z_2$ in the 7-scalar model gives a 6-scalar model in which ${\phi}_3 =-{\phi}_4 \ne0$, but in which the 5d dilaton is non-trivial. This is an expanded version of the $\mathcal{N}=2^*$ model. Note that ${\alpha}_3$ flows non-trivially, but it will not have a source in the UV. The same is true for ${\phi}_3 =-{\phi}_4$, since neither of the corresponding operators will have a mass term for the $\mathcal{N}=2^*$ theory. We comment briefly on its properties in Section \ref{sec:holoN2revisited}.
\item {\bf 6-scalar model for $\mathcal{N}=1^*$ on $\mathbb{R}^4$ with condensate}: 
There exists a limit of our 10 scalar model suitable for constructing supergravity domain walls with flat slicing. In this limit the scalars ${\alpha}_{1,2,3}$ are absent and so is the dilaton. This truncation is obtained as 
\be 
   \label{genGPPZ}
     \text{$\mathcal{N}=1^*$ 6-scalar model on $\mathbb{R}^4$:}~~~
   \tilde{z}_i=-z_i\,
   ~~~~\text{for}~~i=1,2,3,4\,.
\ee  
The model resulting from this is a generalization of the GPPZ model \cite{Girardello:1999bd} to the case of three unequal masses. It was studied before in \cite{Khavaev:2000gb} (see also \cite{Bobev:2010de}) and provides a useful limit for determining the finite counterterms in holographic renormalization.  In particular, it plays a big role in the determination of the finite counterterm in \eqref{Sfinite} using the Bogomolnyi trick.

\end{itemize}
We have  checked explicitly that all of the above truncations are consistent with the BPS equations.

\section{Numerical analysis of the holographic models}
\label{sec:num}

We construct numerical holographic flows for two of the $\mathcal{N}=1^*$ truncations: the one-mass model and the equal-mass model. To do so, we need to impose suitable boundary conditions in the IR\@. The field theory on $S^4$ is expected to become trivial at energies below the scale set by the sphere, so a natural choice of boundary conditions in the bulk IR is that the metric caps off smoothly as flat space 
\be
  \label{IRmetric}
  ds^2 = dr^2 + (r-r_0)^2\,ds_{S^4}^2\,
\ee
with no conical singularity. The same IR boundary condition was also used for the holographic description of $\mathcal{N}=2^*$ on $S^4$ in \cite{Bobev:2013cja}. 
Due to the shift symmetry in the radial coordinate, we are free to choose $r_0 = 0$ in \reef{IRmetric}.  

The BPS equations are  analyzed  in the IR subject to the above boundary conditions. The warp factor $e^{2A}$ and the scalars are simple power expansions in $r$, which we present  in Appendix \ref{app:numdetails}. The general numerical approach is to shoot from the IR and generate the numerical flows of the scalars and warp factor.  At large values of $r$, the flow will approach the UV solution \reef{UVexp2}, with an extra parameter $r_\text{UV}$ introduced by shifting $r \to r - r_\text{UV}$ that accounts for our choice $r_0 = 0$.  The flow can then be fitted to determine the values of the UV parameters $\mu_i$, $v_i$, and $w$. Because the  smooth IR boundary condition leaves fewer IR parameters than UV parameters, this effectively means that $v_i$ and $w$ become functions of $\mu_i$. (The source $s$  of the dilaton $\varphi$ decouples as noted in \reef{shiftsym}.) 

We refer the reader to Appendix \ref{app:numdetails} for details on the numerical analysis. The remainder of this section presents the physical results for the $\mathcal{N}=1^*$ one-mass model (Section \ref{sec:holoLS}) and  equal-mass models (Section \ref{sec:holoEqM}). These two limits allow us to deduce information about the free energy of the  $\mathcal{N}=1^*$ theory with three general masses (Section \ref{sec:compare}). Finally, we discuss the possibility of generating more general ${\cal N}=2^*$ flows in (Section \ref{sec:holoN2revisited}) and argue that the solutions presented in \cite{Bobev:2013cja} capture all of the physically inequivalent flows.

\subsection{One-mass model}
\label{sec:holoLS}

The truncation of the $\mathcal{N}=1^*$ supergravity dual to the model with only one non-zero mass has 3 scalar fields, $z$, $\tz$, and $\beta$. It is obtained as the consistent truncation of the 10 scalar model described in \reef{LStruncSec5}. The BPS equations reduce to three differential equations for $z$, $\zt$,  and $\b$, and an algebraic equation for the warp factor $A$. In this limit, the scalar fields $\phi_4$ and $\varphi$ (dual to the gaugino condensate and the $\text{Tr} F_{\mu\nu}^2$, respectively) decouple, which greatly simplifies the problem. Solving this system of first order equations leads to three integration constants in the UV: a mass term $\mu$, a vev $v$, and the shift $r_\text{UV}$ in the radial coordinate; the UV solution is obtained from truncating the UV expansion in \eqref{UVexp2}, setting for example $\mu \equiv \mu_1$, $v \equiv v_1$, while 
$\mu_2=\mu_3=0$, $v_2=v_3=0$, and  $w=0$.  In the IR, the smoothness condition leaves only one free parameter (see \reef{IRinLS}). Via the numerical analysis, we therefore find $v = v(\mu)$. Due to the symmetry $(z, \tilde z )\to (-z, -\tilde z )$ of the model, $v$ must be an odd functions of $\mu$. In principle, $\mu$ can be any complex number.

The physical relevance of $v$ is that it encodes the source-induced vev of the chiral field 
$\< Z Z \> ~\rightsquigarrow~ \< \mathcal{O}_{\alpha}\> = 2 v$, where $Z$ is the chiral scalar whose mass is turned on in the field theory. This vev is subject to finite counterterms. This is seen on the holographic side by the fact that a shift of $r$ changes $v$ by a term proportional to $\mu$, as can be seen from \reef{UVexp2}. Thus only $v''(\mu)$ is universal. 
Indeed, the universal part of the free energy is proportional to $v''(\mu)$:
\be
     {\cal N}=1^* ~\text{one-mass model}:~~~~~
     \frac{d^3F}{d\mu^3} = - \frac{1}{2}N^2 v''(\mu)\,.
\ee
In the following, we will present the results for the free energy $\frac{d^3F}{d\mu^3}$ as extracted from the numerics. 

\begin{figure}[t!]
\begin{center}
  \raisebox{-3cm}{\includegraphics[width=.45\textwidth]{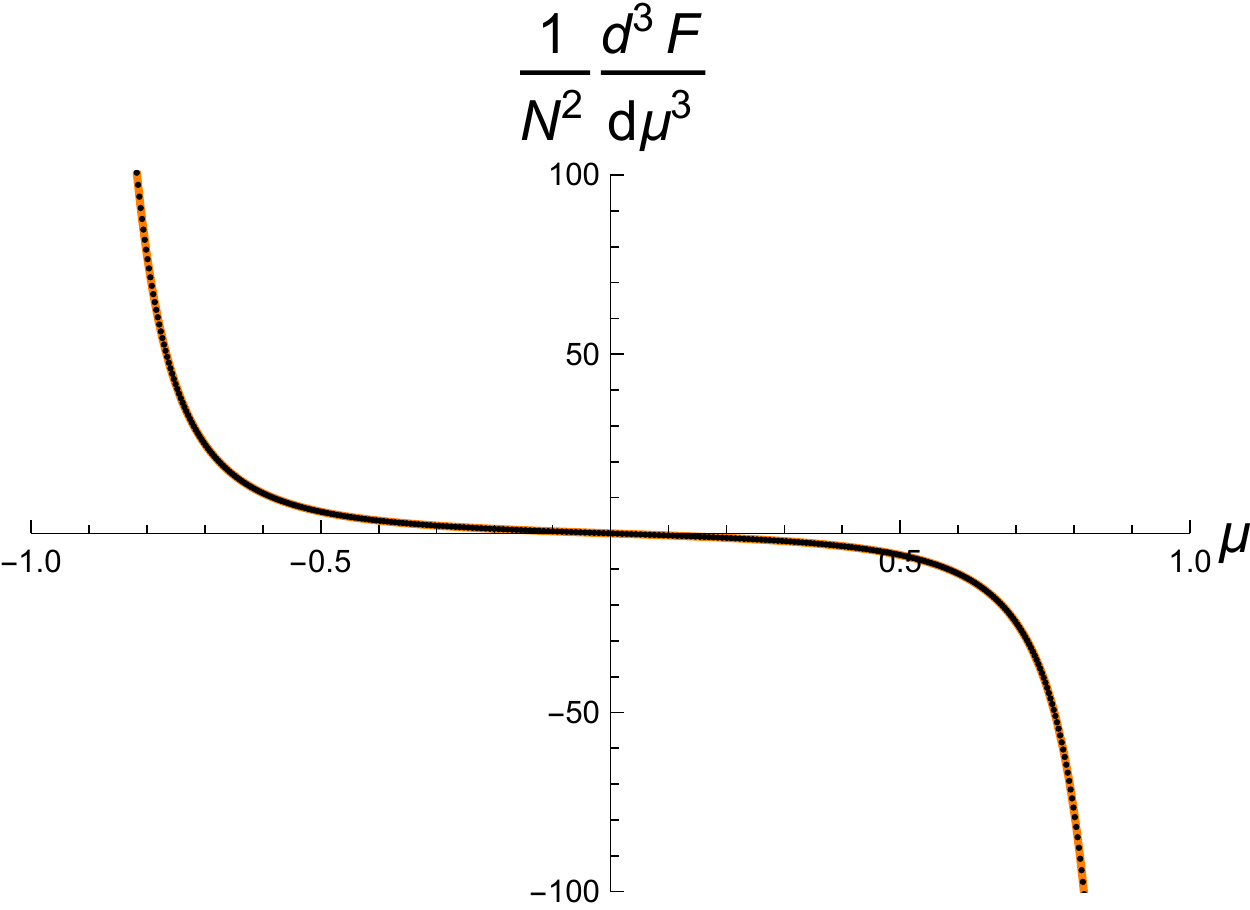}}~~~~      
  \raisebox{-3cm}{\includegraphics[width=.45\textwidth]{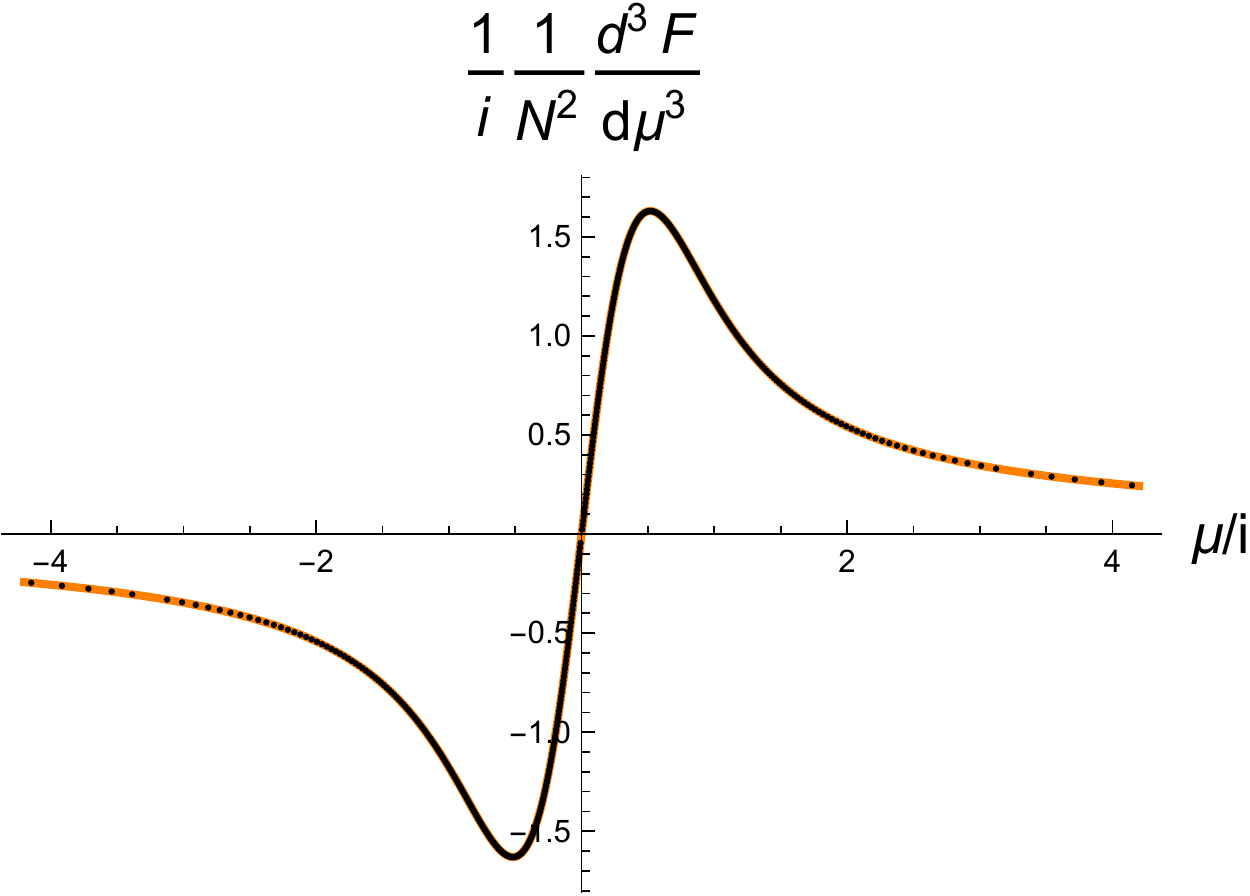}}
\end{center}
  \caption{Plots showing the universal part of the free energy $d^3F/d\mu^3$ as a function of the dimensionless mass $\mu = \pm \im ma$ for the one-mass model. The plot on the left (right) shows the results for real (purely imaginary) values of $\mu$. Orange curves are the interpolation function, the black points indicate the data points the interpolation function is based on.}
  \label{fig:LS}
\end{figure}

\noindent {\bf Real axis}\\
For real values of $\mu$, we find solutions with  $-1<\mu<1$ only. 
The results for $\frac{d^3F}{d\mu^3}$ are shown in Figure \ref{fig:LS}.

At $\mu = \pm 1$, we find a pole of order 3 in $v''(\mu)$ with residue $\mp\frac{1}{2}$:\footnote{Examining the behavior of $v''$ near $\mu=\pm1$, the uncertainty in the power is estimated to be about $0.1\%$.} 
\be
  \frac{d^3 F}{d\mu^3} \sim - \frac{4}{(1-\mu^2)^3} N^2 
  ~~~\text{as}~~~\mu \to \pm1\,. 
\ee
Near $\mu=0$, we have 
\be 
  \label{LSslope}
  \frac{d^3F}{d\mu^3} = \big(-5.653\mu + 14.4 \mu^3 + O(\mu^5) \big) N^2\,,
\ee
where the last digit of each coefficient has an uncertainty of about $\pm 2$. 

\noindent {\bf Imaginary axis}\\
For purely imaginary values of $\mu$, we generate flows with any value of $-\infty < \mu/\im < \infty$. Figure \ref{fig:LS} shows $\frac{d^3F}{d\mu^3}$ vs.~$\mu$. The slope at small $\mu$ agrees with the analytic continuation $\mu \to \im\mu$ of  \reef{LSslope}.
On the imaginary axis we find no poles, but local extrema at $\mu = \pm 0.519\im$ where 
$\frac{d^3F}{d\mu^3} =  \pm1.630\im$.

 As $\mu \to \pm \im \infty$, the value of $\frac{d^3F}{d\mu^3}$ asymptotes to zero as $1/\mu$,  where the numerator has been determined to about $0.1\%$. To check this large-$\mu$ falloff, we have extended the numerics into the complex plane away from the imaginary axis, and the results indicate that the asymptotic behavior persists everywhere except on or near
 the real axis where we have not been able to extend the solution beyond the poles at $\mu = \pm 1$.\footnote{This may at first seem surprising: if $v''$ has isolated poles at $\mu = \pm 1$, one should expect to find solutions ``around the poles". However, recall that what we are directly computing in the numerical solution is $v$, not $v''$. A function like $v'' \sim 1/(1-\mu^2)^3$  has only isolated poles, but $v$ has logarithmic branch cuts and we cannot generate numerical solutions on or near the branch cuts. This was also the case in the analysis of the holographic dual of $\mathcal{N}=2^*$ on $S^4$ in \cite{Bobev:2013cja} where $v(\mu) = -2 \mu - \mu \log(1-\mu^2)$.}

The behavior of $\frac{d^3F}{d\mu^3}$ is qualitatively very similar to that of the same quantity in $\mathcal{N}=2^*$. One might then naively guess that the function of interest has the form $\frac{d^3F}{d\mu^3}  =  \frac{\mu(a + b \mu^2 + c \mu^4)}{ ( 1 - \mu^2 )^3 }$.
 This function has zeros of order 1 at $\mu = 0$ and $\mu = \infty$ and possibly other locations. Its only poles are of order 3 at $\mu = \pm1$. The numerical results listed above overconstrain the three parameters and no miracle occurs to permit a solution. We have tried fitting other functions but the analytic continuation from the real to the imaginary axes places strong constraints and we have not found an analytic function that fits our numerical results for both real and imaginary values of $\mu$. 

In the limit $\mu \to \pm \im\infty$, the flows approach a very simple solution with constant scalars and an AdS warp factor
\be
  \label{LS0scalarsMT}
  \begin{split}
  &z = - \zt = \pm \im \sqrt{7 - 4 \sqrt{3}}\,,
  ~~~~
  e^{3\beta} = \sqrt{2} \,,\\
  &e^{2A(r)} = L_\text{LS}^2 \sinh^2(r/L_\text{LS}) 
    ~~~\text{with}~~~L_\text{LS} = \frac{3}{2^{5/3}}\,,
    \end{split}
\ee
for the entire range $0< r < \infty$. The scalar potential also becomes constant, $V = - \frac{8\, 2^{1/3}}{3} = -\frac{3}{L_\text{LS}^2}$.  Thus the solution corresponds to Euclidean AdS with radius $L_\text{LS}$. It is the dual of the well-known Leigh-Strassler fixed point that is reached in the IR of the RG flow of $\mathcal{N}=1^*$ with one mass turned on in flat space \cite{Leigh:1995ep}. The holographic dual RG flow with flat-sliced domain walls of the flow from $\mathcal{N}=4$ SYM in the UV to this Leigh-Strassler fixed point CFT in the IR was studied by FGPW in \cite{Freedman:1999gp}. Here we do not find the whole FGPW flow, but only the IR Leigh-Strassler fixed point. This is very reasonable: we are studying $S^4$-sliced domain wall dual RG flows and  the limit $|\mu| \to \infty$ corresponds to the sphere-decompactification limit $a \to \infty$. In particular we note that the scalar $z+\tz$, which in the UV is dual to the operator $\frac{\im m}{a}\text{Tr}\, (Z^2 + \tilde Z^2)$ specific to the $S^4$, vanishes.  Since there are no other scales in the problem, our large mass limit $|\mu| \sim |m a| \to \infty$ leaves no room for any RG flow and solution is fixed to be the IR CFT.

\subsection{Equal-mass model}
\label{sec:holoEqM}

The $\mathcal{N}=1^*$ equal-mass supergravity model with 4 scalars $(z_1,\tilde{z}_1,z_2,\tilde{z}_2)$ is obtained from the 10-scalar model via the consistent truncation detailed in \reef{eqmasstrunc}. We impose smooth boundary conditions \reef{IRmetric} in the IR. Solving the BPS equations in a small-$r$ expansion, leaves two free parameters $a_0$ and $b_0$ in the IR solution. Since the BPS equations have a shift symmetry for the dilaton $\varphi \to \varphi + \text{constant}$, only one combination of $a_0$ and $b_0$ is relevant for the UV parameters $\mu$, $v$, and $w$. Solving the BPS equations numerically (see details in Appendix \ref{app:numdetails})  allows us to determine $v$ and $w$ as functions of $\mu$. The symmetry $(z_i, \zt_i) \to (-z_i, -\zt_i)$ of the equal-mass supergravity model implies that both $v$ and $w$ are odd functions of $\mu$.

On the real axis, we find solutions with arbitrarily large $\mu$. On the imaginary axis, however, we only find solutions with $-2.318 \lesssim \im\mu \lesssim 2.318$.  We describe the results of the numerical analysis below.  

\begin{figure}[t!]
\centerline{
 \raisebox{-2cm}{\includegraphics[height=4.6cm]{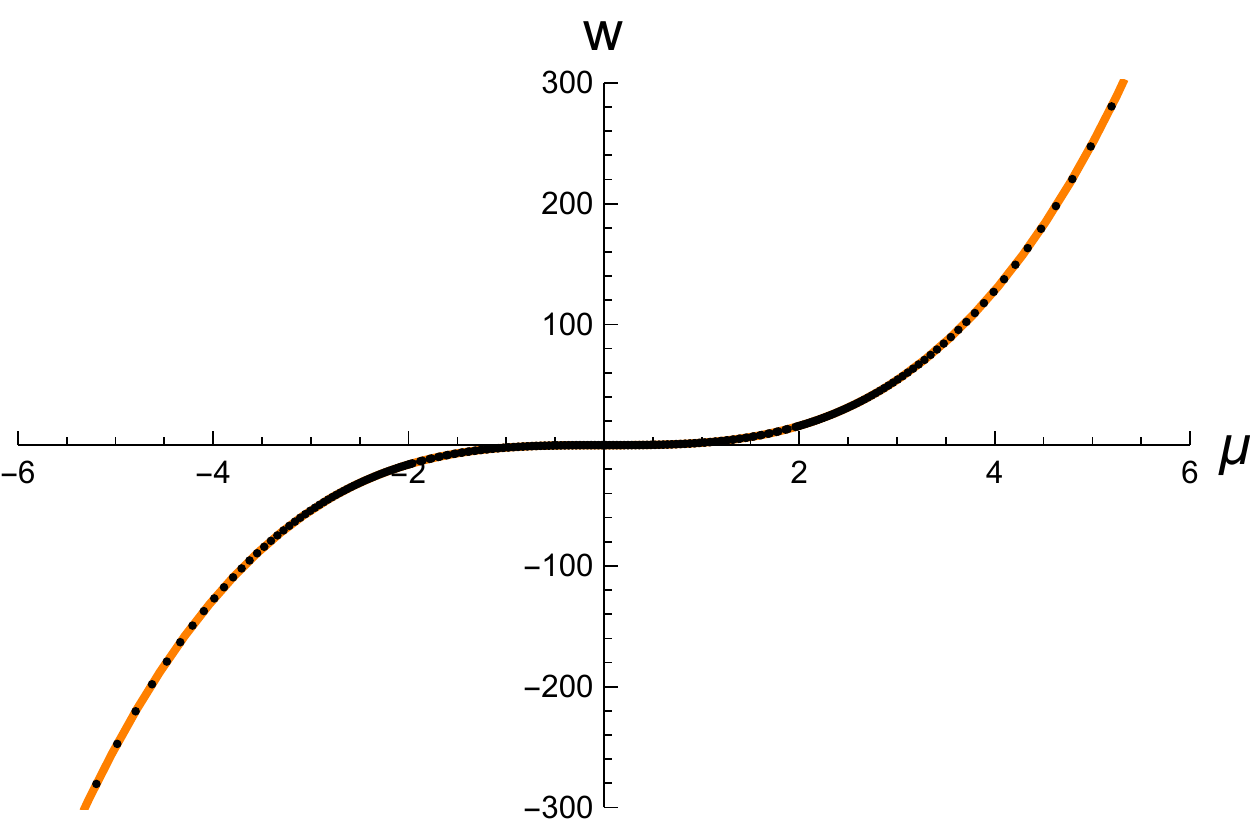}~~~~~~
 \includegraphics[height=4.6cm]{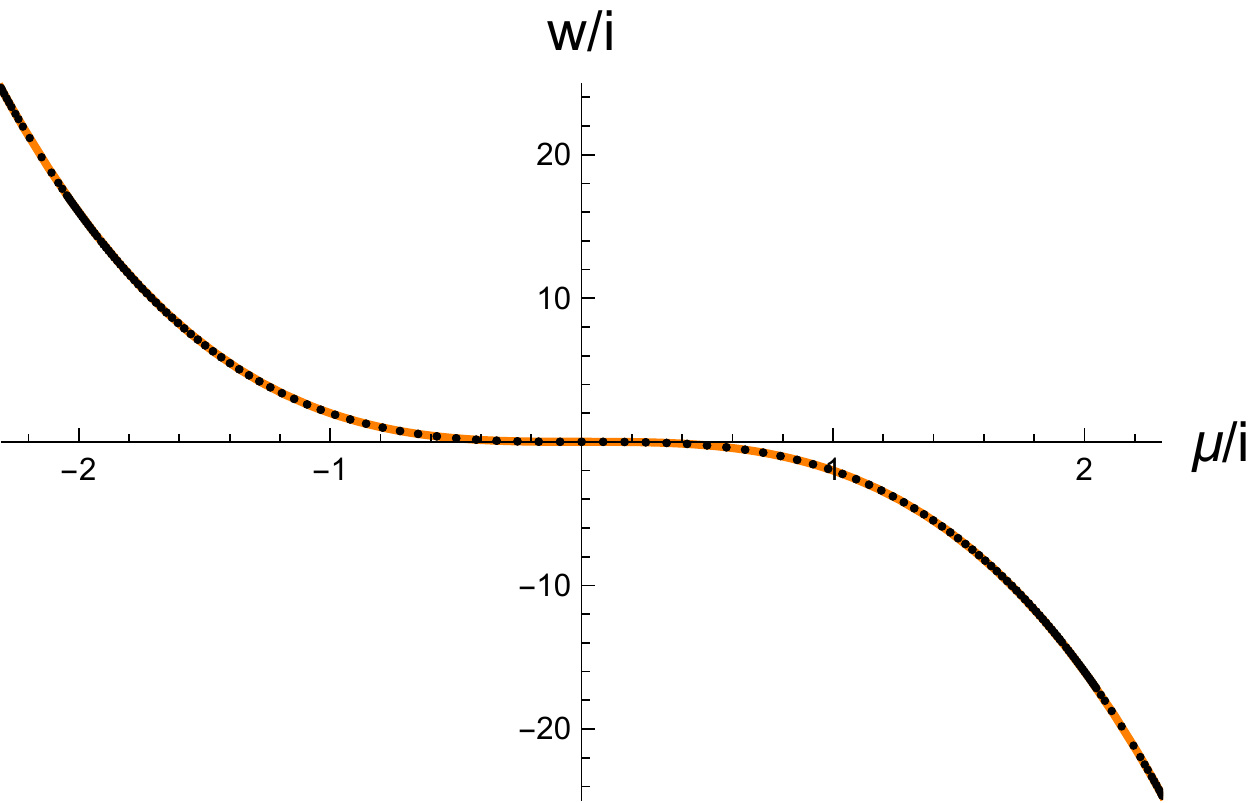}}}
 \caption{The holographic variables $w$ encodes the gaugino condensate $\< \lambda  \lambda \> \sim w(\mu) N^2$ with $\mu = \pm  \im ma$. We find numerically that $w(\mu) = 2\mu^3$; the plots show the results of $w$ vs.~$\mu$ for $\mu$ real and purely imaginary.
  Orange curves are the interpolation function, the black points indicate the data points the interpolation function is based on. The orange curves are indistinguishable from $w(\mu) = 2 \mu^3$ within our precision.}
 \label{fig:wVSmu}
\end{figure}

\noindent {\bf Gaugino condensate.}
By \reef{1ptfct-phi4} the gaugino condensate is determined by $w(\mu)$. The numerical results for $w$ vs.~$\mu$ are shown in Figure \ref{fig:wVSmu}. 
The curves for $w$ vs.~$\mu$ are fitted to the function $2 \mu^3$ for both real and purely imaginary values to better than a relative error of $10^{-5}$; the relative residue is noise in the numerics. This is strong numerical evidence that indeed
\be 
  \label{wresult}
  w(\mu) = 2\mu^3 \,,
\ee 
which via $\mu = \pm \im ma$ and \reef{1ptfct-phi4} predicts $\<\Tr\, ( \lambda  \lambda + \tilde \lambda \tilde \lambda) \> = \frac{4}{\pi^2} m^3 N^2$ (up to a sign) at large $N$ and large 't Hooft coupling in the equal-mass truncation of $\mathcal{N}=1^*$ on $S^4$. For unequal masses, this translates to $\< \Tr\, ( \lambda  \lambda + \tilde \lambda \tilde \lambda) \> = \frac{4}{\pi^2} m_1 m_2 m_3 N^2$ (up to a sign).  In particular, this is independent of the gauge coupling. 

\noindent {\bf Expectation value $\< \Tr\, F_{\mu\nu} F^{\mu\nu} \>$.}
From \eqref{OpDilat}, the field theory dual of the dilaton 1-point function $\< O_\varphi \>$ is  $\frac{1}{2 g_\text{YM}^2} \<\Tr\, F_{\mu\nu} F^{\mu\nu} \>$. By \reef{1ptfct-varphi} we have
\be
  \frac{1}{2 g_\text{YM}^2} \<\Tr\, F_{\mu\nu} F^{\mu\nu} \> =
 \< \mathcal{O}_{\varphi}\> = \frac{3N^2}{\pi^2} (w - 2 \mu_1 \mu_2 \mu_3) \,.
\ee
It follows from our result \reef{wresult} that our holographic model predicts 
$ \<\Tr\, F_{\mu\nu} F^{\mu\nu} \> = 0$ for $\mathcal{N}=1^*$ on $S^4$ at leading order in the supergravity limit.

\begin{figure}[t!]
\centerline{
 \raisebox{-2cm}{\includegraphics[height=5.1cm]{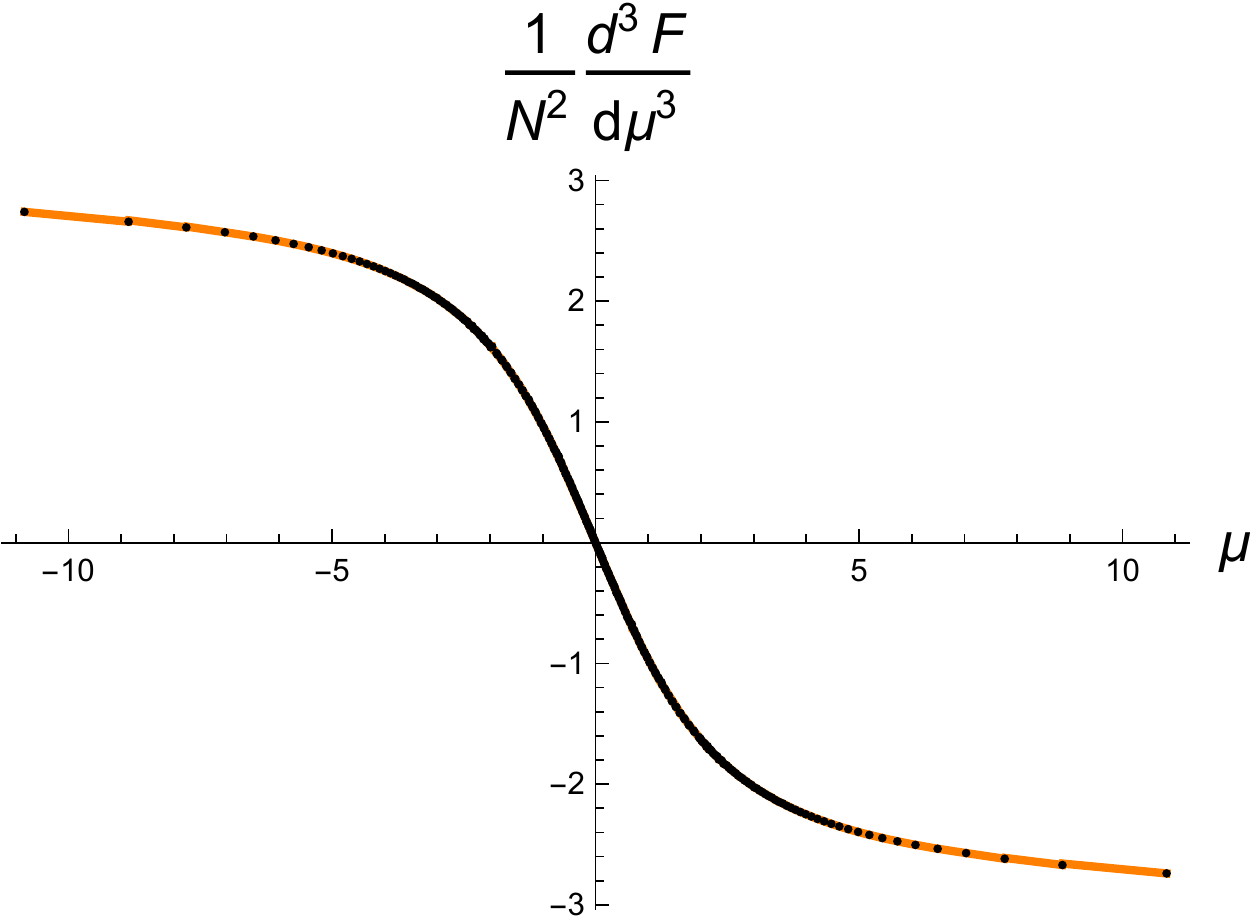}~~~~~~
 \includegraphics[height=5.1cm]{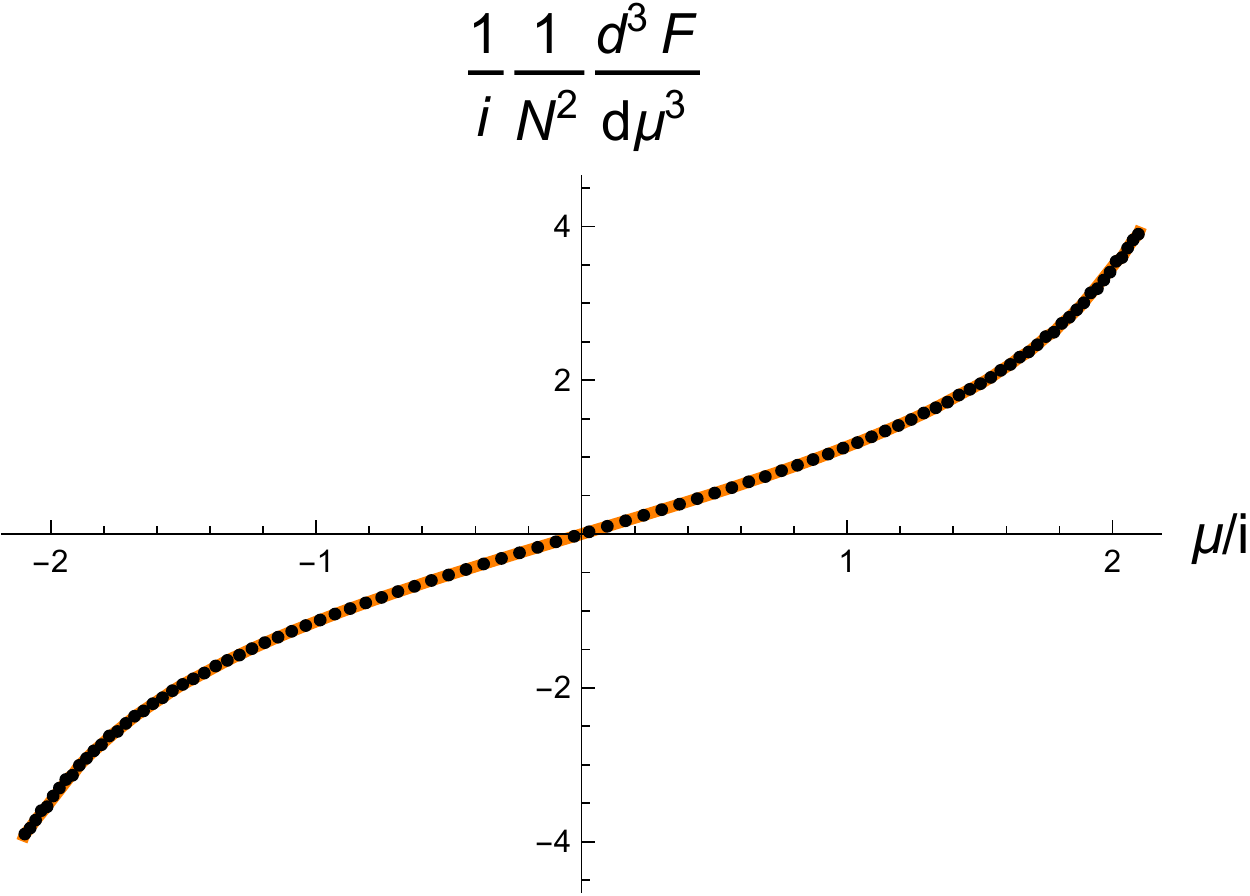}}}
 \caption{Plots showing the universal part of the free energy $d^3F/d\mu^3$ as a function of the dimensionless mass $\mu = \pm \im ma$ for the equal-mass model. For $\mu$ real we find solutions with any value, and $d^3F/d\mu^3 \to \mp 3N^2$ for large $\mu$. When $\mu$ is purely imaginary, we only find solutions with $-2.318 \lesssim \im\mu \lesssim 2.318$; as $\mu$ approaches this values, the interpolation function becomes increasingly noisy and does not appear to be reliably determined beyond $|\mu| \gtrsim 2.1$, hence we restrict the plot to this range. Orange curves are the interpolation function, the black points indicate the data points the interpolation function is based on.}
 \label{fig:vVSmuEqMass}
\end{figure}

\noindent {\bf Free energy $F$.} As found with holographic renormalization in \reef{freeE}, the free energy in the equal-mass model is
\be
  \label{freeEeqm}
  \frac{\pa^3 F}{\pa \mu^3} = - \frac{3}{2}N^2\,  v''(\mu)\,.
\ee
In our numerical analysis, we determine `data points'  $(\mu,v)$. This can be turned into an interpolation function which can then be differentiated twice with respect to $\mu$ to find the result for  $\frac{\pa^3 F}{\pa \mu^3}$ shown in Figure \ref{fig:vVSmuEqMass}.

On the real axis, we find that 
\be
  \frac{\pa^3 F}{\pa \mu^3} \to \mp 3 N^2 + \mathcal{O}\Big(\frac{1}{\mu}\Big)
  ~~~\text{as}~~~
  \mu \to \pm \infty\,.
\ee  
We have verified this asymptotic behavior with $\mu$ up to 100.

For small $\mu$, we find that 
\be
  \label{eqmassslope}
  \frac{\pa^3 F}{\pa \mu^3} = \big(-1.041 \mu + \mathcal{O}(\mu^3)\big) \,N^2\;.
\ee 
We have not been able to reliable determine the coefficient of the $\mu^3$-term in \eqref{eqmassslope}. In the field theory, $\mu$ translates to $\pm \im ma$.

\noindent {\bf Chiral condensate $\<\Tr\, Z^2\>$.} 
 As discussed in Section \ref{sec:holoren}, $\<\Tr\,( Z_i^2 + \tilde Z_i^2) \> =2 \< \mathcal{O}_{\alpha_i}\> = \frac{2N^2}{\pi^2} v_i$. For the equal scalar model, this becomes $ \sum_{i=1}^3 \<  \Tr\, (Z_i^2 + \tilde Z_i^2) \> = 2 \< \mathcal{O}_{\alpha}\> = \frac{6N^2}{\pi^2} v$. In the large-$\mu$ limit (for real $\mu$, which translates into pure imaginary $m$), we therefore have $\sum_{i=1}^3 \<  \Tr\, (Z_i^2 + \tilde Z_i^2) \> \propto   N^2 \mu^2 \, \text{sgn}\, \mu$.  For unequal masses, this should be interpreted in the field theory as $\< \Tr\, (Z_1^2 + \tilde Z_1^2) \> \propto  N^2 \mu_2 \mu_3  \, \text{sgn}\, \mu_1 $ etc. 
 
\subsection{General  $\mathcal{N}=1^*$ result at small mass}
\label{sec:compare}

Our models all have $\tilde{m}_i  =  m_i$. Since the holographic dictionary identifies $\mu_i = \pm \im m_i a$, we write the expectation for the universal part of the free energy as
\be
   \label{Fmus}
   \begin{split}
    F_{S^4}^\text{univ} /N^2
    \,=\,&
    A_1 \Big( \mu_1^4 + \mu_2^4 + \mu_3^4 \Big)
    +
    A_2 \Big(\mu_1^2 + \mu_2^2 + \mu_3^2 \Big)^2
    + 
    \\
    &
    +
    B_1 \Big( \mu_1^6 + \mu_2^6 + \mu_3^6 \Big)
    +
    B_2 \Big(\mu_1^2 + \mu_2^2 + \mu_3^2 \Big)^3
    +
    B_3 \,\mu_1^2 \mu_2^2 \mu_3^2 
    + O(\mu^8) \,.
    \end{split}
\ee
No other combinations are linearly independent from those above. 

For the three truncated models, it follows from \reef{N2freeE}, \reef{LSslope}, and \reef{eqmassslope} that we have
 \be
  \label{freeEmodels}
    \begin{split}
     {\cal N}=2^* ~\text{model}:&~~~~
     F_{S^4}^\text{univ} = N^2 \Big(- \frac{1}{4} \mu^4 - \frac{1}{12} \mu^6 + \mathcal{O}(\mu^8) \Big) \,,\\
     {\cal N}=1^* ~\text{one-mass model}:&~~~~
     F_{S^4}^\text{univ} = N^2 \Big(-0.2355 \mu^4 - 0.12 \mu^6 + \mathcal{O}(\mu^8)\Big) \,,\\
     {\cal N}=1^* ~\text{equal-mass model}:&~~~~
     F_{S^4}^\text{univ} = N^2 \Big(-0.0433 \mu^4 + \mathcal{O}(\mu^6)\Big) \,,
  \end{split}
\ee
where $\mu$ means the following: in  ${\cal N}=2^*$, $\mu=\mu_1=\mu_2$ and $\mu_3=0$; in ${\cal N}=1^*$ one-mass $\mu_1=\mu_2=0$ and $\mu=\mu_3\ne0$; and in ${\cal N}=1^*$ equal-mass $\mu=\mu_1=\mu_2=\mu_3\ne0$.

The three results for $\mathcal{O}(\mu^4)$ overconstrain the two coefficients $A_1$ and $A_2$ in \reef{Fmus}, but there is a consistent solution given by $A_1\approx -0.346$ and $A_2\approx0.1105$. This is good consistency check on our analyses. 

We are not able to determine $B_3$ since we did not have sufficiently good precision in our numerics to compute the $\mu^3$-term in $\frac{\pa^3 F}{\pa \mu^3}$ in the equal-mass model. However, the $\mathcal{N}=2^*$ and one-mass-model results imply that 
$(B_1 + 4 B_2)= - 1/24$ and $(B_1 +  B_2)=  -0.12$, from which we find that 
$B_1 \approx -0.146$ and $B_2 \approx 0.026$.

\subsection{Is there a more general $\mathcal{N}=2^*$ flow?}
\label{sec:holoN2revisited}
As described in Section \ref{sec:truncs}, there is a consistent truncation of the 10-scalar model to a 6-scalar model compatible with $\mathcal{N}=2$ supersymmetry.  It has a scalar
$\phi_3 =-\phi_4 \ne 0$ dual to an $\mathcal{N}=2$ gaugino condensate and contains also the five-dimensional dilaton. Moreover, the scalar $\alpha_3$ flows non-trivially, but in such a way that it has no source; its vev-rate $v_3$ would  correspond to the vev of the boson bilinear of the scalar in the $\mathcal{N}=2$ vector multiplet. These statements are all about the UV parameters. 

This 6-scalar model is a generalization of the 3-scalar model $\mathcal{N}=2^*$  \cite{Bobev:2013cja}, which it contains. One may ask if this extended holographic dual of $\mathcal{N}=2^*$ on $S^4$ gives other flows than the ones in \cite{Bobev:2013cja}. It would be puzzling if such more general flows exist with smooth boundary conditions in the IR\@. At least they should not give new results for the free energy: after all, the free energy of $\mathcal{N}=2^*$ on $S^4$ was calculated via supersymmetric localization in the field theory, and matched with the 3-scalar model in \cite{Bobev:2013cja}. 

We have constructed numerical flows for the $\mathcal{N}=2^*$ 6-scalar model. When imposing the smooth boundary conditions \reef{IRmetric} in the IR, one finds in the UV that the vev rate $v_3$ of the scalar $\alpha_3$ vanishes and so does the vev-rate $w$ of the scalar $\phi_3 =-\phi_4$ dual to the gaugino bilinear. Finally one finds that $v(\mu)$ exactly matches the function found in  \cite{Bobev:2013cja}, so the free energy is the same and therefore matches the localization result. We conclude that with smooth cap-off to $\mathbb{R}^5$ in the IR, the 6-scalar model does not contain new physics compared to the 3-scalar model for $\mathcal{N}=2^*$ on $S^4$ studied in \cite{Bobev:2013cja}.

\section{Summary, discussion, and outlook}
\label{sec:discuss}

In this work we explored properties of the $\mathcal{N}=1^*$ gauge theory on a round $S^4$ using holography.  Let us now summarize our main results and list a number of open questions.  

We have explicitly constructed, as a consistent truncation of 5d $\mathcal{N}=8$ gauged supergravity, a model with 10 scalars that is the holographic dual of $\mathcal{N}=1^*$ on $S^4$ at large $N$ and large 't Hooft coupling $\lambda$. It describes the situation where the complex masses of the three chiral multiplets in $\mathcal{N}=1^*$ are taken to be equal to their conjugates,  $m_i = \tilde{m}_i$, $i=1,2,3$. 

The 10-scalar supergravity model includes the $\mathcal{N}=2^*$ model analyzed previously in \cite{Bobev:2013cja} as well as two  $\mathcal{N}=1^*$ limits that we study in particular detail, namely the one-mass model where two of the masses vanish and the equal-mass model where the three masses are equal. In each of the latter two models, we construct numerical RG flows dual to the $\mathcal{N}=1^*$ theory on $S^4$, imposing as the IR boundary condition that the metric caps off smoothly and the scalar fields do not diverge.  This is a natural IR boundary condition since the $S^4$ provides an IR cutoff of the physics in the field theory. 

Via careful holographic renormalization, we compute the free energy numerically as a function of the dimensionless parameter, $ma$, for the one-mass and the equal-mass limits of $\mathcal{N}=1^*$. Since the known techniques for supersymmetric localization do not apply for theories with less than $\mathcal{N}=2$ supersymmetry on $S^4$, our approach is currently the only available insight into the behavior of the universal part $\frac{d^3 F_{S^4}}{d (ma)^3}$ of the $S^4$ free energy at strong coupling.  The numerical results are displayed in Figures \ref{fig:LS} and \ref{fig:vVSmuEqMass}. 

When the masses $m_i$ are small compared to the scale set by the radius  $a$ of the $S^4$, i.e.~$m_i a \ll 1$, it follows from the limits studied that the general result for the universal part of the free energy is
\be
   \label{Funiv2}
    F_{S^4}
    \,=\, (\ldots) 
    -0.346 N^2 a^4 \Big( m_1^4 + m_2^4 + m_3^4 \Big)
    +
    0.1105 N^2 a^4 \Big(m_1^2 + m_2^2 + m_3^2 \Big)^2
    + \mathcal{O}(m^6 a^6) \,,
\ee 
when $m_i = \tilde{m}_i$. In \eqref{Funiv2}, the ellipses denote terms with lower powers of $m_i$ that are not universal.  On the field theory side, the numerical factors encode integrated correlation functions that can in principle be calculated in conformal perturbation theory. This however is no trivial task as it would involve integrated 4-point functions, suitably regularized. We have not attempted such a calculation. Using our numerical results we were able also to extract two of the three independent coefficients in the  $\mathcal{O}(m^6 a^6)$ terms in \eqref{Funiv2}. To reproduce these coefficients from conformal perturbation theory one will have to compute integrated 6-point functions.

In addition to the solutions with real and purely imaginary masses shown in Figures \ref{fig:LS} and \ref{fig:vVSmuEqMass}, we have also constructed holographic RG flows with general complex values of the mass. However, we have not been able to access the full complex plane. For example, for the one-mass model, we find evidence for a third order pole at $\mu =  \pm 1$ (recall that $\mu = \pm \im m a$). If this is truly an isolated pole, one would expect to be able to find solutions close to the real $\mu$-axis also at values beyond this pole. In our approach we seem to be restricted in finding numerical solutions to a region that does not get close to the real axis for  $|\text{Re}\mu|>1$. It is not clear if this restriction is a parameterization problem of our scalar manifold or whether it hides potentially interesting physics. This issue requires further investigation, and we hope to come back to it in the future.

An interesting feature of the equal-mass model is that the flow of the dilaton and of the scalar dual to the gaugino bilinear cannot be consistently set to zero. This novel feature is in contrast with the situation in flat space, where in the GPPZ flow \cite{Girardello:1999bd,Pilch:2000fu} the dilaton is decoupled, and the scalar  
dual of the gaugino bilinear is optional and controlled by an arbitrary parameter, in conflict with the field theory.  However, the GPPZ flow is also singular in the IR, so it is not entirely clear how to interpret it in field theory. This problem was addressed by Polchinski and Strassler, who showed how to resolve the physically acceptable singularities by uplifting the supergravity solution to ten dimensions \cite{Polchinski:2000uf}. Our flows with $S^4$ radial slicing offer an alternative resolution of the flat space singularities: the five-dimensional solutions are smooth everywhere, and, rather than being arbitrary, the gaugino condensate is fixed by the masses as 
$\<\Tr\,( \lambda  \lambda + \tilde{\lambda}  \tilde{\lambda}) \> = \frac{4}{\pi^2} m_1 m_2 m_3 N^2$.  Note that we find this result to be valid for all masses, not just in the small- or large-mass limit.  

We also compute results for expectation values of other bilinear operators. Specifically, we find $\<\Tr\, F_{\mu\nu} F^{\mu\nu} \> = 0$. And at large purely imaginary masses, $|m_i a| \gg 1$, we find  $\<\Tr ( Z_i^2 + \tilde Z_i^2) \> \propto N^2 m_2 m_3  \, \text{sgn}(\im  m_i)$. Note for the purpose of the subsequent discussion that these expectation values are independent of the gauge coupling. 
 
It follows from the shift symmetry in the 5d dilaton, $\varphi \to \varphi + \text{constant}$, that no physical quantity computed in our model will depend non-trivially on the dilaton UV source-term $s$. The analysis in Appendix \ref{app:uplift} shows via the uplift to Type IIB in 10d that $s$  encodes the UV value of the  Yang-Mills coupling
\be
  \label{gYMs}
  \frac{(1+s)^2}{(1 - s)^2}  = \frac{4\pi}{g_\text{YM}^2} \,.
\ee
The supergravity model has two field redefinition symmetries which help in understanding this formula.  The first is $(z_i, \tilde{z}_i) \to (1/z_i, 1/\tilde{z}_i)$, which has the effect of mapping $s \to 1/s$ while leaving \reef{gYMs}, as well as any other physical observable, invariant. This symmetry of the bulk theory implies that the inside, ($z_i \tilde{z}_i < 1$), and outside, ($z_i\tilde{z}_i > 1$), of the Poincar\'e disk target space of the scalars $(z_i, \tilde{z}_i)$ describes the same physics. More interesting is the second symmetry $(z_i, \tilde{z}_i) \to (-z_i, -\tilde{z}_i)$ under which $s \to -s$. This symmetry amounts to $4\pi/g_\text{YM}^2 \to g_\text{YM}^2 /(4\pi)$, which is an incarnation of S-duality.\footnote{Note that the supergravity approximation is valid as long as $N$ is large; the supergravity limit need not be taken as a 't Hooft limit, but is valid more generally, including for finite value of $g_\text{YM}$ since large $N$ is sufficient to ensure that the 't Hooft coupling is large.}

The fact that our results for the physical observables do not depend on $g_\text{YM}$ --- or, more generally, on the complexified gauge coupling $\tau$ --- appears to be a non-trivial statement from the field theory perspective. We should emphasize that this is not a general statement about the observables in the $\mathcal{N}=1^*$ theory: it applies only to physical quantities in $\mathcal{N}=1^*$ theory on $S^4$ at large $N$ and large 't Hooft coupling. In contrast, it is argued in the field theory by the authors of \cite{Aharony:2000nt,Dorey:2000fc,Dorey:2001qj} that the gaugino condensates $\< \Tr\, \lambda\lambda\>$ and the bosonic bilinear $\<\Tr\, ZZ\>$ of $\mathcal{N}=1^*$ in {\em  flat space} depend holomorphically on $\tau$ and transform as modular forms under an $SL(2, \mathbb{Z})$ action. This property relies on the rich vacuum structure of $\mathcal{N}=1^*$ in flat space, which was described in detail by Polchinski and Strassler \cite{Polchinski:2000uf}. Under  $SL(2, \mathbb{Z})$, the vacua are permuted. Hence the formulas of \cite{Aharony:2000nt,Dorey:2000fc,Dorey:2001qj} for the chiral condensates also depend explicitly on the choice of vacuum. When the $\mathcal{N}=1^*$ theory is put on $S^4$, the vacuum structure is likely lifted,\footnote{What `vacuum structure'  means is also a little subtle since all we can talk about are the stationary fixed points of the Euclidean action.} but in the limit $ma \to \infty$ one could hope to be able to make contact with the flat space results of \cite{Aharony:2000nt,Dorey:2000fc,Dorey:2001qj}. The holographic analysis performed above predicts that the chiral condensates in the large $S^4$-radius limit (and at large $N$ and large 't Hooft coupling) are independent of $\tau$. It would be interesting to understand these statements from a field theory perspective.

Our analysis of the field theory shows that the free energy can only depend on certain combinations of the masses, namely 
\be 
  \label{msymsAgain}
  m_i \tilde{m}_i ~~\text{for each $i = 1, 2, 3$} \,, 
  ~~~\text{and}~~~
  m_1 m_2 m_3\,,
  ~~~\text{and}~~~
  \tilde{m}_1 \tilde{m}_2 \tilde{m}_3\,.
\ee
Also, it must be invariant under the $\mathbb{Z}_3$-symmetry that cyclically exchanges the masses.  The 10-scalar model has the symmetry $(z_i, \tilde{z}_i) \to (-z_i, -\tilde{z}_i)$ which along with the S-duality discussed above  implies that the free energy computed in this model must be even as a function of the masses (see Section \ref{sec:num}). This excludes odd-powers of $m_1 m_2 m_3$ or  $\tilde{m}_1 \tilde{m}_2 \tilde{m}_3$ which does not seem to be a restriction in the field theory (unless one can invoke modular invariance in some form). To investigate this further, one could use a supergravity model with $m_i \ne \tilde{m}_i$. Indeed, as we discussed in Section \ref{sec:SG}, the 10-scalar supergravity model is part of a larger consistent truncation with 18 scalars which would be needed to describe $\mathcal{N}=1^*$ on $S^4$ with $m_i \ne \tilde{m}_i$. 
This model in turn has a consistent truncation dual to the $\mathcal{N}=1^*$ theory with $m_1 = m_2 = m_3$ and $\tilde{m}_1 = \tilde{m}_2 = \tilde{m}_3$. It contains 8 real scalars which parametrize the coset space $G_{2(2)} / SO(4)$. The 8 scalars encode the fermion bilinears $\sum_{i=1}^3 \Tr\, \chi_i \chi_i$ and $\sum_{i=1}^3 \Tr\, \tilde{\chi}_i \tilde{\chi}_i$, the bosonic chiral operators $\sum_{i=1}^3 \Tr\, Z_i^2$ and $\sum_{i=1}^3 \Tr\, \tilde{Z}_i^2$, the gaugino condensates $\<\Tr\, \lambda \lambda\>$ and $\<\Tr\, \tilde{\lambda} \tilde{\lambda} \>$, and $\Tr\, F_{\mu\nu} F^{\mu\nu}$ and $\eps^{\mu\nu\rho \sigma}\Tr\, F_{\mu\nu} F_{\rho \sigma}$ (i.e.~encoding the couplings $\tau$ and $\tilde\tau$). 
The 8-scalar model was discussed in Section 4 of \cite{Pilch:2000fu}, where a 4-scalar subsector of it was constructed explicitly. 
It would be interesting to construct solutions of the full 8-scalar model for the purpose of understanding better if the free energy can depend on the mass combinations $m_1 m_2 m_3$ and $\tilde{m}_1 \tilde{m}_2 \tilde{m}_3$ as well as $\tau$ and $\tilde\tau$. 

To extract field theory results from our supergravity solutions we need to be able to fix finite counterterms using supersymmetry to complete the holographic renormalization procedure. It is clear from this analysis, as well as from the recent discussions in \cite{Cassani:2015upa,Assel:2015nca,Assel:2014tba,Cassani:2014zwa}, that it would be useful to have a first-principles holographic renormalization scheme which implements supersymmetry. We circumvented this technical obstacle by employing the Bogomolnyi trick.  However, this trick is valid in flat space and may involve subtleties when the results are adopted to curved space. Thus, a more systematic treatment is highly desirable, and it constitutes a relevant open problem in holography.

 An interesting future direction is also the construction of the full type IIB uplifts of the five-dimensional solutions presented in this paper. One should be able to construct these ten-dimensional solutions using the formulas derived in \cite{Lee:2014mla, Baguet:2015sma}, but due to the reduced global symmetry preserved in the field theory, we expect the 10d backgrounds to be of considerable complexity. Nevertheless, their explicit form will allow for holographic calculations of some quantities, such as expectation values of line and surface operators, that are not accessible from the five-dimensional solutions constructed here. An explicit form of the ten-dimensional uplift will also allow for a comparison with the solutions studied by Polchinski and Strassler \cite{Polchinski:2000uf} and may offer some insight into the vacuum selection mechanism when the $\mathcal{N}=1^*$ theory is placed on $S^4$. 
 
As mentioned before, there are currently no techniques for exact calculation of the supersymmetric partition function of an $\mathcal{N}=1$ supersymmetric theory on $S^4$. This was one reason we studied the $\mathcal{N}=1^*$ theory on $S^4$: it is an opportunity for holography to yield insights into field theory observables for which we have no other means to access. One can of course also take an approach to test holography in cases where field theory results are available, such as it was done for $\mathcal{N}=2^*$  on $S^4$ in \cite{Bobev:2013cja}. Such tests may be feasible for $\mathcal{N}=1$ theories on other manifolds, such as $S^3\times S^1$ or $S^2\times T^2$. The $\mathcal{N}=1^*$ theory, with its rich physics and well-behaved UV limit, naturally lends itself to such future tests.

\section*{Acknowledgments}
We would like to thank Dan Freedman for early collaboration on part of this work. We are also grateful to Marco Baggio, Davide Cassani, Thomas Dumitrescu, Fridrik Gautason, Marios Hadjiantonis, Finn Larsen, Zohar Komargodski, Juan Maldacena, M\'ark Mezei, Matt Strassler, Joe Polchinski, Balt Van Rees, and Ran Yacoby for useful discussions. We would also like to thank each other's institutions for kind hospitality during various stages of this project.

The work of NB is supported in part by the starting grant BOF/STG/14/032 from KU Leuven, by the COST Action MP1210 The String Theory Universe, and by the European Science Foundation Holograv Network. 
HE is supported in part by NSF CAREER Grant PHY-0953232. HE is a Cottrell Scholar from the Research Corporation for Science Advancement. 
UK acknowledges partial support from the Research Corporation for Science Advancement. 
TO was supported by NSF Graduate Research Fellowship under Grant \#F031543.
SSP is supported by the US NSF under Grant No.~PHY-1418069.

\appendix

\section{Details of the $\mN=1^*$ field theory on $S^4$}
\label{app:FT}
The full Lagrangian of the massive ${\cal N}=1^*$ theory on the 4-sphere can be written as the following sum
\begin{equation}
\mL = \mL_{\text{kinetic}}
+\mL_{\text{2}}
+\mL_{\text{Yukawa}}
+\mL_{\text{3}}
+\mL_{\text{4}}\,.
\end{equation}
The kinetic terms and the mass terms are respectively given by 
 \es{LagAppendix}{
   \mathcal{L}_{\text{kinetic}}  &= 
     \frac{1}{4 g_\text{YM}^2} \left(  F_{\mu\nu}^{a}  \right)^2 
    + \frac{\theta}{ 16 \pi^2} \epsilon^{\mu\nu\rho\sigma} F_{\mu\nu}^a F_{\rho \sigma}^a \\
   &\qquad{}- \lamt^{a\; T}_{i} \sigma_2 \bar{\sigma} ^{\mu} D_{\mu} \lam_i^a 
   + D^{\mu}\Zt ^a _i D_{\mu} Z ^a _i 
   - \chit^{a\; T}_{i} \sigma_2 \bar{\sigma} ^{\mu} D_{\mu} \chi_i^a \,, \\
 \mathcal{L}_{\text{2}} &=
\mt_i m_i \Zt_i^a Z_i^a
 -\frac{1}{2} m_i \left(  \chi^{a\; T}_i \s2 \chi_i^a  \right)
 - \frac{1}{2} \mt_i \left(  \chit^{a\; T}_i \s2 \chit_i^a  \right) \\ 
   &\qquad {}+ \frac{2}{a^2} \Zt _i^a  Z_i^a
  \pm\, \frac{\im}{2a} \left[  m_i Z_i^a Z_i^a +\mt_i \Zt_i^a \Zt_i^a  \right] 
 \,.
 }
We refer the reader to \cite{Bobev:2013cja} for details about how to write the $\mN=4$ SYM Lagrangian in the $\mN=1$ formulation and the modifications needed to put it on a 4-sphere.  The first of those is simply the conformal mass associated with the curvature of sphere; this also arises when a conformal theory is placed on the conformally flat sphere. The second term is induced from the mass term in the superpotential, as described in Section \ref{sec:FT}.

The Yukawa terms and the quartic interaction terms are the same as in the $\mN=4$ theory,
\begin{eqnarray}
\nonumber
\mathcal{L}_{\text{Yukawa}} 
&=& \sqrt{2} g_\text{YM} f^{abc} \bigg[
\big( \lam^{a\; T} \sigma_2 \chi^b_i \big) \Zt ^c_i
+ \big( \tilde{\lam}^{a\; T} \sigma_2 \tilde{\chi}^b_i \big) Z ^c_i
\\
&&\hspace{1.6cm}+\frac{1}{2} \ep_{ijk}
\big( \chi_i^{a\; T} \sigma_2 \chi^b_j \big) Z ^c_k
+\frac{1}{2} \ep_{ijk}
\big( \tilde{\chi}_i^{a\; T} \sigma_2 \tilde{\chi}^b_j \big) \tilde{Z}^c_k
\bigg] 
\\
\mathcal{L}_{\text{4}} &=& 
\frac{g_\text{YM}^2 }{2} f^{abc}f^{ade}
\left[
-\Zt _i^b Z_i^c \Zt _j^d Z_j^e
+2 \Zt_j^b \Zt_i^c Z_j^d Z_i^e
\right],
\end{eqnarray}
but  cubic interaction terms are induced by the masses, 
\begin{equation}
\mathcal{L}_{\text{3}} =
-\frac{g_\text{YM}}{\sqrt{2}} f^{abc}\ep_{ijk} \left[
\mt_i \Zt_i^a Z_j^b Z_k^c +m_i Z_i^a \Zt_j^b \Zt_k^c
\right].
\end{equation}

The theory is invariant under the following $\mathcal{N}=1$ supersymmetry transformations
\begin{equation}
\begin{aligned}
\delta Z_i^a  &= - \chi_i^{a \; T} \sigma_2 \ep \,, 
&
\delta \Zt_i^a  &=  - \chit_i^{a \; T} \sigma_2 \ept \,,
\\
\delta \chi_i^a &=   \sigma^{\mu} D_{\mu}Z_i^a  \ept   + \left( F_i^a  { \pm\frac{\im}{a} Z_i^a} \right)  \ep \,,
&
\delta \chit_i^a  &=   \sigmab^{\mu} D_{\mu}\Zt_i^a  \ep   +  \left(   \Ft_i^a  { \pm\frac{\im}{a} \Zt_i^a}  \right) \ept \,,
\\
\delta F_i^a  &=   D_{\mu} \chi_i^{a \; T} \s2 \sigma^{\mu} \ept   +   \sqrt{2}  g_\text{YM} f^{abc}   Z_i^b  (\lamt^{c\; T} \s2 \ept) \,,
&
\delta \Ft_i^a  &= D_{\mu} \chit_i^{a \; T} \s2 \sigmab^{\mu} \ep   +   \sqrt{2}  g_\text{YM} f^{abc}  \Zt_i^b  (\lam^{c\; T} \s2 \ep)   \,,
\\
\delta A_{\mu}^a  &=  \frac{1}{\sqrt{2}}   \left[   \lamt^{a\; T} \s2 \sigmab_{\mu} \ep  +\lam^{a\; T} \s2 \sigma_{\mu} \ept \right]   \,, 
&
\delta D^a &=   \frac{\im}{\sqrt{2}}   \left[  D_{\mu} \lam^{a \; T} \s2 \sigma^{\mu} \ept     -  D_{\mu} \lamt^{a \; T} \s2 \sigmab^{\mu} \ep          \right]  \,, 
\\
\delta \lam^a  &=   - \frac{1}{\sqrt{2}}  \left[   \frac{1}{2} \sigma^{[ \mu}  \sigmab ^{\nu ] } F_{\mu\nu}^a  +\im D^a        \right]  \ep \,, 
&
\delta \lamt^a  &=   - \frac{1}{\sqrt{2}}  \left[   \frac{1}{2} \sigmab^{[ \mu}  \sigma ^{\nu ] } F_{\mu\nu}^a  - \im D^a        \right]  \ept \,, 
\\
\end{aligned}
\end{equation}
where $\ep$ is a Killing spinor on $S^4$
\begin{eqnarray}
\nabla_{\mu}\ep =  { \pm \frac{\im}{2a} \sigma_{\mu}\ept} \,,
\hspace{1.5cm}
\nabla_{\mu}\ept =  { \pm \frac{\im}{2a} \sigmab_{\mu}\ep}\,.
\end{eqnarray}

The $U(1)^2$ mass symmetries described in \eqref{TransfSubgroup} arise as follows. We rescale the masses as
\be
  m_i \to e^{-2 \im \alpha_i} m_i 
  ~~~~\text{and}~~~~
  \tilde{m}_i \to \tilde{m}_i e^{2 \im \alpha_i}\,.
\ee
When $\alpha_1 + \alpha_2 + \alpha_3 = 0$, this can be compensated by the field redefinition
\be
   \begin{split}
   Z_i \to e^{\im \alpha_i} Z_i\,,~~~~~
   &  \tilde{Z}_i \to e^{-\im \alpha_i} \tilde{Z}_i \,,
   \\
   \chi_i \to e^{\im \alpha_i} \chi_i\,, ~~~~~
   &  \tilde{\chi}_i \to e^{-\im \alpha_i} \tilde{\chi}_i \,,
   \end{split}
\ee
while leaving the gauge and gaugino fields $A_\mu$ and $\lambda$ invariant. The result is that the free energy can only depend on the mass-combinations \reef{msyms}.

\section{Truncating the five-dimensional $\mathcal{N}=8$ $SO(6)$-gauged supergravity}
\label{app:SG}

Here we present some details on the truncation of the five-dimensional $\mathcal{N}=8$ $SO(6)$ supergravity to a 10 scalar model of Section~\ref{LORENTZIAN}. We follow the conventions and notation of \cite{Gunaydin:1985cu} to derive the Lorentzian theory.  We also present more details on the derivation of the BPS equations for this model in both Lorentzian and Euclidean signature.

\subsection{Truncation}

The ${\cal N} = 8$ gauged supergravity theory is described in \cite{Gunaydin:1985cu} in terms of $42$ scalar fields, $15$ gauge fields, $12$ two-form fields, the metric, $8$ symplectic Majorana gravitini, and $48$ symplectic Majorana spin-$1/2$ fields.  Here we will only be concerned with a subset of the scalars, the metric, and the supersymmetry variations of the gravitini and of the spin-$1/2$ fields.

The $42$ scalars parameterize the coset space $E_{6(6)} / USp(8)$.  In \cite{Gunaydin:1985cu}, an element of $E_{6(6)}$ is written as the matrix exponential of a $27 \times 27$ matrix that represents an algebra element of $E_{6(6)}$ as acting in the fundamental representation.  An element of the coset $E_{6(6)} / USp(8)$ is then represented also as a matrix exponential of a $27 \times 27$ matrix that not only lies within the $E_{6(6)}$ Lie algebra but is also orthogonal to the $USp(8)$ subalgebra.   Such an algebra element can be parameterized by the scalar fields $\Lambda^I{}_J$ (symmetric traceless in $I,J$), $\Lambda^\alpha{}_\beta$ (symmetric traceless in $\alpha, \beta$), and $\Sigma_{IJK\alpha}$ (anti-symmetric in $I, J, K$ obeying the self-duality condition $\Sigma_{IJK\alpha} = \frac 16 \epsilon_{\alpha\beta} \epsilon_{IJKLMN} \Sigma_{LMN\beta}$), where $I, J = 1, \ldots 6$ are $SL(6)$ indices, raised and lowered with the Kronecker symbol, and $\alpha, \beta = 1, 2$ are $SL(2, \R)$ indices raised and lowered with the two-index Levi-Civita symbol.  Here, $SL(6) \times SL(2, \R)$ is a subalgebra of $E_{6(6)}$, and $\Lambda^I{}_J$ and $\Lambda^\alpha{}_\beta$ are precisely the non-compact generators of $E_{6(6)}$ that belong also to $SL(6) \times SL(2, \R)$;  the $\Sigma_{IJK\alpha}$ are the non-compact generators of $E_{6(6)}$ that do not belong to $SL(6) \times SL(2, \R)$.  

The bulk gauge symmetry is the $SO(6)$ subgroup of $SL(6)$, and, according to the AdS/CFT dictionary, this gauge symmetry should be identified with the global $SO(6)_R$ symmetry of the boundary ${\cal N} = 4$ SYM theory.  Consequently, we have the following (schematic) identification between fields and operators built out of the $6$ adjoint scalars $X_I$, fermions $\lambda_{A\alpha}$ and $\tilde \lambda^{A\dot{\alpha}}$, and gauge field strength $F_{\mu\nu}$ of the ${\cal N} =4$ SYM theory:
\begin{table}[h]
\begin{center}
\begin{tabular}{c|c|c}
 bulk field & operator (schematic) & $SO(6)_R$ \\
 \hline
 $\Lambda^I{}_J$ & $\Tr  \left[ X^I X_J - \frac 16 \delta^I_{J} X^K X_K \right] $ & ${\bf 20}'$ \\
 $\Lambda^\alpha{}_\beta$ & $\Tr \left[ F_{\mu\nu} F^{\mu\nu} \right] , \ \epsilon^{\mu\nu\rho\sigma} \Tr \left[ F_{\mu\nu} F_{\rho\sigma}\right]$ & ${\bf 1}$ \\
 $\Sigma_{IJK1} + \im \Sigma_{IJK2}$ & $\Tr \left[ C_{IJK}^{AB} \lambda_A \lambda_B + \frac 16 \epsilon_{IJKLMN} X_L[X_M, X_N] + \im X_I[X_J, X_K]  \right]$ & ${\bf 10}$ \\
 $\Sigma_{IJK1} - \im \Sigma_{IJK2}$ & $\Tr \left[ \tilde C^{IJK}_{AB} \tilde \lambda^A \tilde \lambda^B + \frac 16 \epsilon_{IJKLMN} X_L[X_M, X_N] - \im X_I[X_J, X_K]  \right]$ & $\overline{\bf 10}$ \\
\end{tabular}
\caption{Scalar fields of ${\cal N} = 8$ gauged supergravity theory, their dual operators in ${\cal N} = 4$ SYM, as well as their transformation properties under the $SO(6)_R$ R-symmetry. \label{OperatorTable}} 
\end{center}
\end{table}
Note that because of the self-duality property of $\Sigma_{IJK\alpha}$, the combinations $\Sigma_{IJK1} + \im \Sigma_{IJK2}$  ($\Sigma_{IJK1} - \im \Sigma_{IJK2}$) are imaginary self-dual (ISD) and imaginary anti-self-dual (IASD), respectively, with respect to the $IJK$ indices.  The Clebsch-Gordan coefficients $C_{IJK}^{AB}$ ($\tilde C^{IJK}_{AB}$) transform from the symmetric product of two ${\bf 4}$'s (two $\overline{\bf 4}$'s) of $SU(4)_R$ to the ISD (IASD) product of three ${\bf 6}$'s of SO(6).

We take the relation between the fields used in Appendix~\ref{app:FT} to describe the ${\cal N} = 4$ SYM theory and those in the table above to be
 \es{RelationX}{
  Z_1 &= \frac{X_1 + \im X_2}{\sqrt{2}} \,, \qquad \tilde Z_1 = \frac{X_1 - \im X_2}{\sqrt{2}} \,, \\
  Z_2 &= \frac{X_3 + \im X_4}{\sqrt{2}} \,, \qquad \tilde Z_2 = \frac{X_3 - \im X_4}{\sqrt{2}} \,, \\
  Z_3 &= \frac{X_5 + \im X_6}{\sqrt{2}} \,, \qquad \tilde Z_3 = \frac{X_5 - \im X_6}{\sqrt{2}} \,, \\
  \chi_{i\alpha} &= \lambda_{i \alpha} \,, \qquad\qquad \tilde \chi_i^{\dot \alpha} = \lambda^{i \dot \alpha} \,, \\
  \lambda_\alpha &= \lambda_{4 \alpha} \,, \qquad\qquad \tilde \lambda^{\dot \alpha} = \lambda^{4 \dot \alpha} \,,
 }
where we suppressed all of the gauge indices.

To obtain the relevant supergravity truncation corresponding to the 18-scalar model in Section~\ref{sec:SG}, one can keep only the fields of the ${\cal N} = 8$ gauged supergravity theory that are invariant under a $(\Z_2)^3$ subgroup of $SO(6) \times SL(2, \R)$ and set to zero all the other fields.  Let us take the generators of $(\Z_2)^3$ to be products $P_1 Q$, $P_2 Q$, and $P_3 Q$ of $\Z_2$ generators $P_i$, $i=1, 2, 3$, that lie within $SO(6)$ and a $\Z_2$ generator $Q$ that lies within $SL(2, \R)$, as follows:  
 \es{PQ}{
P_1 &= \text{diag}\{ -1,-1,1,1,1,1\} \,, \\
P_2 &= \text{diag}\{1,1,-1,-1,1,1\}\,, \\
P_3 &= \text{diag}\{1,1,1,1,-1,-1\}\,, \\
Q &= \text{diag} \{ -1, -1\} \,.
 }
(Here, these generators are represented as matrices that act in the fundamental representation of $SO(6)$ for the $P_i$ and of $SL(2, \R)$ for $Q$.) The action of these generators leaves the following 8 scalars in the ${\bf 20}'$ invariant (the overall trace should be subtracted)
\es{20Invariants}{
\Lambda^1{}_1\,, \quad \Lambda^2{}_2\,, \quad &\Lambda^3{}_3\,, \quad \Lambda^4{}_4\,, \quad 
 \Lambda^5{}_5\,, \quad \Lambda^6{}_6\,, \quad \Lambda^{1}\,_{2}\,, \quad \Lambda^{3}\,_{4}\,, \quad \Lambda^{5}\,_{6} \,, \quad \sum_{I=1}^6 \Lambda^I{}_I  = 0 \,.
 }
In addition to that there are the following $8$ real scalars in  $\bf10 \oplus \overline{10}$
\begin{equation}
\Sigma_{135\alpha}\;, ~~~ \Sigma_{136\alpha}\;, ~~~\Sigma_{145\alpha}\;, ~~~\Sigma_{146\alpha}\;, ~~~\Sigma_{235\alpha}\;, ~~~\Sigma_{236\alpha}\;, ~~~\Sigma_{245\alpha}\;, ~~~\Sigma_{246\alpha}
\end{equation}
(obeying $\Sigma_{IJK\alpha} = \frac 16 \epsilon_{\alpha\beta} \epsilon_{IJKLMN} \Sigma_{LMN\beta}$), as well as the axion and dilaton $\Lambda^{\alpha}\,_{\beta}$.  As is mentioned in the main text and can be seen explicitly from the identification between bulk fields and operators presented above, the $18$ supergravity scalars invariant under $(\Z_2)^3$ correspond to the 14 operators\footnote{Supergravity modes do not capture the operator $\abs{Z_1}^2+\abs{Z_2}^2+\abs{Z_3}^2$.}  $\Tr\, \abs{Z_i}^2$, $\Tr\, \chi_i \chi_i$, $\Tr\, \tilde \chi_i \tilde \chi_i$, $\Tr \, Z_i^2$, $\Tr\, \tilde Z_i^2$ that participate in the mass deformation of the ${\cal N} = 4$ theory on $S^4$, as well as to the $4$ operators $\Tr\, F_{\mu\nu} F^{\mu\nu}$, $\epsilon^{\mu\nu\rho\sigma} \Tr\, F_{\mu\nu} F_{\rho\sigma}$, $\Tr\, \lambda \lambda$, and $\Tr\, \tilde \lambda \tilde \lambda$ that may acquire expectation values in the presence of this mass deformation.

After imposing the $(\Z_2)^3$ invariance above at the level of the full ${\cal N} = 8$ gauged supergravity theory, one can show that the invariant sector consists of an $\mathcal{N}=2$ supergravity theory coupled to two vector multiplets and four hyper multiplets.  There are two real scalars in the two vector multiplets and $16=4\times 4$ real scalars in the four hypers, for a total of $18$ scalars as discussed above.  Explicit computation shows that the scalar manifold of this theory is that given in \eqref{M18}, namely
\begin{equation}
\mathcal{M}_{18} = \left[SO(1,1)\times SO(1,1)\right] \times \frac{SO(4,4)}{SO(4)\times SO(4)}\,.
\end{equation}

In the present work we use a further truncation of this $18$ scalar model to $10$-scalars and less by only studying holographic RG flows with $m_i = \tilde m_i$.  On the supergravity side, one can truncate the $18$-scalar model further by keeping all the fields that are invariant under a further $\Z_2$ parity symmetry of the supergravity theory generated by $P_4 Q'$, with
 \es{P4Qp}{
P_{4}=\text{diag}\{1,-1,1,-1,1,-1\}\,, \qquad
 Q' = \text{diag}\{1, -1\} \,.
 }
The careful reader may notice that the matrix $P_4$ above does not lie within the $SO(6)$ symmetry group of the ${\cal N} = 8$ supergravity theory, but it lies instead in $O(6)$;  similarly, $Q'$ also does not lie within the $SL(2, \R)$ symmetry group of the ${\cal N} = 8$ supergravity theory, but in $GL(2)$.  It is thus not clear whether the $\Z_2$ action generated by $P_4 Q'$ is in fact a symmetry of the gauged ${\cal N} = 8$ supergravity theory, whose symmetry group is usually stated as $SO(6) \times SL(2, \R)$.  However, as explained in \cite{Pilch:2000fu}, the symmetry group of ${\cal N} = 8$ gauged supergravity is actually $\left[ O(6) \times SL^\pm(2, \R) \right] / \Z_2 $, where $SL^\pm(2, \R)$ is the subgroup of $GL(2, \R)$ matrices with determinant $\pm 1$, and where the $\Z_2$ in the denominator requires the determinants of the $O(6)$ group element written as a $6\times 6$ matrix and that of the $SL^\pm(2, \R)$ matrix to be equal.  The action of $P_4 Q'$ therefore generates a symmetry of the ${\cal N} = 8$ theory, and can thus be used to truncate to the sector of fields invariant under it.\footnote{It would be interesting to understand if the action of $P_4 Q'$ also represents a symmetry of the dual field theory at finite $N$ and Yang-Mills coupling, or whether it is only an approximate symmetry in the large $N$ and strong coupling limit where the supergravity description is accurate.}

Keeping the fields of the $18$-scalar model that are invariant under $P_4 Q'$, one is left with the following 5 scalars in the ${\bf 20}'$ (the overall trace should be subtracted)
 \es{20Left}{
   \Lambda^1{}_1\,, \quad \Lambda^2{}_2\,, \quad &\Lambda^3{}_3\,, \quad \Lambda^4{}_4\,, \quad 
 \Lambda^5{}_5\,, \quad \Lambda^6{}_6\,, \qquad \sum_{I=1}^6 \Lambda^I{}_I  = 0 \,,
 }
4 scalars in $\bf 10 \oplus \overline{10}$
\begin{equation}\label{4Sigma}
\Sigma_{1351} = -\Sigma_{2462}\,, \qquad \Sigma_{1461} = -\Sigma_{2352}\,, \qquad \Sigma_{2361} = -\Sigma_{1452}\,, \qquad \Sigma_{2451} =- \Sigma_{1362} \,,
\end{equation}
as well as the components 
\begin{equation}\label{axdildef}
\Lambda^{1}\,_1 = -\Lambda^{2}\,_2 \;,
\end{equation}
of the axion-dilaton $\Lambda^\alpha{}_\beta$.  All other components of $\Lambda^I{}_J$, $\Lambda^\alpha{}_\beta$, and $\Sigma_{IJK\alpha}$ are set to zero.  Thus we are left with a truncation with 10 scalars.  As we will see, an explicit computation of the scalar kinetic term using the formalism of ${\cal N} = 8$ gauged supergravity shows that the scalar manifold is
\begin{equation}
\mathcal{M}_{10} =  \left[SO(1,1)\times SO(1,1)\right] \times \left[\ds\frac{SU(1,1)}{U(1)} \right]^4\,.
\end{equation}

In order to have an explicit parameterization of the scalar manifold, let us write the non-zero components of $\Lambda^I{}_J$ as
 \es{ScalarDefs}{
  \Lambda^1{}_1 &= \bar \alpha_1 + \beta_1 + \beta_2 \,, \\
  \Lambda^2{}_2 &= -\bar \alpha_1 + \beta_1 + \beta_2 \,, \\
  \Lambda^3{}_3 &= \bar \alpha_2 + \beta_1 - \beta_2 \,, \\
  \Lambda^4{}_4 &= -\bar \alpha_2 + \beta_1 - \beta_2 \,, \\
  \Lambda^5{}_5 &= \bar \alpha_3 -2 \beta_1  \,, \\
  \Lambda^6{}_6 &= -\bar \alpha_3 -2 \beta_1  \,;
 }
the non-zero components of $\Sigma_{IJK\alpha}$ as
 \es{SigmaComp}{
  \Sigma_{1351} &= -\Sigma_{2462} = \frac 12 \left(\bar \phi_1 + \bar \phi_2 + \bar \phi_3 - \bar \phi_4 \right) \,, \\
  \Sigma_{1461} &=-\Sigma_{2352} = \frac 12 \left(-\bar \phi_1 + \bar \phi_2 + \bar \phi_3 + \bar \phi_4 \right) \,, \\
  \Sigma_{2361} &=-\Sigma_{1452} = \frac 12 \left(\bar \phi_1 - \bar \phi_2 + \bar \phi_3 + \bar \phi_4 \right) \,, \\
  \Sigma_{2451} &=-\Sigma_{1362} = \frac 12 \left(\bar \phi_1 + \bar \phi_2 - \bar \phi_3 + \bar \phi_4 \right) \,;
 }
and the non-zero components of $\Lambda^\alpha{}_\beta$ as
 \es{LabComp}{
  \Lambda^1{}_1 = -\Lambda^2{}_2 = \bar \varphi \,.
 }

The change of variables
\begin{equation}
\label{physfieldsderiv}
\begin{split}
\bar \alpha_1 + \im \bar \phi_1 &= \frac{1}{4}(r_1 e^{\im \zeta_1}+r_2 e^{\im \zeta_2} +r_3 e^{\im \zeta_3} +r_4 e^{\im \zeta_4})\,, \\
\bar \alpha_2 + \im \bar \phi_2 &= \frac{1}{4}(r_1 e^{\im \zeta_1}-r_2 e^{\im \zeta_2} +r_3 e^{\im \zeta_3} -r_4 e^{\im \zeta_4})\,, \\
\bar \alpha_3 + \im \bar \phi_3 &= \frac{1}{4}(r_1 e^{\im \zeta_1}+r_2 e^{\im \zeta_2} -r_3 e^{\im \zeta_3} -r_4 e^{\im \zeta_4})\,, \\
\bar \varphi - \im \bar \phi_4 &= \frac{1}{4}(r_1 e^{\im \zeta_1}-r_2 e^{\im \zeta_2} -r_3 e^{\im \zeta_3} +r_4 e^{\im \zeta_4})
\end{split}
\end{equation}
followed by
\begin{equation}
z_j = \tanh \left(r_j/2\right) e^{\im\zeta_j }\;,
\end{equation}
yields the Lorentzian signature Lagrangian given in Eqs.~\eqref{bulkLag}--\eqref{Ppot} in the main text.  The derivation of this Lagrangian follows from a careful evaluation of the formulas presented in \cite{Gunaydin:1985cu}.

It is worth noting that close to the asymptotic boundary, the fields defined in \eqref{ScalarDefs}--\eqref{LabComp} approach those defined in \eqref{ztophysUV}, namely $\bar \alpha_i \approx \alpha_i$, $\bar \phi_i \approx \phi_i$, and $\bar \varphi \approx \varphi$ to the first non-vanishing order, but this relation ceases to be true away at higher orders in the UV expansion.  From the dictionary between supergravity fields and field theory operators in Table~\ref{OperatorTable}, the correspondence in \eqref{Scalars} follows.


\subsection{Supersymmetry variations: Lorentzian signature}
\label{app:susyvarL}

In the notation of \cite{Gunaydin:1985cu}, the supersymmetry variations of the 5d $\mathcal{N}=8$ theory, parameterized by variation parameters $\epsilon^a$, with $a = 1, \ldots, 8$ a $USp(8)$ index, take the form
\begin{equation}\label{5dN8susy}
\begin{split}
\delta\psi_{\mu a} &= \nabla_{\mu}\epsilon_a + Q_{\mu a}\;^{b}\epsilon_{b} - \frac{g}{6} W_{ab} \gamma_{\mu}\epsilon^{b}\,, \\
\delta \chi_{abc} &= \sqrt{2} \left[\gamma^{\mu}P_{\mu abcd} - \frac{g}{2} A_{dabc}\right] \epsilon^{d}\,,
\end{split}
\end{equation}
where the indices $g = 2/L$ and where the various quantities appearing on the right-hand side can be computed from \cite{Gunaydin:1985cu} in terms of the scalar fields in \eqref{ScalarDefs}--\eqref{LabComp}.  Note that these are variations derived assuming Lorentzian (mostly minus) signature. 

The matrix $W_{ab}$ has 4 sets of complex conjugate eigenvalues $\{\lambda_{1,2,3,4},\bar{\lambda}_{1,2,3,4}\}$. Only one of them has a holomorphic structure and can be written as
\begin{equation}
\lambda_{4} = e^{\mathcal{K}/2} \mathcal{W}\,,
\end{equation}
with ${\cal K}$ and ${\cal W}$ given in \eqref{Kahler} and \eqref{suppotdef}, respectively.  Restricting $\epsilon^a$ to lie within the space spanned by the eigenvectors of $W_{ab}$ with eigenvalues $\lambda_4$ and $\bar \lambda_4$, we obtain the supersymmetry variations given in \eqref{Vars}.

\subsection{Supersymmetry variations: Euclidean signature}
\label{app:susyvarE}

As explained at the beginning of Section~\ref{EUCLIDEAN}, the SUSY variations can be continued to Euclidean signature.  Upon using the $SO(5)$-invariant Ansatz in  \eqref{MetricAnsatz1}--\eqref{ScalarsR}, one finds that the vanishing of the SUSY variations of the spin-$1/2$ fields in this case reduce to 
\be
 \label{matrixBPS1}
 \left(
   \begin{array}{ll}
    \im \pa_r z^a 
    & -\frac{1}{2} e^{\mathcal{K}/2}\mathcal{K}^{a\bar b} \nabla_{\bar b} \widetilde{\mathcal{W}}
    \\[2mm]
    \frac{1}{2} e^{\mathcal{K}/2}\mathcal{K}^{a\bar b} \nabla_{a} {\mathcal{W}}~~~
    &
    \im \pa_r \tilde{z}^{\bar b} 
    \\[2mm]
    3\im \pa_r \beta_1 
    & -\frac{1}{4} e^{\mathcal{K}/2} \pa_{\beta_1} \widetilde{\mathcal{W}}
    \\[2mm]
    \frac{1}{4} e^{\mathcal{K}/2} \pa_{\beta_1} {\mathcal{W}}
    & 3\im \pa_r \beta_1 
    \\[2mm]
    \im \pa_r \beta_2
    & -\frac{1}{4} e^{\mathcal{K}/2} \pa_{\beta_2} \widetilde{\mathcal{W}}
    \\[2mm]
    \frac{1}{4} e^{\mathcal{K}/2} \pa_{\beta_2} {\mathcal{W}}
    & \im \pa_r \beta_2
   \end{array}
 \right)
 \left(
   \begin{array}{l}
   \varepsilon_1 
   \\ 
   \varepsilon_2
    \end{array}
 \right)
 ~=~
 0\,.
\ee 

In order to write the condition that the spin-$3/2$ SUSY variations vanish, we need to define the vielbein $e^5 = e^r$ and $e^i = e^{A(r)}\hat{e}_i$, where $\hat{e}_i$ is the vielbein for the unit 4-sphere. The spin connection is
$\omega^{ij}=\hat{\omega}^{ij}$ and $\omega^{i5}=-\omega^{5i} = A' e^i$. It then follows that
\be
  \nabla_\mu \varepsilon_1 = \pa_\mu \varepsilon_1 + \frac{1}{4} \omega_\mu^{ab} \gamma_{[a} \gamma_{b]}\varepsilon_1
  ~~~\implies~~~
  \left\{ 
  \begin{array}{l}
    \nabla_r \varepsilon_1 = \pa_r \varepsilon_1 \,, \\[1mm]
    \nabla_i \varepsilon_1 = \hat{\nabla}_i \varepsilon_1 + \frac{1}{2} A' e^A \gamma_{[i} \gamma_{5]}\varepsilon_1\,,
  \end{array}  
  \right.
\ee
and similarly for $\varepsilon_2$.  Taking $\varepsilon_1 = a(r) \zeta_{\pm}$ and $\varepsilon_2 = b(r) \zeta_{\pm}$, where $\zeta_{\pm}$ is a Killing spinor of the 4-sphere, satisfying the Killing equation 
$\hat{\nabla}_i \zeta_{\pm} = \pm \gamma_5 \hat{\gamma}_i \zeta_{\pm}$, we then obtain that the vanishing of the spin-$3/2$ variations in the $S^4$ directions take the form
\be
 \label{matrixBPS2}
  \left(
     \begin{array}{ll}
         A' e^A \mp 1 
         & -\frac{\im}{3} e^{\mathcal{K}/2} \widetilde{\mathcal{W}} e^A
         \\
         -\frac{\im}{3} e^{\mathcal{K}/2} \mathcal{W} e^A
         & - A' e^A \mp 1 
     \end{array}
  \right)
   \left(
   \begin{array}{l}
   \varepsilon_1 
   \\ 
   \varepsilon_2
    \end{array}
 \right)
 ~=~
 0\,.
\ee

The SUSY variations \reef{matrixBPS1} and \reef{matrixBPS2} must hold simultaneously, and that requires that all $2 \times 2$ minors of the matrices in these two sets of questions vanish.  The vanishing of these minors reduces to the BPS equations \eqref{BPSAll}.

\section{Uplift of the dilaton-axion to type IIB supergravity}
\label{app:uplift}

Given a solution of the five-dimensional maximal gauged supergravity of \cite{Gunaydin:1984qu,Pernici:1985ju,Gunaydin:1985cu} one can find the dilaton $\Phi_{10}$ and axion $C_{10}$ of the corresponding IIB supergravity solution by using the uplift formulas in \cite{Pilch:2000ue,Pilch:2000fu}.  The interest in the 10d dilaton-axion comes from the fact that it is the 10d complex field
 \es{tauIIB}{
  \tau_\text{10} = C_{10} + \im e^{-\Phi_{10}}
 }
that can be identified with the complexified gauge coupling 
 \es{tauApproach}{
  \tau = \frac{\theta}{2 \pi} + \frac{4 \pi \im }{g_\text{YM}^2} 
 }
of the gauge theory.   Indeed, for a 10d background that approaches $AdS_5 \times S^5$ asymptotically, the AdS/CFT dictionary states that the boundary value of the field $\tau_{10}$ equals the UV gauge coupling $\tau$;  Away from the space-time boundary, $\tau_{10}$ captures some information about the RG flow of the complexified gauge coupling.  

To be concrete about our normalizations, we take the type IIB two-derivative supergravity action in Einstein frame to be
 \es{IIBSUGRA}{
  S_\text{IIB} = \frac{1}{16 \pi G_{10}} \int d^{10}x\, \sqrt{g} \left[ R_{10} - \frac{\partial_\mu \tau_{10} \partial^\mu \tilde \tau_{10}}{2 (\text{Im}\, \tau_{10})^2} - \frac{1}{480} F^{(5)}_{\mu\nu\rho\sigma \tau} F^{(5)\mu\nu\rho\sigma\tau}+ \ldots \right] \,,
 }
where $G_{10}$ is the 10d Newton constant, related to the 5d Newton constant $G_5$ via $G_{10} = G_5 / (\pi^3 L^5)$ (with $G_5 = \pi L^3 / (2 N^2)$), $R_{10}$ is the Ricci scalar in 10 dimensions, and $F_{(5)}$ is the self-dual five-form of type IIB supergravity.  The ellipses in \eqref{IIBSUGRA} denote the kinetic terms and interaction terms involving the other fields of type IIB supergravity.  Instead of using the complex field $\tau_{10}$, which takes values in the upper half-plane, it is sometimes convenient to use a field $B_{10}$ that takes values in the interior of the unit disk.  The two fields are related by:
 \es{BDef}{
  \tau_{10} = \im \frac{1-B_{10}}{1+B_{10}} \qquad \Longleftrightarrow \qquad B_{10} = \frac{1+\im\tau_{10}}{1-\im\tau_{10}}  \,.
 }
In terms of $B_{10}$, \eqref{IIBSUGRA} becomes
  \es{IIBSUGRA2}{
  S_\text{IIB} = \frac{1}{16 \pi G_{10}} \int d^{10}x\, \sqrt{g} \left[ R_{10} - \frac{2 \partial_\mu B_{10} \partial^\mu \bar B_{10}}{(1 - \abs{B_{10}}^2)^2 } - \frac{1}{480} F^{(5)}_{\mu\nu\rho\sigma \tau} F^{(5)\mu\nu\rho\sigma\tau} + \ldots \right] \,.
 }

From \eqref{IIBSUGRA} and \eqref{IIBSUGRA2} we can see that 10d axion-dilaton parameterizes the hyperbolic space $\HH^2$, represented either as the upper-half plane in terms of $\tau_{10}$ in \eqref{IIBSUGRA} or as the Poincar\'e disk in terms of $B_{10}$ in \eqref{IIBSUGRA2}.  Since $\HH^2$ can be written as a coset, $\HH^2 = SU(1, 1) / U(1) = SL(2, \R) / SO(2)$, the axion-dilaton kinetic term can be derived from a coset space construction.  In the $SU(1, 1)/U(1)$ presentation of the coset, it is customary to represent the coset elements by $SU(1, 1)$ matrices
 \es{VParam}{
  V^{(\theta)} = \frac{1}{\sqrt{1 - \abs{B_{10}}^2}} \begin{pmatrix}
    1 & B_{10} \\
    \bar B_{10} & 1 
    \end{pmatrix}
     \begin{pmatrix} e^{\im \theta} & 0 \\
     0 & e^{-\im \theta} \end{pmatrix} \,,
 }
modulo the identification $V^{(\theta_1)} \sim V^{(\theta_2)}$.  The $SL(2, \R) / SO(2)$ presentation can be obtained by a conjugation with the matrix 
 \es{UMat}{
  U = \frac{1}{\sqrt{2}} \begin{pmatrix}
   1 & 1 \\
   \im & -\im
   \end{pmatrix} \,,
 }
so that the $SL(2, \R)/ SO(2)$ coset element is
 \es{SParam}{
  S^{(\theta)} = U V^{(\theta)} U^\dagger = 
   \frac{1}{2 \sqrt{ 1- \abs{B_{10}}^2}}
    \begin{pmatrix}
     2 + B_{10} + \bar B_{10} & \im (B_{10} - \bar B_{10} ) \\
     \im (B_{10} - \bar B_{10} ) & 2 - B_{10} - \bar B_{10}
    \end{pmatrix} 
    \begin{pmatrix}
     \cos \theta & \sin \theta \\
     -\sin \theta & \cos \theta
    \end{pmatrix} 
 }
defined up to the identification $S^{(\theta_1)} \sim S^{(\theta_2)}$.  One can of course choose to represent each coset element by a representative $V^{(\theta)}$ or $S^{(\theta)}$ with $\theta = 0$; the coset elements are parameterized by $B_{10}$ and $\bar B_{10}$.

In the 10d uplift of any 5d asymptotically-AdS background of ${\cal N} = 8$ gauged SUGRA, the 10d axion-dilaton $\tau_{10}$ is of course a function of the 5d coordinates parameterizing the asymptotically $AdS_5$ space (or $\HH^5$ in Euclidean signature) as well as of the other five coordinates parameterizing the internal space.  The uplift formulas \cite{Pilch:2000ue,Pilch:2000fu} give $\tau_{10}$ as a rather complicated function of the $42$ scalars of the gauged ${\cal N} = 8$ theory that parameterize $E_{6(6)}/USp(8)$.  The expression for $\tau_{10}$ simplifies, however, at the boundary of $AdS_5$, where most 5d scalars vanish.  Let us study this limit first before we present the more complicated formulas of \cite{Pilch:2000ue,Pilch:2000fu}.  

To get started, let us truncate the ${\cal N} = 8$ supergravity theory by setting to zero all the fields other than the metric and the 5d dilaton-axion $\Lambda^\alpha{}_\beta$, for which we choose the parameterization
 \es{DrasticTrunc}{
  \Lambda =  \begin{pmatrix}
   r \cos \zeta & -r \sin \zeta \\
   -r \sin \zeta & -r \cos \zeta
  \end{pmatrix} \,.
 } 
With the definition $B = e^{\im \zeta} \tanh r $, we have
 \es{eLam}{
  \exp \Lambda = \frac{1}{2 \sqrt{ 1- \abs{B}^2}}
    \begin{pmatrix}
     2 + B + \bar B & \im (B - \bar B ) \\
     \im (B - \bar B ) & 2 - B - \bar B
    \end{pmatrix} \,,
 }
and one can show using the formalism of \cite{Gunaydin:1985cu} that the 5d Lagrangian reduces to 
 \es{bulkLagDrastic}{
  \mathcal{L}_\text{5d} = \frac{1}{16 \pi G_5} \left[ R - \frac{2 \partial_\mu B \partial^\mu \bar B}{(1 - \abs{B}^2)^2 }   +\frac{12}{L^2} \right] \,.
 }
This is precisely the 5d Lagrangian one expects from the 10d action \eqref{IIBSUGRA2} if we set to zero all the fields other than the metric, the five-form, and the 10d dilaton-axion, and take the 10d metric and dilaton-axion to be 
 \es{B10ToB}{
   ds_{10}^2 = ds_5^2 + L^2 ds_{S^5}^2\,, \qquad F^{(5)} = \frac{4}{L} \left[ \text{vol}_5 + L^5 \text{vol}_{S^5} \right]  \,,  \qquad B_{10} = \frac{a B + b}{b^* B + a^*} \,,
 }
where $ds_{S^5}^2$ and $\text{vol}_{S^5}$ are the line element and the volume form on a five-sphere of unit radius, $ds_5^2$ and $\text{vol}_5$ are the 5d metric used in \eqref{bulkLagDrastic} and the corresponding volume form, $B$ depends only on the 5d coordinates, and $a$ and $b$ are arbitrary complex numbers obeying $a a^* - b b^* = 1$.  The numbers $a$ and $b$ parameterize an $SU(1, 1)\cong SL(2, \R)$ matrix\footnote{The matrix $\begin{pmatrix} a & b \\ b^* & a^* \end{pmatrix}$ is an $SU(1, 1)$ matrix.  Upon conjugation with $U$, one obtains an $SL(2, \R)$ matrix given by $\begin{pmatrix}
 \text{Re}\, (a+b) & \text{Im}\, (a - b) \\
 - \text{Im}\, (a + b) & \text{Re}\, (a - b)
\end{pmatrix}$.  } and encode a possible mismatch of $SL(2, \R)$ frames between the 5d and 10d supergravity theories.  To match the formulas in \cite{Pilch:2000ue,Pilch:2000fu}, we will henceforth take $B_{10} = -B$.  While the relation $B_{10} = - B$ (or $S^{(0)} = \exp \left[ - \Lambda\right]$ as follows from \eqref{eLam} and \eqref{SParam}) holds exactly in the consistent truncation containing only the metric and the dilaton-axion, this relation is in general only true when evaluated at the boundary of the asymptotically AdS space:
 \es{BdyRel}{
  \exp [- \Lambda]  \big\vert_\text{bdy} = S^{(0)} \big\vert_\text{bdy} \,.
 }

In the 10-scalar model, we have that the boundary value of $\Lambda^\alpha{}_\beta$ is
 \es{LBdy}{
  \exp [-\Lambda] \big\vert_\text{bdy} = \exp \begin{pmatrix}
     -\varphi \big\vert_\text{bdy} & 0 \\
     0 & \varphi \big\vert_\text{bdy}
   \end{pmatrix} = \begin{pmatrix}
     \frac{1-s}{1+s} & 0 \\
     0 & \frac{1+s}{1 - s}
   \end{pmatrix} \,,
 }
where we used $\varphi \big\vert_\text{bdy} = 2\, \text{arctanh}(s)$, as follows from the UV asymptotics in \eqref{UVexp2}.  From \eqref{BdyRel} and \eqref{SParam}, we find
 \es{BBdy}{
  B_{10} \big\vert_\text{bdy} = -\frac{2s}{1 + s^2} \,.
 }
Extracting the boundary value of $\tau_{10}$ from \eqref{BDef} and writing it in terms of the complexified gauge coupling $\tau$ of the gauge theory, we infer that in the 10-scalar model the relation between the parameter $s$ and the complexified gauge coupling is 
 \es{gtheta10}{
  \tau = \im \frac{(1+s)^2}{(1 - s)^2}  \,.
 }
In particular, this relation implies that in the 10-scalar model we have
 \es{gTosAppendix}{
  \frac{4\pi}{g_\text{YM}^2} = \frac{(1+s)^2}{(1 - s)^2} \,, \qquad \theta = 0 \,.
 }

Away from the boundary, one can identify the 10d axion-dilaton from
 \es{diluplift}{
 {\cal M}^{\alpha\beta} \equiv \left(S^{(\theta)} S^{(\theta)T}\right)^{\alpha\beta} = C \epsilon^{\alpha\gamma}\epsilon^{\beta\delta} \mathcal{V}_{I\gamma}^{ab}\mathcal{V}_{J\delta}^{cd} y^I  y^J\Omega_{ac}\Omega_{bd}\;,
 }
where $y^I$ are coordinates on $\mathbb{R}^6$ obeying $\sum_{I}(y^I)^2=1$, ${\cal V}_{I \alpha}^{ab}$ is a function of the scalars defined in  \cite{Gunaydin:1985cu}, $\Omega_{ab}$ is the symplectic form, and $C$ is determined from the condition that the expression on the RHS should be unimodular.  Note that the matrix ${\cal M} = S^{(\theta)} S^{(\theta)T}$ is equal to
 \es{MMatrix}{
   \mathcal{M} =
   \frac{1}{1-\abs{B_{10}}^2}
    \begin{pmatrix}
      (1+B_{10})(1+\bar B_{10}) & \im(B_{10}-\bar B_{10})  \\
       \im(B_{10}-\bar B_{10}) & (1-B_{10})(1-\bar B_{10})
    \end{pmatrix}
 }
and is by construction independent of the unphysical angle $\theta$ in \eqref{SParam}. 

Applying the formula \eqref{diluplift} to the equal mass model, we obtain
\begin{equation}\label{cMgeneral}
\begin{split}
\mathcal{M}^{11} &= \frac{1}{\xi}\dfrac{(1+z_2)(1+\tilde{z}_2)}{(1-z_2\tilde{z}_2)}\left[\dfrac{(1+z_2)(1+\tilde{z}_2)}{(1-z_2\tilde{z}_2)} u^2  + \dfrac{(1-z_1)(1-\tilde{z}_1)}{(1-z_1\tilde{z}_1)} v^2\right]\;, \\
\mathcal{M}^{12} &=  \mathcal{M}^{21} = \frac{1}{\xi}\dfrac{(\tilde{z}_2-z_2)}{(1-z_2\tilde{z}_2)}\left[\dfrac{(z_1-\tilde{z}_1)}{(1-z_1\tilde{z}_1)} + \dfrac{(z_2-\tilde{z}_2)}{(1-z_2\tilde{z}_2)} \right] (u\cdot v)\;, \\
\mathcal{M}^{22} &= \frac{1}{\xi}\dfrac{(1-z_2)(1-\tilde{z}_2)}{(1-z_2\tilde{z}_2)}\left[\dfrac{(1+z_1)(1+\tilde{z}_1)}{(1-z_1\tilde{z}_1)} u^2  + \dfrac{(1-z_2)(1-\tilde{z}_2)}{(1-z_2\tilde{z}_2)} v^2\right] \;.
\end{split}
\end{equation}
where we used the renaming of the coordinates $y^I$ as $y^I = (u_1, v_1, u_2, v_2, u_3, v_3)$.  Imposing that the determinant of this matrix is equal to 1 leads to an explicit expression for the warp factor $\xi$ which is quite lengthy so we will not present it here. 

As a check, we note that the boundary values of the $z_i$ and $\bar z_i$ are 
 \es{zBdy}{
  z_1 \big\vert_\text{bdy} = \tilde z_1 \big\vert_\text{bdy}
   = -z_2 \big\vert_\text{bdy} = -\tilde z_2 \big\vert_\text{bdy} = s \,.
 }
Plugging this expression into \eqref{cMgeneral}, we see that
 \es{MBdyEqual}{
  {\cal M} \big\vert_\text{bdy} = 
   \begin{pmatrix}
    \frac{(1-s)^2}{(1 + s)^2} & 0 \\
     0 & \frac{(1+s)^2}{(1 - s)^2}
   \end{pmatrix} \,,
 }
which is the square of the matrix on the RHS of \eqref{LBdy}, as it should be.

As a last comment, recall that the supergravity action and supersymmetry transformation rules are invariant under $z_j \to -z_j$ and $\tilde z_j \to -\tilde z_j$.  An interpretation of this symmetry is obtained when we combine it with a transformation of the internal coordinates $u_i$ and $v_i$,
 \es{SGMsym}{
   z_j \to -z_j \,, \qquad\qquad \tilde{z}_j \to -\tilde{z}_j \,, \qquad\qquad u_i \to v_i \,, \qquad\qquad v_i \to -u_i \,.
 }
Under \eqref{SGMsym}, the matrix in \eqref{cMgeneral} is \textit{not} invariant but undergoes the map ${\cal M} \to {\cal M}^{-1} $, which can also be represented as ${\cal M} \to Q'' {\cal M} (Q'')^T$ with
 \es{QppDef}{
  Q'' \equiv \begin{pmatrix}
   0 & 1 \\
   -1 & 0
  \end{pmatrix} \,.
 } 
Using \eqref{MMatrix} and \eqref{BDef}, we see that ${\cal M} \to {\cal M}^{-1}$ implies $B_{10} \to -B_{10}$ and $\tau_{10} \to -1/\tau_{10}$.  In \eqref{SGMsym}, this transformation is combined with a R-symmetry rotation by $90$ degrees in the $y^1 y^2$-, $y^3 y^4$-, and $y^5 y^6$-planes represented by the $SO(6)$ matrix 
 \es{SO6Matrix}{
  P_5 \equiv \begin{pmatrix}
   0 & 1 & 0 & 0 & 0 & 0 \\
   -1 & 0 & 0 & 0 & 0 & 0 \\
   0 & 0 & 0 & 1 & 0 & 0 \\
   0 & 0 & -1 & 0 & 0 & 0 \\
   0 & 0 & 0 & 0 & 0 & 1 \\
   0 & 0 & 0 & 0 & -1 & 0 
  \end{pmatrix} \,.
 }
The combined transformation $P_5 Q''$ commutes with the action of $P_i Q$ in \eqref{PQ} and $P_4 Q'$ in \eqref{P4Qp} used to construct the 10-scalar consistent truncation of the ${\cal N} =8$ supergravity theory.  In the field theory, \eqref{SGMsym} can be interpreted as a combined S-duality transformation and an R-symmetry rotation by the matrix \eqref{SO6Matrix}.

\section{Holographic renormalization}
\label{app:holoren}

\subsection{Hamiltonian approach to infinite counterterms}

One starts with an Ansatz for the divergent terms on-shell action 
\be 
  S_\text{on-shell,div} = \int d^4x \,\sqrt{\gamma} \, U \,,
\ee
where $\gamma_{ij} = \frac{1}{\rho} (g_0)_{ij} = e^{-2r} (g_0)_{ij} $ is the boundary metric, $g_0$ is the metric on a round $S^4$ with radius $1/2$, and  
$U$ is a function of the fields (generically denoted $\Phi^I$) and it can also have explicit dependence on $r$. 
The basic idea is to use the Hamilton-Jacobi equation 
\be 
   \frac{\partial S_\text{on-shell,div}}{\partial r} + H(\Phi^I,p_I, \gamma^{ij},p_{ij}) = 0\,,
\ee
where the canonical momenta are calculated as 
\be
  p_I =  \frac{1}{\sqrt{\gamma}} \frac{\delta S_\text{on-shell,div}}{\delta \Phi^I}\,,
  ~~~~~
  p_{ij}  =  \frac{\delta S_\text{on-shell,div}}{\delta \gamma^{ij}}
   = \frac{1}{4} \sqrt{\gamma} (K_{ij} - K \gamma_{ij})
  \,,
\ee
where $K_{ij}$ is the extrinsic curvature. 

For a theory with Lagrangian
\be
 \mathcal{L} = -\frac{1}{4}R + \frac{1}{2} G_{IJ} \partial_\mu \Phi^I \partial^\mu \Phi^J + V\,,
\ee
the Hamiltonian density can be written
\be
 \mathcal{H} = \frac{1}{4} \sqrt{\gamma} 
 \Big[ 
    K_{ij} K^{ij} - K^2 + R + 2 G^{IJ} p_I p_J 
    - 2 \gamma^{ij} \partial_i \Phi^I \partial_j \Phi^I 
    -4 V
 \Big]\,.
\ee

Suppose we just focus on the zero-derivative terms $U_0$ in the Ansatz $U$. Noting that we have a symmetry in our theory that takes $z_i \to - z_i$ and $\tilde z_i \to - \tilde z_i$, the result can only depend on even powers of the physical variables $\phi_i$, $\alpha_i$, and $\phi$. Moreover, our potential \reef{Sbulkexpanded} is symmetric in the permutations of the $\phi_i$'s or $\alpha_i$'s respectively. The cosmological term and the mass-terms are model independent, thus generically fixed, so this  leaves only three unfixed coefficients in our Ansatz:
\be
  \begin{split}
 U_0 =& -\frac{3}{2} - \frac{1}{2} \sum_{i=1}^4 \phi_i^2 
 - \bigg( 1+ \frac{1}{\log\rho} \bigg) \Big(  6\beta_1^2 +2 \beta_2 + \sum_{k=1}^3 \alpha_k^2 \Big) \\
 &
 + A_1  \sum_{i=1}^4 \phi_i^4 
 + A_2    \sum_{1\le  i<j\le 4} \phi_i^2 \phi_j^2 
 + A_3 \,\phi_1 \phi_2 \phi_3 \phi_4\,.
 \end{split}
\ee
The coefficients $A_{1,2,3}$ can be functions of $r$ as well as have dependence on the dilaton $\varphi$. Any higher powers of the fields would not correspond to divergent terms in the on-shell action. 

At zero-derivative order, one easily finds that the extrinsic curvature is simply fixed by $U_0$:   $K_0 = -\frac{8}{3} U_0$ and $K_0{}^i{}_j = \frac{1}{4} K_0 \delta^i{}_j$.
This means that at zeroth order in derivatives, the Hamilton-Jacobi equation becomes
\be
  \frac{8}{3} U_0^2  - 2 G^{IJ} p_I p_J 
  + 2 V - 2 \frac{\partial U_0}{\partial r} ~=~0\,,
\ee
with $p^I = dU_0/d\Phi^I$ and 
$\Phi^I = \{\beta_1,\beta_2,\alpha_1,\alpha_2,\alpha_3,\phi_1,\phi_2,\phi_3,\phi_4,\phi\}$.
To solve this equation, one sets the coefficient of each field to zero. This gives three constraints for the three unknown functions $A_{1,2,3}$:
\be
  \label{solveA1A2}
  \begin{array}{rclccrcl}
    2\displaystyle \frac{\partial A_1}{\partial r}  - \frac{8}{3}&=& 0  
    &\longrightarrow& A_1 &=& \displaystyle \frac{4}{3} r + \text{constant} \;,
    \\[4mm]
    2\displaystyle \frac{\partial A_2}{\partial r}  + \frac{8}{3}&=& 0  
    &\longrightarrow& A_2 &=& \displaystyle -\frac{4}{3} r + \text{constant} \;,
  \end{array}
\ee 
and $A_3 = 0$.
The  integration constant in \reef{solveA1A2} can be dropped since that would correspond to finite terms in the on-shell action and we are here  focusing only on the divergent ones. 
Thus we conclude that
\be
  A_1 = \frac{2}{3} \log\rho\,,~~~~~
  A_2 = -\frac{2}{3} \log\rho\,,~~~~~
  A_3 = 0\,.
\ee
One can similarly repeat this analysis for the 2-derivative terms and 4-derivative terms. The only possible Ansatz (assuming that the fields do not depend on the $S^4$-coordinates) is
\be
  U_2 = B_1  R[\gamma] + B_2 R[\gamma] \sum_{i=1}^4 \phi^2 \,,~~~~
  U_4 = C_1 R[\gamma]_{ij}R[\gamma]^{ij} + C_2 R[\gamma]^2\,.
\ee
One finds 
\be
  B_1 = -\frac{1}{8}\,,~~~~
  B_2 = -\frac{1}{24} \log\rho\,,~~~~
  C_1= \frac{1}{32}\,,~~~~
  C_2 = -\frac{1}{96}\,.
\ee
The results for $B_1$, $C_1$, and $C_2$ can be directly adopted from the generic case with no matter; the result for $B_2$ is also generic for a dimension-3 scalar. Putting everything together, the result is 
\be
  \begin{split}
   U = &
      - \frac{3}{2} - \frac{1}{8}R[\gamma] 
       - \frac{1}{2} \sum_{i=1}^4 \phi_i^2 
      - \bigg( 1+ \frac{1}{\log\rho} \bigg) 
         \Big(  6\beta_1^2 +2 \beta_2 + \sum_{k=1}^3 \alpha_k^2 \Big) 
 \\[1mm]
 &
   +\log\rho 
   \bigg[
     \frac{1}{32} \bigg(  R[\gamma]_{ij}R[\gamma]^{ij} - \frac{1}{3} R[\gamma]^2\bigg) 
     - \frac{1}{24} R[\gamma]  \sum_{i=1}^4 \phi_i^2 
     - \frac{2}{3}  \sum_{i=1}^4 \phi_i^4 
     + \frac{2}{3}  \sum_{1\le  i<j\le 4} \phi_i^2 \phi_j^2 
   \bigg]
  \end{split}  
\ee
All in all, we can conclude that the infinite counterterms for our analysis is then minus the result of the infinite counterterms found above. The result is given in \reef{Sctinf}.

\subsection{Bogomolnyi approach to finite counterterms}
The Bogomolnyi method for determining the finite counterterms is valid in flat space, so we take the limit of our 10-scalar model to flat space. As described in \reef{genGPPZ}, this limit is obtained as the consistent truncation 
\be 
   \label{truncz}
   \tilde{z}_i = - z_i\,,
\ee
which sets the three scalars $\alpha_{1,2,3}= 0$ and the dilaton $\varphi=0$.

The resulting model is described by the Euclidean bulk Lagrangian 
\be
  \mathcal{L} = -\frac{1}{4}R + 3(\partial \beta_1)^2 + (\partial \beta_2)^2
  + \frac{1}{2} G_{ij} \partial_\mu z^i \partial^\mu z^j + V\,,
\ee
where the scalar target space metric $G_{ij}$ is diagonal and can be read off after performing the truncation \reef{truncz} on the K\"ahler scalar term 
$ \frac{1}{2}\mathcal{K}_{\alpha\bar{\beta}}\partial_{\mu}z^{\alpha}\partial^{\mu}\bar{z}^{\bar{\beta}}$ from \reef{bulkLag}. The scalar potential $V$ is equal to our scalar potential $\mathcal{P}$ in \reef{Ppot} after applying \reef{truncz}.

For the bulk space metric Ansatz $ds^2 = dr^2 + e^{2A(r)} ds_{\mathbb{R}^4}^2$, the Ricci tensor is $R=-8A'' - 20 A'^2$ and $\sqrt{g} = e^{4A}$. After partial integration we can use this to replace $-\frac{1}{4}\sqrt{g} R$ in the action by $-3 (A') e^{4A}$.  With the radial Ansatz for all the scalars, the Lagrangian then takes the form
\be
  \sqrt{g}\mathcal{L} =e^{4A} \bigg\{ -3 (A')  + 3( \beta_1')^2 + ( \beta_2')^2
  + \frac{1}{2} G_{ij} z^i{}' z^j{}' + V \bigg\}\,.
\ee
Let us now suppose that there is a superpotential\footnote{Since we do not have complex variables, this is not a holomorphic superpotential, but can perhaps be thought of as a fake superpotential. As we shall see, it is nonetheless closely related to our actual superpotential.} $W$ which is related to the scalar potential $V$ by
\be
  \label{bogoVfromW}
  V= \frac{1}{12} (\partial_{\beta_1} W)^2
  + \frac{1}{4} (\partial_{\beta_2} W)^2
  +\frac{1}{2} G^{ij} \partial_{i} W\partial_{j} W 
  - \frac{4}{3} W^2\,.
\ee
Here $\partial_{i}= \partial/\partial z_i$.

The Lagrangian can then be written as
\be
  \begin{split}
  \sqrt{g}\mathcal{L} \,=\,&
  e^{4A} \bigg\{ -3 \Big(A' \pm \frac{2}{3} W\Big)^2  
  + 3\Big( \beta_1' \mp \frac{1}{6} \partial_{\beta_1} W \Big)^2 
  + \Big( \beta_2' \mp \frac{1}{2} \partial_{\beta_2} W \Big)^2
  \\
  &\hspace{1cm}
  + \frac{1}{2} G_{ij} \Big(z^i{}' \mp G^{ik} \partial_k W \Big)
      \Big(z^j{}'  \mp G^{jl} \partial_l W \Big) \bigg\}
 \pm \partial_r \Big( e^{4A} W\Big)   \,. 
  \end{split}
\ee
The Bogomolnyi trick is to extremize the action by setting each of the squared terms to zero. This gives BPS equations
\be
  \label{bogoBPS}
  A' = \mp \frac{2}{3} W\,,~~~~~
  \beta_1' =\pm \frac{1}{6} \partial_{\beta_1} W\,,~~~~~
  \beta_2' =\pm \frac{1}{2} \partial_{\beta_2} W\,,~~~~~
  z^i{}'  =\pm G^{ij} \partial_j W\,.
\ee
And it tells us what the boundary term is: 
$S_W = \int_\text{bdr} e^{4A} W= \int_\text{bdr} \sqrt{\gamma} W$ will be the counterterm action for the supersymmetric theory in flat space. It will include both infinite and finite counterterms.

Let us now compare with the BPS equations in our model. When the metric in the bulk is flat-sliced, the spin-3/2 supersymmetry variations  \reef{matrixBPS2} become
\be
  \label{matrixBPS2flat}
  \left(
     \begin{array}{ll}
         A' e^A 
         & -\frac{\im}{3} e^{\mathcal{K}/2} \widetilde{\mathcal{W}} e^A
         \\
         -\frac{\im}{3} e^{\mathcal{K}/2} \mathcal{W} e^A
         & - A' e^A
     \end{array}
  \right)
   \left(
   \begin{array}{l}
   \eps_1 
   \\ 
   \tilde{\eps}_2
    \end{array}
 \right)
 ~=~
 0\,.
\ee
This means that there is no longer an analytic solution for $e^{2A}$, but instead on has to handle the $A'$ BPS equation directly. Setting the determinant of the $2 \times 2$ matrix in \reef{matrixBPS2flat} to zero, we find 
\be
  \label{Aprimeflat1}
  A'^2 = \frac{1}{9}e^{\mathcal{K}} \mathcal{W} \widetilde{\mathcal{W}}\;.
\ee
Now in the truncation limit \reef{truncz}, we have $\mathcal{W} = \widetilde{\mathcal{W}}$ and $e^{\mathcal{K}} = \frac{1}{(1+z_1^2)(1+z_2^2)(1+z_3^2)(1+z_4^2)}$. Therefore we can write \reef{Aprimeflat1} as
\be
  \label{bogoW}
  A' = \pm \frac{2}{3} W\,,
  ~~~~~
  W = \frac{1}{2} e^{\mathcal{K}/2} \mathcal{W}  \bigg|_{\tilde{z}_i = - z_i}\;.
\ee
Thus we have identified the superpotential $W$ for the Bogomolnyi superpotential. It is  easy to verify that upon truncating with \reef{truncz}, the other BPS equations reduce to the gradient flow form
\reef{bogoBPS} and that one obtains the scalar potential $V$ from $W$ in \reef{bogoW} via the relation \reef{bogoVfromW}.

Performing the change of variables \reef{ztophysUV} (now with $\alpha_{1,2,3} = \varphi = 0$ due to the truncation; see the comments around \reef{genGPPZ}) and expanding near the UV boundary to keep terms only up to finite contributions, we find
\be
   \label{Wexpanded}
   \begin{split}
    W =& \,
    \frac{3}{2} + \frac{1}{2} \big( \phi_1^2 +  \phi_2^2  + \phi_3^2 \big) 
    + \frac{3}{2} \phi_4^2
    + 2 (3 \beta_1^2 + \beta_2^2) \\[2mm]
    & 
    -2 \beta_1 \big(\phi_1^2 +  \phi_2^2  -2 \phi_3^2 \big)
    - 2\beta_2 \big(\phi_1^2 -  \phi_2^2 \big) \\[2mm]
    & 
    +\frac{1}{3} \big( \phi_1^4 +  \phi_2^4  + \phi_3^4 \big) + \phi_4^4
    + 2\big( \phi_1^2 +  \phi_2^2  + \phi_3^2 \big) \phi_4^2
    + \big( \phi_1^2  \phi_2^2+  \phi_2^2 \phi_3^2 + \phi_3^2 \phi_1^2\big) 
    - 6\phi_1\phi_2\phi_3\phi_4\,.
    \end{split}
\ee
This is not symmetric in the four $\phi_i$'s, despite the scalar potential having this symmetry up to the same order (but not at higher orders), and there is a good reason for that. Consider the quadratic terms. In the  scalar potential, quadratic terms encode the mass of the scalar and this dictates the falloff near the boundary. The EOM are second order and given a mass term, there are therefore two possible falloff solutions: the source rate and the vev-rate. For a dimension-3 scalar in 5d, these are the $e^{-r}$ and $e^{-3r}$ falloffs, respectively.  The BPS equations are first order, so there the distinction between the two falloff solutions comes from the superpotential. Therefore the superpotential $W$ has two possible coefficients for the quadratic terms that give the {\em same} mass term in  $V$; this is easy to show in general. For a dimension-3 scalar $\phi$, the options are $\frac{1}{2}\phi^2$ or $\frac{3}{2}\phi^2$; the former gives the source rate, the latter gives the vev-rate. Our BPS equations `know' that $\phi_4$ cannot be sourced: it would break supersymmetry if all four fermion bilinears in the field theory were added to the Lagrangian as sources. So therefore the quadratic term in the Bogomolnyi superpotential $W$ for $\phi_4$ must be $\frac{3}{2} \phi_4^2$, because if the coefficient had been $1/2$ it would have allowed for a source for the gaugino condensate. From these arguments we can conclude that there must be an asymmetry between the four scalars $\phi_i$ already at the level of the quadratic terms in the superpotential. 

A few further comments on the form of the superpotential \reef{Wexpanded} are in order. Note that the terms that mix $\beta_{1,2}$ and the $\phi_i$'s mimic the identification of $\beta_{1,2}$ in the field theory (see \reef{Scalars}). These terms do not treat $\phi_{1,2,3}$ equally. This makes sense, because if we exchange $\phi_1$ and $\phi_2$, which are dual of the fermion bilinears, supersymmetry requires that we also exchange the dual  mass-terms for the field theory scalars, i.e.~$Z_1 \tilde Z_1$ and $Z_2 \tilde Z_2$. But by \reef{Scalars}, this amounts to changing the sign of $\beta_2$. Thus we see that the combination $\beta_2 \big(\phi_1^2 -  \phi_2^2 \big)$ that appears in $W$ is indeed invariant under $\{ \phi_1 \lra \phi_2, \beta_2 \to -\beta_2\}$ as it should be, but not just under $\phi_1 \lra \phi_2$. It is easy to check that indeed the full $W$ has the symmetry $\{ \phi_1 \lra \phi_2, \beta_2 \to -\beta_2\}$.

The conclusion of this somewhat lengthy analysis is that the finite counterterms are the ones that appear in the last line of \reef{Wexpanded}. These counterterms are the ones we present and use in the main text in \eqref{Sfinite}.

\section{IR expansions and numerics}
\label{app:numdetails}

In this appendix, we provide some details of the IR expansion and numerical analysis for the $\mathcal{N}=1^*$ models with one mass and three equal masses.

\subsection{One-mass model}
\label{sec:IRLS}

Imposing the smoothness condition \reef{IRmetric} on the IR solution of the one-mass model reduces the number of non-trivial integration constants to just one, $z_0$. Using the freedom to shift $r$ to set $r_0 = 0$, the IR expansion of the fields  is then 
\als{
\label{IRinLS}
z &= z_0 + z_1 r^2 +\mathcal{O}\left(r^4\right),\\[1mm]
\zt &=     - \frac{3 z_0^4+14 z_0^2-1}{z_0^4-14 z_0^2-3} z_0 \left[    1+ \frac{\left(9 z_0^8-292 z_0^6-458 z_0^4-292 z_0^2+9\right) }{5 z_0 \left(z_0^4-14 z_0^2-3\right) \left(3 z_0^4+14 z_0^2-1\right)}\, z_1 r^2+\mathcal{O}\left(r^4\right) \right],  \\[1mm]
e^{3\b} &= 
\sqrt{1- \frac{8 z_0^2}{z_0^4+ 6z_0^2+1}}
\left[ 1+ \frac{16 z_0 }{3  \left(z_0^4+6 z_0^2+1\right)} \frac{z_0^2+1}{z_0^2-1}  \,   z_1 r^2 +\mathcal{O}\left(r^4\right) \right],
}
where
\be
z_1 = -\frac{z_0 \left(z_0^2+1\right) \left(z_0^2+3\right) \left(3 z_0^2+1\right) \left(z_0^4+14 z_0^2+1\right)}{9 (1-z_0^2)^{1/3} \left(z_0^4+6 z_0^2+1\right)^{7/3}}\,.
\ee
The Euclidean fields, and hence the integration constants, are generally complex valued. However, if we restrict the IR input $z_0$ to the real or imaginary $z_0$-axes, all the fields --- and hence all UV constants --- become pure real or imaginary, respectively. 

\noindent {\bf Real axis}\\
On the real axis, the IR parameter  $z_0$ can be restricted to the range $-1<z_0<1$ via the symmetry $z \to 1/z$. This generates numerical solutions with $-1<\mu<1$. 

\noindent {\bf Imaginary axis}\\
For purely imaginary values of $z_0$, we only find solutions for  $-z_*< z_0/\im<z_*$, where $z_* = \sqrt{7 - 4 \sqrt{3}} \approx 0.2679\dots$. As $z \to \pm \im z_*$, the numerical solutions have $\mu/\im \to \pm \infty$. Thus in the allowed range, we generate flows with any value of $-\infty < \mu/\im < \infty$. The solutions with $-z_*<z_0/\im <   z_*$ are all non-trivial RG flows from Euclidean AdS in the UV to the smooth flat-space solution imposed in the IR. 

The values $\pm \im z_*= \pm \im \sqrt{7 - 4 \sqrt{3}}$ are roots of the 
polynomial $1 + 14 z_0^2 + z_0^4$ which appears in the numerator of $z_1$ in the IR expansion \reef{IRinLS}. (The two other roots are $\pm \im \sqrt{7 + 4 \sqrt{3}} \approx 3.73$.)
When $z_0= \pm \im z_*$, all $r$-dependence of all three scalars goes away and instead they take constant values
\be
  \label{LS0scalars}
  z = - \zt = \pm \im z_*
  ~~~\text{and}~~~
  e^{3\beta} = \sqrt{2} \,. 
\ee
The scalar potential also becomes constant, $V = - \frac{8\, 2^{1/3}}{3}$. (Recall that we are setting $L=1$.)
 Solving for the warp factor, one finds that over the entire range of $0< r < \infty$, it takes the form
\be
   e^{2A(r)} = L_\text{LS}^2 \sinh^2(r/L_\text{LS}) 
    ~~~\text{with}~~~L_\text{LS} = \frac{3}{2^{5/3}}\,.
\ee
Thus the solutions with $z_0 = \pm i z_*$ correspond to Euclidean AdS with radius $L_\text{LS}$. This is the dual of the well-known Leigh-Strassler fixed point \cite{Leigh:1995ep} that is reached in the IR of the RG flow of $\mathcal{N}=1^*$ with one mass turned on in flat space. (Its holographic RG flow was described in  \cite{Freedman:1999gp}.) 

\subsection{Equal-mass model}

To obtain the IR expansion of the equal mass model we again will assume that the space caps off smoothly as $r$ approaches some finite value $r_0$ which we are free to set to $0$. This is implemented by requiring that the metric function approaches $e^{2A} = r^2 + \dots$. Solving the BPS equation iteratively in the small-$r$ expansion  then yields
\begin{equation}
\label{IReqmass}
\begin{split}
z_1 &= -b_0 - \frac{3(a_0-b_0)(1-b_0^2)}{20(1-a_0b_0)} r^2 +\mathcal{O}(r^4)\;, \\
\tilde{z}_1 &= \frac{2b_0-a_0(3-b_0^2)}{1+2a_0b_0-3b_0^2} + \frac{3(1-a_0^2)(a_0-b_0)(1-b_0^2)^2}{4(1-a_0b_0)(1+2a_0b_0-3b_0^2)^2} r^2 +\mathcal{O}(r^4)\;, \\
z_2 &= a_0+ \frac{9(a_0-b_0)^3-(a_0-b_0)(1-a_0b_0)^2}{20(1-a_0b_0)(1-b_0^2)} r^2 +\mathcal{O}(r^4)\;, \\
\tilde{z}_2 &= b_0 + \frac{(a_0-b_0)(1-b_0^2)}{4(1-a_0b_0)} r^2 +\mathcal{O}(r^4)\;, \\
e^{2A} &= r^2  + \frac{(1-a_0^2)(1-b_0^2)}{3(1-a_0b_0)^2} r^4 + \mathcal{O}(r^6)\;.
\end{split}
\end{equation}

For $b_0=a_0$ one finds that the space is simply $H_5$, namely
\begin{equation}
z_2 = \tilde{z}_2 = -z_1 = -\tilde{z}_1 = a_0\;, \qquad e^{2A} = \sinh^2(r)\;.
\end{equation}
In this limit, $a_0$ parametrizes the constant value of the dilaton in the 5d $\mathcal{N}=8$ theory and it is clearly a zero mode.

\begin{figure}[th]
\centerline{
\includegraphics[width=8cm]{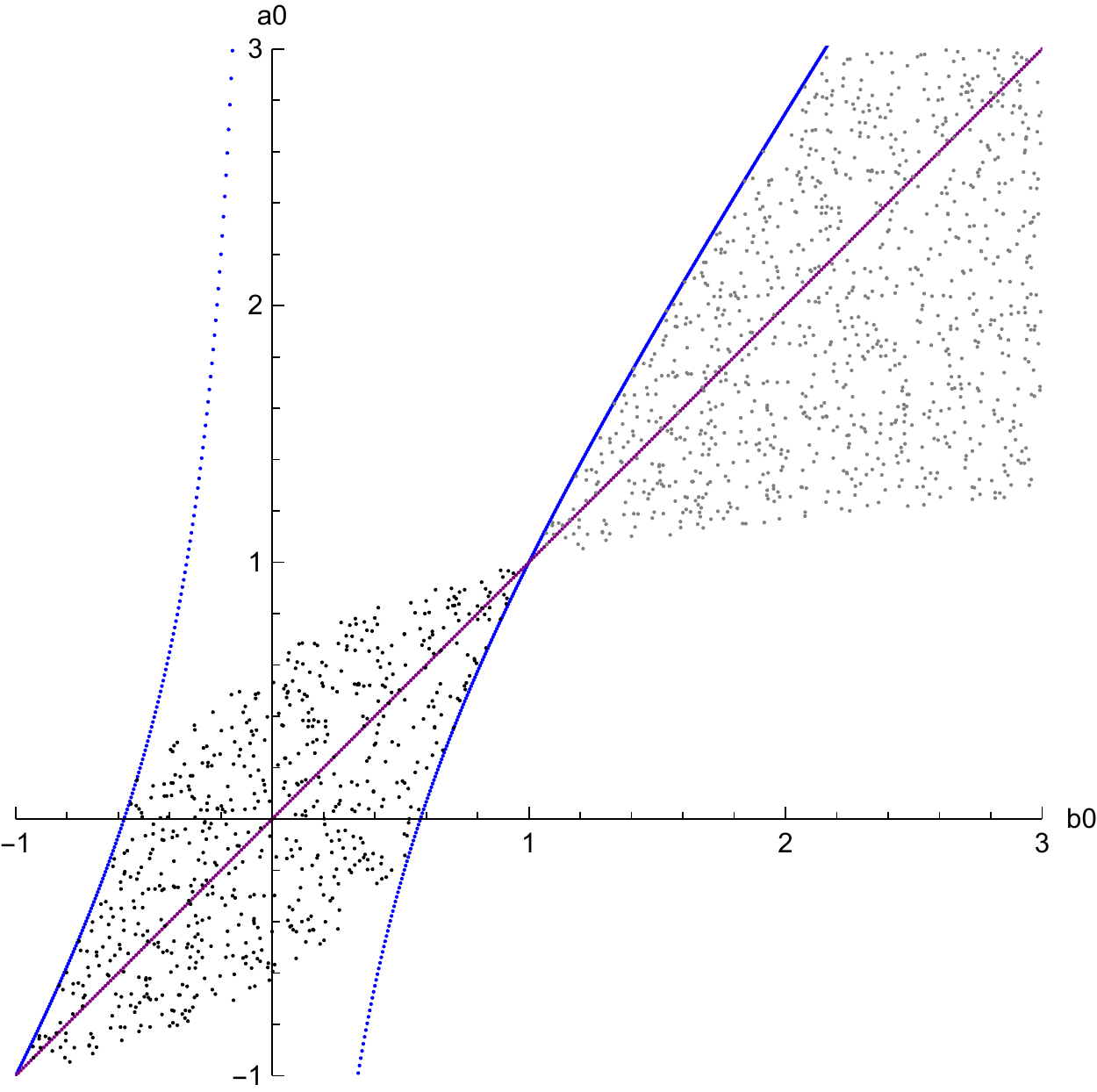}
\includegraphics[width=8cm]{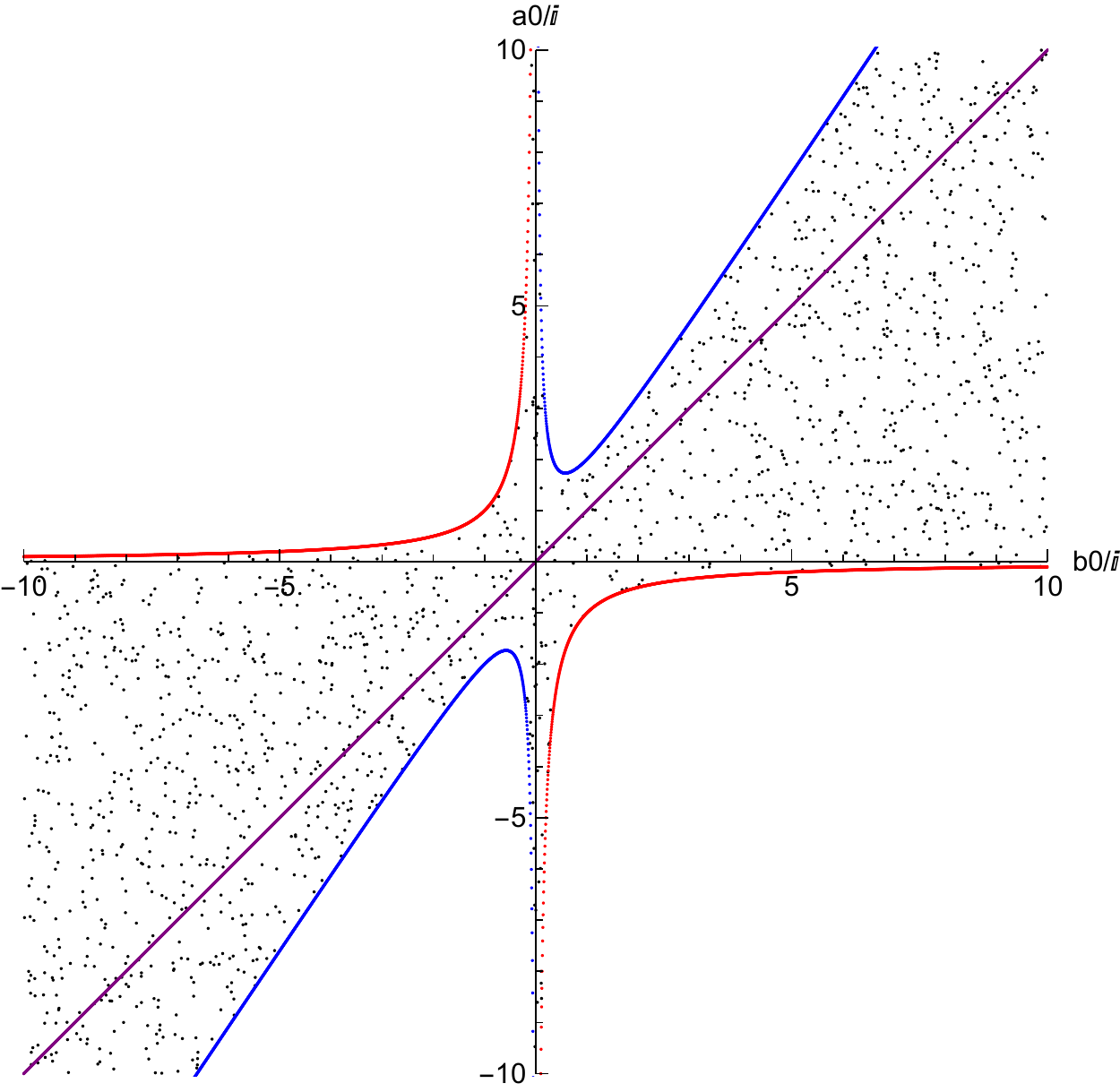}}
\caption{Scatter plots showing the regions in which we find solutions with smooth IR boundary conditions. On the left, we take real values for $a_0$ and $b_0$. The bounds from \reef{IRbnds} are the red and blue curves. Note that for real $a_0$ and $b_0$, there is another bound in play too: the region of smooth solutions is bounded by curves that correspond to $s=\pm1$; we do not know an analytic from of these curves in the $b_0,a_0$-plane.  The diagonal (purple) are all trivial Euclidean AdS solutions with $\mu=0$. In the plot on the left, the region with $-1< b_0,a_0<-1$ has $-1<s<1$, while the region with $b_0,a_0>1$ has $s<-1$; there is an equivalent region with $b_0,a_0<-1$ in which $s>1$. By the symmetry $(z_i,\zt_i) \to (1/z_i,1/\zt_i)$, the solutions with $-1<a_0,b_0<1$ are equivalent to those with $a_0,b_0 > 1$ or $<-1$.}
\label{fig:IRa0b0}
\end{figure}

Note that the IR expansion is divergent when a denominator in \reef{IReqmass} vanish. 
This happens if
\be
  \label{IRbnds}
  a_0 b_0 =1~~\text{(red)}
  ~~~~~\text{or}~~~~~
  a_0 = \frac{3b_0^2 -1}{2b_0}
  ~~\text{(blue)}\,.
\ee
The first case, $a_0 b_0 =1$ is equivalent to the scalars hitting the boundary of the Poincar\'e disk, $z_i \tilde{z}_i =1$ for $i=1$ or $2$. The second case is less intuitive. Numerically, we find that these conditions turn out to be bounds between existence and non-existence of solutions with the given smooth IR boundary conditions. This is illustrated for the cases of $a_0,b_0$ both real or both purely imaginary in Figure \ref{fig:IRa0b0}. In \reef{IRbnds}, ``red" and ``blue" refer to the color of these lines in the plots in Figure \ref{fig:IRa0b0}.

Since the dilaton decouples, only one parameter in the IR determines the UV variables $\mu$, $v$, and $w$; this way the latter two become functions of $\mu$. One can take the scatter data in the plots above and plot $v$ vs.~$\mu$ or $w$ vs.~$\mu$: in both cases one finds that the scatter plots collapse to curves, clearly shoving that the three parameters $\mu$, $v$, $w$ are independent of one of the two IR parameters. We  generate numerical solutions  efficiently by keeping $b_0=$fixed and letting $a_0$ vary between the bounds in the plots in Figure \ref{fig:IRa0b0}. The simplest is just to set $b_0=0$ and that is what we used to generate the plots in Section \ref{sec:holoEqM}.

Note that when $a_0, b_0$ are both real, then the fields $z_i,\tilde{z}_i$ are real throughout the flow, in particular $\mu, v, w, s$ are also real. When $a_0, b_0$ are both purely imaginary, then so are the $z_i,\tilde{z}_i$'s and hence also  $\mu, v, w, s$.

There is one technical point that we should comment on. The field transformation from $z_i,\tilde{z}_i$'s to the physical fields in the UV involve $\text{arctanh}({z}_i)$ and $\text{arctanh}(\tilde{z}_i)$. For real values of the fields, the arctanh function generates an imaginary value when any $z_i ,\tilde{z}_i >1$. For this reason, it is not useful to solve the BPS equations numerically in terms of the physical UV fields where the dilaton decouples directly; in fact doing so limits the $\mu$ UV parameter to lie in the range $-1.149 < \mu < 1.149$ since the IR value $\tilde{z}_1$ is only less than 1 for $\mu$ in that range. This is just a technical problem that we get around by solving the BPS equations in $z_i,\tilde{z}_i$-variables.

\bibliographystyle{ssg}
\bibliography{N1star}

\end{document}